\newif\ifconfver
\confverfalse      
\confvertrue        

\newif\ifonecoltab
\onecoltabtrue        

\newif\ifplainver  
\plainvertrue
\plainverfalse

\ifplainver
    \confverfalse                         
\fi

\ifconfver
     \documentclass[10pt,journal]{IEEEtran}
\else
    \ifplainver
        \documentclass[11pt]{article}
        \usepackage{fullpage}
    \else
        \documentclass[11pt,draftcls,onecolumn]{IEEEtran}
    \fi
\fi

\usepackage{calc,amsfonts,amssymb,amsmath,bm,url,color,theorem,graphicx,cite,epstopdf,nicefrac}
\usepackage{float}
\usepackage[algoruled,linesnumbered]{algorithm2e}
\usepackage{multirow}
\usepackage[normalem]{ulem}
\usepackage{framed}
\usepackage{amsfonts, amsmath, color, graphicx,framed,bbm,epstopdf}
\usepackage{amsmath,amssymb,bm,bbold}
\usepackage{stmaryrd,tikz,wrapfig}

\definecolor{orange}{RGB}{255,107,0}

\newtheorem{Def}{Definition}

\theorembodyfont{\rmfamily}

\definecolor{shadecolor}{RGB}{220,220,220}

\newcommand{\W}{\boldsymbol{W}}
\newcommand{\Z}{\boldsymbol{Z}}
\newcommand{\G}{\boldsymbol{G}}
\newcommand{\Q}{\boldsymbol{Q}}
\newcommand{\X}{\boldsymbol{X}}
\newcommand{\C}{\boldsymbol{C}}
\newcommand{\E}{\boldsymbol{E}}
\newcommand{\U}{\boldsymbol{U}}

\renewcommand{\H}{\boldsymbol{H}}
\newcommand{\M}{\boldsymbol{M}}
\newcommand{\A}{\boldsymbol{A}}
\newcommand{\B}{\boldsymbol{B}}
\newcommand{\x}{\boldsymbol{x}}

\renewcommand{\a}{\boldsymbol{a}}

\newcommand{\cone}[1]{\textsf{cone}\left\{#1\right\}}
\newcommand{\conv}[1]{\textsf{conv}\left\{#1\right\}}
\newcommand{\T}{{\!\top\!}}

\newcommand{\bOmega}{\bm{\varOmega}}
\newcommand{\bTheta}{\bm{\varTheta}}
\newcommand{\bM}{\bm{M}}

\newcommand{\bbR}{\mathbb{R}}

\DeclareMathOperator*{\minimize}{\textrm{minimize}}
\DeclareMathOperator{\rank}{rank}

\DeclareMathOperator*{\maximize}{\rm maximize}

\begin{document}

	\newcommand{\papertitle}{
		{Nonnegative Matrix Factorization for Signal and Data Analytics:}\\
		{[Identifiability, Algorithms, and Applications] }
	}
	
	\newcommand{\paperabstract}{
Nonnegative matrix factorization (NMF) has become a workhorse for signal and data analytics, triggered by its model parsimony and interpretability. Perhaps a bit surprisingly, the understanding to its model identifiability---the major reason behind the interpretability in many applications such as topic mining and hyperspectral imaging---had been rather limited until recent years. Beginning from the 2010s, the identifiability research of NMF has progressed considerably: Many interesting and important results have been discovered by the signal processing (SP) and machine learning (ML) communities. NMF identifiability has a great impact on many aspects in practice, such as avoiding ill-posed problem formulations and coming up with performance-guaranteed algorithms. 
This tutorial will help researchers and graduate students grasp the essence and insights of NMF identifiability, thereby avoiding typical `pitfalls' that are often times due to unidentifiable NMF formulations. This paper also aims at helping practitioners pick/design suitable factorization tools for their own problems.
	}

	
	\ifplainver
	
	\date{\today}
	
	\title{\papertitle}
	
	\author{
		Xiao Fu, Kejun Huang, Nicholas D. Sidiropoulos, and Wing-Kin Ma
		\thanks{X. Fu is with the School of Electrical Engineering and Computer Science, Oregon State University, Corvallis, OR 97331, USA. email: xiao.fu@oregonstate.edu. K. Huang is with the Department
			of Electrical and Computer Engineering, University of Minnesota, Minneapolis,
			MN, 55455 USA email: huang663@umn.edu. N. D. Sidiropoulos is with the Department of Electrical and Computer Engineering, University of Virginia, Charlottesville, VA 22904, USA email: nikos@virginia.edu. W.-K. Ma is with the Department of Electronic Engineering, The Chinese University of Hong Kong, Shatin, NT, Hong Kong. email: wkma@cuhk.edu.hk}
		\thanks{This work is supported in part by This work was supported in part by National Science Foundation under
			Project NSF ECCS-1608961 and Project NSF IIS-1447788.}
	}
	
	\date{}

	\maketitle

	
	\else
	\title{\papertitle}
	
	\author{
		Xiao Fu, Kejun Huang, Nicholas D. Sidiropoulos, and Wing-Kin Ma
		
		\thanks{X. Fu is with the School of Electrical Engineering and Computer Science, Oregon State University, Corvallis, OR 97331, USA. email: xiao.fu@oregonstate.edu
			
		K. Huang is with the Department
				of Computer and Information Science and Engineering, University of Florida, Gainesville, FL 32605, USA. email: kejun.huang@ufl.edu

			N. D. Sidiropoulos is with the Department of Electrical and Computer Engineering, University of Virginia, Charlottesville, VA 22904, USA email: nikos@virginia.edu
			
			W.-K. Ma is with the Department of Electronic Engineering, The Chinese University of Hong Kong, Shatin, NT, Hong Kong. email: wkma@cuhk.edu.hk}
		\thanks{This work was supported in part by the National Science Foundation under
			Project NSF ECCS-1608961 and Project NSF IIS-1447788.}
	}

	\ifconfver \else {\linespread{1.1} \rm \fi

		\maketitle
		

		\ifconfver \else
		\begin{center} \vspace*{-2\baselineskip}
		\end{center}
		\fi

		\ifconfver \else \IEEEpeerreviewmaketitle} \fi
	
	\fi
	
	\ifconfver \else
	\ifplainver \else
	\fi \fi

	\section{Introduction}
	Nonnegative matrix factorization (NMF) aims at factoring a data matrix into low-rank latent factor matrices with nonnegativity constraints. Specifically, given a data matrix $\bm X\in\mathbb{R}^{M\times N}$ and a target rank $R$, NMF seeks a factorization model 
	\begin{equation}\label{eq:classic_nmf}
	{\bm X}\approx{\bm W}{\bm H}^\T,~\bm W\in\mathbb{R}^{M\times R},~\bm H\in\mathbb{R}^{N\times R}, 
	\end{equation}
	to `explain' the data matrix $\bm X$, where ${\bm W}\geq{\bm 0}$, ${\bm H}\geq{\bm 0}$, and $R\leq \min\{M,N\}$. At first glance, NMF is nothing but an alternative factorization model to existing ones such as the \emph{singular value decomposition} (SVD) \cite{GHGolub1996} or \emph{independent component analysis} (ICA) \cite{Comon1994} that have different constraints (resp. orthogonality or statistical independence) on the latent factors. 
	However, the extraordinary effectiveness of NMF in analyzing real-life nonnegative data has sparked a substantial amount of research in many fields.
	The linear algebra community has shown interest in nonnegative matrices and nonnegative matrix factorization (known as nonnegative rank factorization) since more than thirty years ago \cite{chen1984nonnegative}. In the 1990s, researchers in analytical chemistry and remote sensing (earth science) already noticed the effectiveness of NMF---which was first referred to as `positive matrix factorization' \cite{craig1994minimum,paatero1994positive}. In 1999, Lee and Seung's seminal paper published in \emph{Nature} \cite{lee1999learning} sparked a tremendous amount of research in computer science and signal processing. Many types of real-life data like images, text, and audio spectra can be represented as nonnegative matrices---and factoring them into nonnegative latent factors yields intriguing results. 
	Today, NMF is well-recognized as a workhorse for signal and data analytics.

	\noindent
	{\bf Interpretability and Model Identifiability} 
	Although NMF had already been tried in many application domains before the 2000's, it was the `interpretability' argument made in the \textit{Nature} article \cite{lee1999learning} that really brought NMF to center stage. Lee and Seung underscored this point in \cite{lee1999learning} with a thought-provoking experiment in which NMF was applied to a matrix $\X$ whose columns  are vectorized human face images. Interestingly, the columns of the resulting ${\W}$ are clear parts of human faces, e.g., nose, ears, and eyes (see details in Sec.~\ref{sec:face}). Interpretable NMF results can be traced back to earlier years.
	As mentioned, back in the 1990s, in remote sensing and analytical chemistry, researchers started to notice that when factoring remotely captured hyperspectral images and laboratory-measured spectral samples of chemical compounds using techniques that are now recognized as NMF algorithms, the columns of ${\bm W}$ are spectra of materials on the ground \cite{craig1994minimum} and constituent elements of the chemical samples, respectively \cite{paatero1994positive}. 
	NMF also exhibits strong interpretability in machine learning applications. In text mining, NMF returns prominent topics contained in a document corpus \cite{arora2012practical,huafusid2016nips}. In community detection, NMF discovers groups of people who have similar activities from a social graph \cite{huang2014non,mao2017mixed}. 
	
	The interpretability of NMF is intimately related to its model uniqueness, or, latent factor identifiability. That is, unlike some other factorization techniques (e.g., SVD), NMF is able to identify the ground-truth generative factors $\W$ and $\H$ up to certain trivial ambiguities---which therefore leads to strong interpretability. In \cite{paatero1994positive} that appeared in 1994, Paatero \textit{et al.} conjectured that the effectiveness of NMF might be related to the uniqueness of the factorization model under certain realistic conditions \cite{paatero1994positive}. 
	This conjecture was confirmed in theory later on by a number of works from different fields \cite{huang2014non,donoho2003does,laurberg2008theorems,CANMS,fu2016robust,fu2017on,lin2014identifiability}. 
	The connection between identifiability and interpretability is intuitively pleasing---if the data really follows a generative model, then identifying the ground-truth generative model becomes essential for explaining the data.
	The understanding of NMF identifiability also opens many doors for a vast array of applications. For example, in signal processing, NMF and related techniques have been applied to speech and audio separation \cite{fu2015blind,fevotte2009nonnegative}, cognitive radio \cite{fu2015power}, and medical imaging \cite{CANMS}; in statistical learning, recent works have shown that NMF can identify topic models coming from the {\it latent semantic analysis} (LSA) model and {\it latent dirichlet allocation} (LDA) model \cite{arora2012practical}, the \textit{mixed membership stochastic blockmodels} (MMSB) \cite{mao2017mixed}, and the \textit{hidden Markov model} (HMM) \cite{lakshminarayanan2010non,huang2018hmm}---with provable guarantees.

	\noindent
	{\bf Our Goal} 
	In this feature article, we will review the recent developments in theory, methods and applications of NMF. We will use model identifiability of NMF as a thread to connect these different aspects of NMF, and show the reason why understanding identifiability of NMF is critical for engineering applications.
	We will introduce different identification criteria for NMF, reveal their insights, and discuss their pros and cons---both in theory and in practice.
	We will also introduce the associated algorithms and applications, and discuss in depth the connection between the nature of problems considered,
	NMF identification criteria, and the associated algorithms.

	{ We should mention that there are a number of existing tutorials on NMF, such as those in \cite{zhou2014nonnegative,gillis2014and,gillis2017introduction,wang2013nonnegative,Ma2013,huang2014putting}. Many of these tutorials focus on computational aspects, and do not cover the identifiability issues thoroughly (an exception is \cite{gillis2014and} where an NMF subclass called separable NMF was reviewed). In addition, many developments and discoveries on NMF identifiability happened after the appearances of \cite{zhou2014nonnegative,gillis2014and,gillis2017introduction,wang2013nonnegative,Ma2013,huang2014putting}, and there is no tutorial summarizing these new results. This article aims at filling these gaps and introducing the latest pertinent research outcomes.}

	\noindent
	{\bf Notation} We mostly follow the established conventions in signal processing. For example, we use boldface $\X\in\mathbb{R}^{M\times N}$ to denote a matrix of size $M\times N$; $\X(m,n)$ denotes the element in the $m$th row and $n$th column; $\X(:,\ell)$ and $\x_\ell$ can both denote the $\ell$th column of $\X$; ``${-1}$'', ``$\dag$'', and ``$\T$'' denote matrix inversion, pseudo-inverse, and transpose, respectively; ${\rm vec}(\bm Y)$ reshapes the matrix $\bm Y$ as a vector by concatenating its columns; ${\bm 1}$ and ${\bm 0}$ denote all-one and all-zero vectors with proper length;
	{ $\X\geq {\bm 0}$ means that every element in $\X$ is nonnegative; ``$\X\succeq{\bm 0}$'' and ``$\X\succ{\bm 0}$'' denote that $\X$ is positive semidefinite and positive definite, respectively}; ${\rm nnz}(\X)$ counts the number of nonzero elements in $\X$; { $\mathbb{R}^{n}_+$ denotes the nonnegative orthant in the $n$-dimensional Euclidean space.}

   \begin{figure}
\centering
\includegraphics[width=0.7\linewidth]{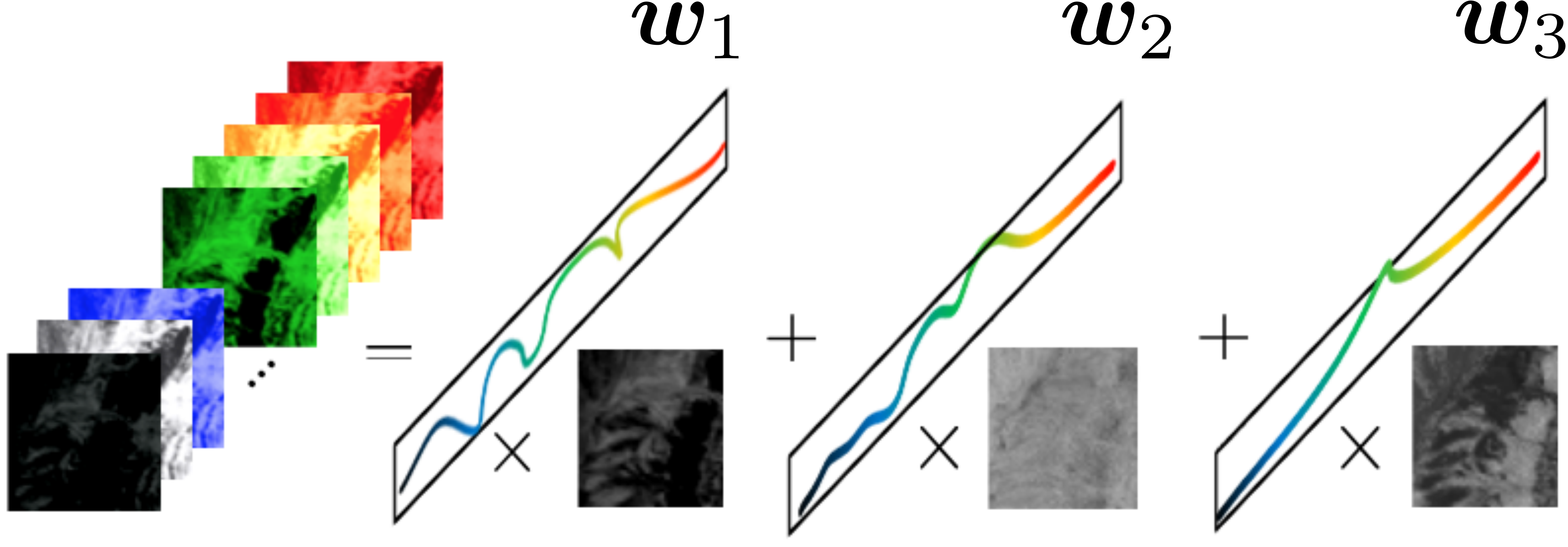}

\vspace{.5cm}

\includegraphics[width=0.7\linewidth]{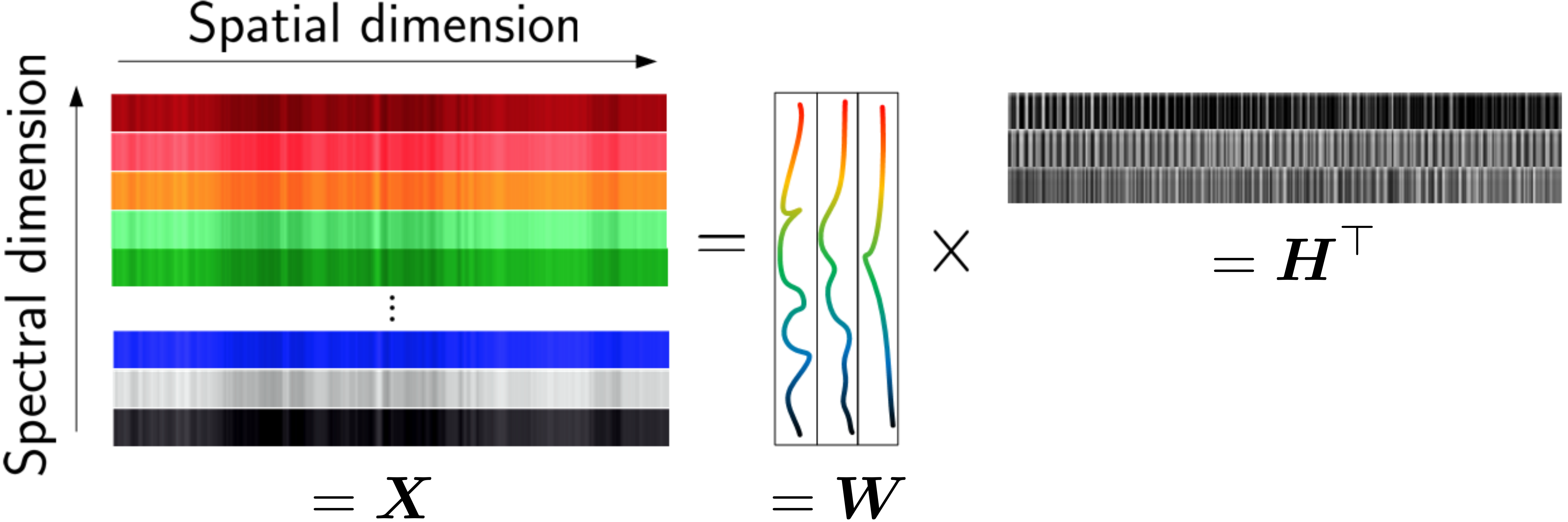}
\caption{{ Classic application of NMF---hyperspectral unmixing in remote sensing. Top: the LMM for spectral unmixing in remote sensing; 
		every slab on the left represents a band of the hyperspectral image; 
		each vector $\bm w_i$ describes the spectral signature of a pure material, or endmember;
		the maps on the right hand side are called the abundance maps of the endmembers, which are re-arranged columns of $\H$. Bottom: the corresponding NMF model after re-arranging the pixels.}}
\label{fig:HU1}

\vspace{-.5cm}
\end{figure}

\section{Where Does NMF Arise?}
\label{sec:app_intro}
Let us first take a look at several interesting and important applications of NMF. These applications will shed light on the importance of the NMF identifiability issues.

\subsection{Early Pioneers: Analytical Chemistry, Remote Sensing, and Image Processing}\label{sec:face}
	The concept of {\it positive matrix factorization} was mentioned in 1994 by Paatero \textit{et al.} \cite{paatero1994positive} for a class of spectral unmixing problems in analytical chemistry. Specifically, the paper aimed at analyzing key chemical components that are present in a collection of observed chemical samples measured at different wavelengths. 
    This problem is very similar to the \textit{hyperspectral unmixing} (HU) problem in hyperspectral imaging   \cite{Ma2013},
    which is a timely application in remote sensing. 
    We give an illustration of the HU data model in Fig.~\ref{fig:HU1}.
    There, $\x_\ell$ is a { spectral vector (or, a spectral pixel)} that is measured via capturing electromagnetic (EM) patterns carried by the ground-reflected sunlight using a remote sensor at $M$ wavelengths. The \textit{linear mixture model} (LMM) employed in hyperspectral imaging, i.e., $\x_\ell \approx \W \H(\ell,:)^\T\in\mathbb{R}^M$, assumes that every pixel is a convex combination (i.e., weighted sum { with weights summing up to one}) of spectral signatures of different pure materials on the ground. Under the LMM, HU aims at factoring $\X$ to recover the `endmembers' (spectral signatures) $\W$ and the `abundances' $\H$. In fact, in remote sensing and earth science, the use of NMF or related ideas has a very long history. Some key concepts in geoscience that are heavily related to modern NMF can be traced back to the 1960s \cite{imbrie1964vector}, although the word `NMF' was not spelled out.
	
\begin{figure}[!h]
		\centering
		
		\includegraphics[width=0.7\linewidth]{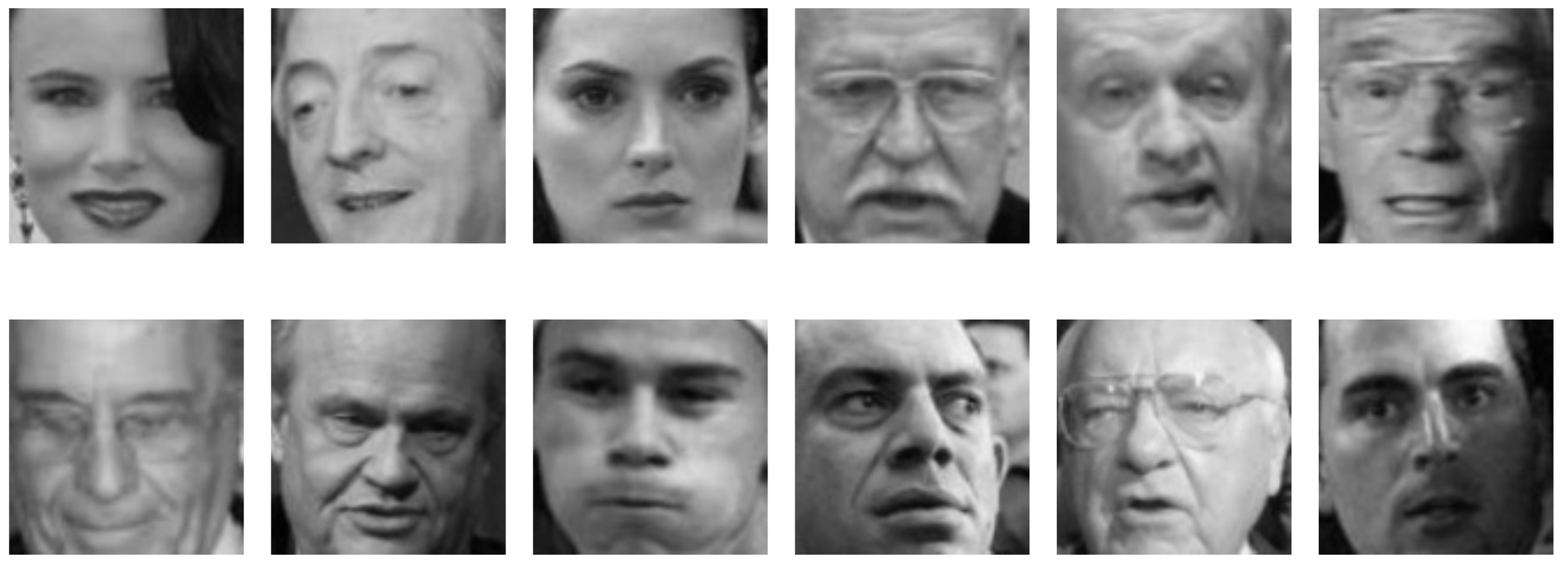}
		
		\includegraphics[width=0.7\linewidth]{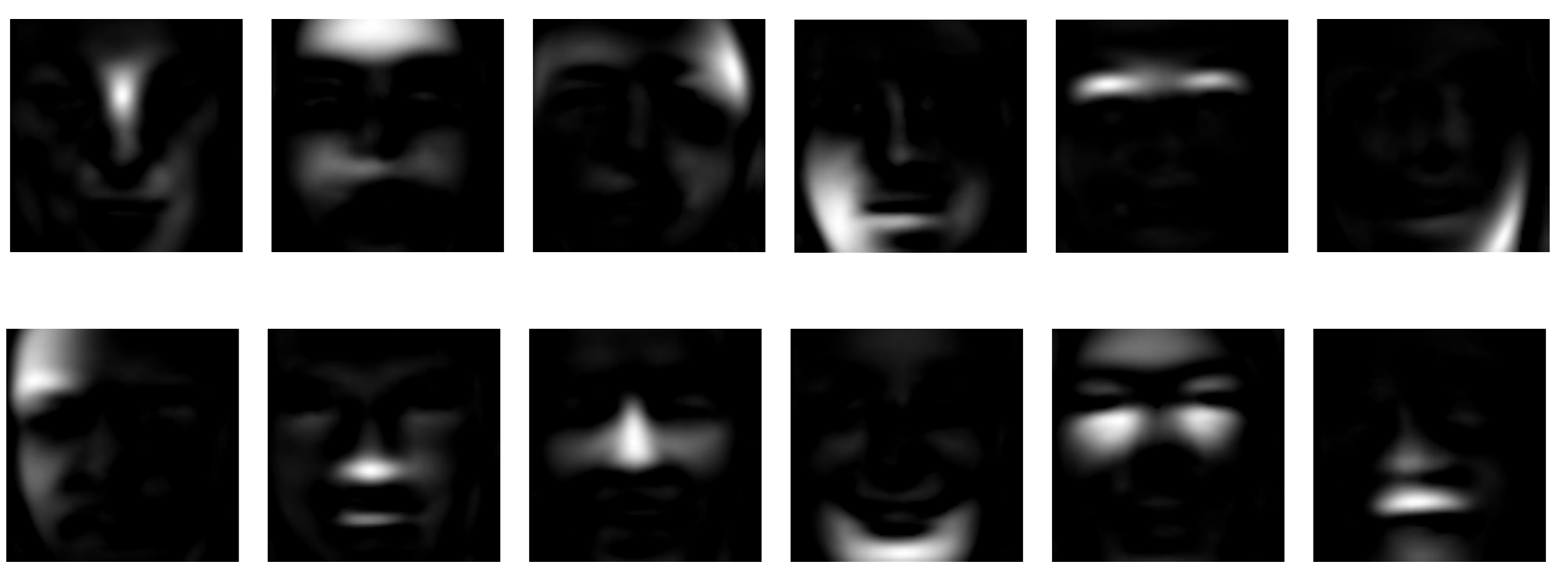}
		\caption{{ Top two rows: face images ($\bm x_\ell$'s). Bottom two rows: learned representation basis ($\bm w_r$'s) for human faces. Here we have $R=49$}.}
		\label{fig:face}
	\end{figure}

	Perhaps the most well-known (and pioneering) application of NMF is representation learning in image recognition \cite{lee1999learning} ({ see Fig.~\ref{fig:face} and the next insert}).
	There, $\x_\ell$ is a vectorized image of a human face, and $\X$ is a collection of such vectors.
	 Learning a nonnegative $\H$ from the factorization model $\X\approx \W\H^\T$ can be understood as learning nonnegative `embeddings' of the faces in a low-dimensional space (spanned by the basis $\W$); i.e., the learned $\H(\ell,:)$ can be used as a low-dimensional representation of face $\ell$. It was in this particular paper that Lee and Seung reported a very thought-provoking empirical result: the columns of $\W$ are interpretable---every column corresponds to a part of a human face. 
	Correspondingly, the embeddings are meaningful---the embedding of a person's face $\H(\ell,:)$ contains weights of the learned parts { and $\H(\ell,:)$ is typically sparse, meaning that every face can be synthesized from few of the available parts}.
	Such meaningful embeddings are often found effective in subsequent processing such as clustering and classification since they reflect the `true nature' of the faces.
	{ In fact, the ability of generating sparse and part-based embedding is considered the major driving-force behind the popularity of NMF in machine learning. See more discussions on this particular point in \cite{gillis2017introduction,gillis2014and}.}

	\begin{shaded*}

			\noindent
			{\bf NMF for Representation Learning.}
			An example of NMF for face representation learning is illustrated in Fig.~\ref{fig:face}. Here, we use a dataset containing 13,232 images (see the LFW dataset at \url{http://vis-www.cs.umass.edu/lfw}), each image has a size of $101\times 101$ pixels. Hence, the $\X$ matrix has a size of $10,201\times 13,232$. We set $R=49$ and run the Lee-Seung multiplicative update  NMF algorithm \cite{lee2001algorithms} with 5,000 iterations. Some of the face images and the learned $\bm w_r$'s are presented in Fig.~\ref{fig:face}.
			One can see that the learned basis $\W$ contains parts of human faces as its columns.
			
	\end{shaded*}

	\subsection{Topic Modeling}
	
	\begin{figure*}
		\begin{minipage}{.5\linewidth}
			\centering
			\includegraphics[width=1\linewidth]{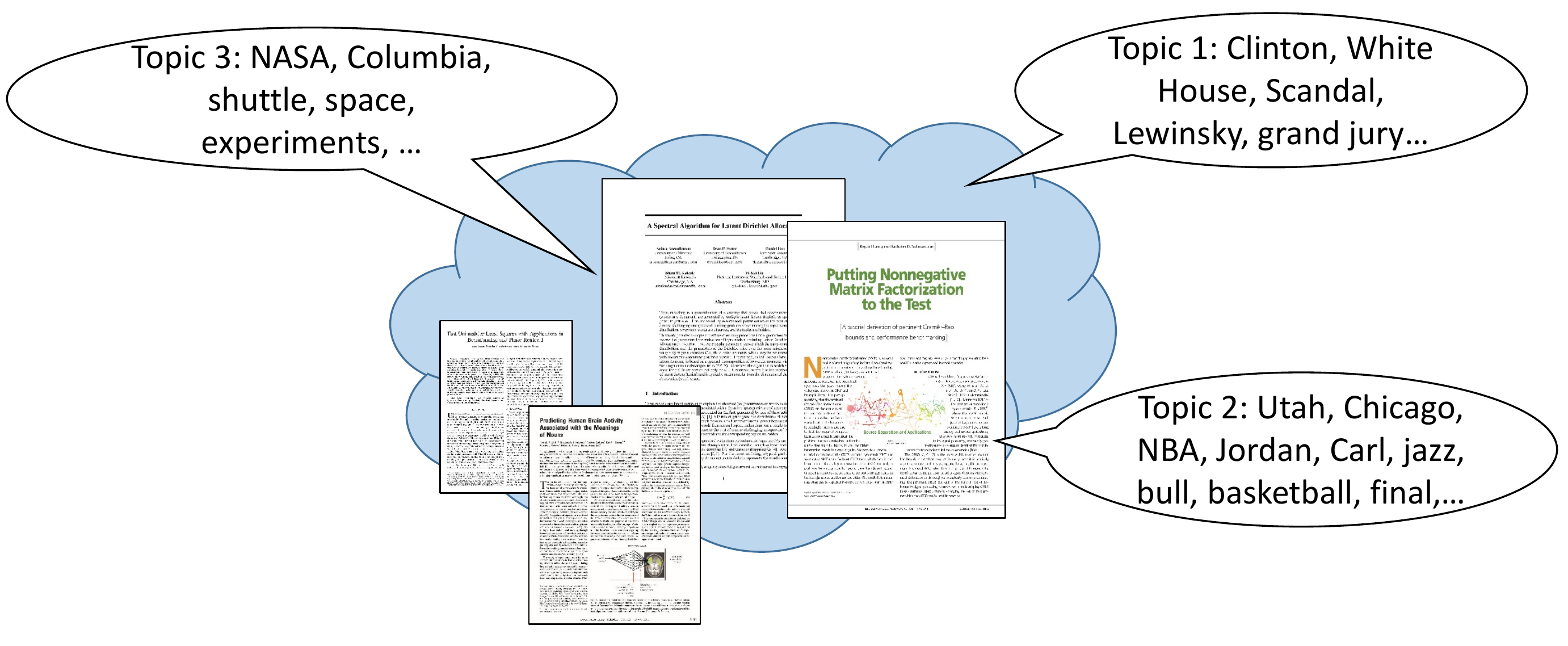}
		\end{minipage}
		\begin{minipage}{.45\linewidth}
			\centering
			\includegraphics[width=1\linewidth]{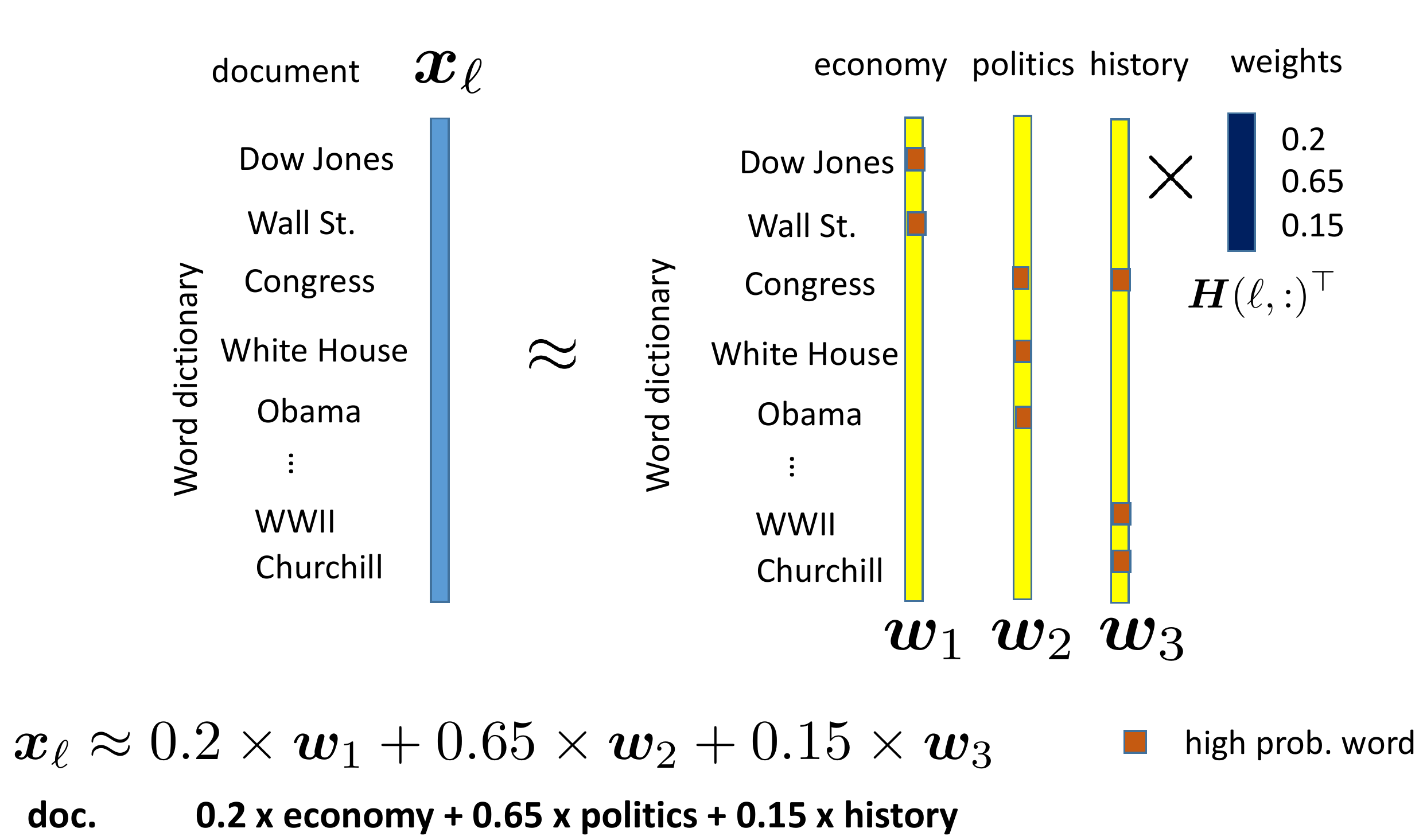}
		\end{minipage}
					\caption{Topic mining and nonnegative representation $\x_\ell=\W\bm H(\ell,:)^\top$ for $\ell=1,\ldots,N$. }	\label{fig:docmine}
	\end{figure*}
	
	In machine learning, NMF has been an important tool for learning topic models \cite{huafusid2016nips,arora2012practical,recht2012factoring,arora2012learning}. 
	The task of topic mining is to discover prominent topics (represented by sets of words) from a large set of documents; see Fig.~\ref{fig:docmine}.
	In topic mining using the celebrated `bag-of-words' model, $\X$ is a word-document matrix that summarizes the documents over a dictionary. Specifically, $\X(m,n)$ represents the $m$-th word feature of the $n$th document, say, \textit{Term-Frequency} (TF) or \textit{Term-Frequency Inverse-Document-Frequency} (TF-IDF) \cite{cai2009probabilistic}.
	Take the TF representation as an example. There, we have $\X(m,n)\approx {\rm Pr}(m,n)$, where ${\rm Pr}(m,n)$ denotes the joint probability of seeing word $m$ in document $n$. One simple way to connect NMF with topic modeling is to consider the {\it probabilistic latent semantic indexing} (pLSI) approach (see detailed introduction in \cite{blei2003latent}). Suppose that given a topic $r$, the probabilities of seeing a word $m$ and a document $n$ are conditionally independent. We have
	\[  {\rm Pr}(m,n)=\sum_{r=1}^R{\rm Pr}(m|r){\rm Pr}(r){\rm Pr}(n|r), \]
	{ where ${\rm Pr}(m|r)$ and ${\rm Pr}(n|r)$ denote the probabilities of seeing word $m$ and document $n$ conditioned on topic $r$.}
	This is equivalent to the following factorization model
	\begin{equation}\label{eq:topic}
		\X =\C \bm \varSigma \bm D^\T 
	\end{equation}
	where $\C(m,r)={\rm Pr}(m|r)$, $\bm \varSigma$ is a diagonal matrix with $\bm \Sigma(r,r)={\rm Pr}(r)$ and $\bm D(n,r)={\rm Pr}(n|r)$. 
	Note that under this model, $\C(:,r)$ is a conditional { probability mass function} (PMF) that describes the probabilities of seeing the words in the dictionary given topic $r$---which naturally represents topic $r$. 
	If one denotes $\H =  \bm D\bm \varSigma$ and $\W = \C$, the model $\X\approx \W\H^\T$ is obtained. Note that $\W$ and $\H$ are naturally nonnegative since they are PMFs \cite{cai2009probabilistic}.  { This nonnegative representation of $\x_\ell$ is illustrated in Fig.~\ref{fig:docmine} (right), which can be understood as that a document is a weighted combination of several topics, in layman's terms.}

	The topic mining problem can also be formulated using the second-order statistics. Specifically, consider the correlation matrix of all documents:
	\begin{equation}\label{eq:topicec}
	\bm P = \mathbb{E}\left[\x_\ell \x_\ell^\T \right]=\C\E \C^\T, 
	\end{equation}
	where $\E= \mathbb{E}[\bm H(\ell,:)^\T \H(\ell,:)]$ denotes the correlation matrix of the topics. The reason for considering the alternative formulation in \eqref{eq:topicec} is that the model \eqref{eq:topic} is a noisy approximation especially  when  the  length  of  each  document  is  short  (e.g., tweets),  so  that  the  number  of  words  is  not  enough  to  estimate ${\rm Pr}(m,n)$ reliably. But if we have a large number of documents, the second-order statistics $\bm P$ can be estimated reliably. The model \eqref{eq:cec } is an NMF model wherein  $\W =\C\E$ and $\H=\C$, if one ignores the underlying structure of $\W$. Nevertheless, we will see that exploiting the symmetric structure of $\bm P$ may lead to better performance or simpler algorithms.
	
%
%
	\color{black}
	
	
	\subsection{Hidden Markov Model Identification}
	HMMs are widely used in machine learning for modeling time series, e.g., speech and language data.
	The graphical model of an HMM is shown in Fig.~\ref{fig:hmm}, where one can see a sequence of observable data $\{Y_t\}_{t=0}^T$ emitted from
	an underlying unobservable Markov chain $\{X_t\}_{t=0}^T$. The HMM estimation problem aims at identifying the transition probability $\Pr(X_{t+1}|X_{t})$ and the emission probability $\Pr(Y_t|X_t)$---{ assuming that they are time-invariant}.
	
	Consider a case where one can accurately estimate the co-occurrence probability between two consecutive emissions, i.e., $\Pr(Y_t,Y_{t+1})$. According to the graphical model shown in Fig.~\ref{fig:hmm}, it is easy to see that given the values of $X_t$ and $X_{t+1}$, $Y_t$ and $Y_{t+1}$ are conditionally independent.
	Therefore, we have the following model \cite{lakshminarayanan2010non,huang2018hmm}:
	\begin{figure}[t]
		\centering
		\begin{tikzpicture}[]
\node at (   0,1.5) [circle,draw,minimum size=8mm,inner sep=-5pt] (Xt) {$X_t$};
\node at (-1.5,1.5) [circle,draw,minimum size=8mm,inner sep=-5pt] (Xt-1) {$X_{t-1}$};
\node at ( 1.5,1.5) [circle,draw,minimum size=8mm,inner sep=-5pt] (Xt+1) {$X_{t+1}$};
\node at (   0,0) [circle,draw,minimum size=8mm,inner sep=-5pt,fill=gray!30] (Yt) {$Y_t$};
\node at (-1.5,0) [circle,draw,minimum size=8mm,inner sep=-5pt,fill=gray!30] (Yt-1) {$Y_{t-1}$};
\node at ( 1.5,0) [circle,draw,minimum size=8mm,inner sep=-5pt,fill=gray!30] (Yt+1) {$Y_{t+1}$};
\node at (-2.9,1.5) (before) {\dots};
\node at ( 2.9,1.5) (after)  {\dots};
\draw [->] (before) -- (Xt-1);
\draw [->] (Xt+1) -- (after);
\draw [->] (Xt-1) -- (Xt);
\draw [->] (Xt) -- (Xt+1);
\draw [->] (Xt-1) -- (Yt-1);
\draw [->] (Xt) -- (Yt);
\draw [->] (Xt+1) -- (Yt+1);
\end{tikzpicture}
		\caption{The graphical model of a HMM.}
		\label{fig:hmm}

	\end{figure}
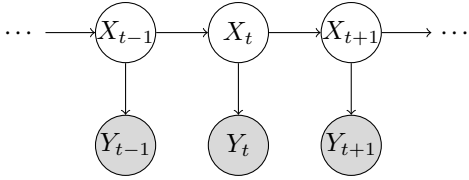
	\begin{align}\label{eq:hmm_fac2}
		\Pr(Y_t,Y_{t+1}) 
		= &\sum_{k,j=1}^{R}\Pr(Y_t|X_t=x_k)\times\\
		&\Pr(Y_{t+1}|X_{t+1}=x_j)\Pr(X_t=x_k,X_{t+1}=x_j)\nonumber
	\end{align}
	Let us define $\bOmega\in\bbR^{M\times M}$, $\bM\in\bbR^{M\times R}$, and $\bTheta\in\bbR^{R\times R}$ such that
	$\bm \varOmega(m,\ell) = \Pr(Y_t=y_m,Y_{t+1}=y_\ell)$,
	$\M(m,r) = \Pr(Y_t=y_m|X_t=x_r)$,
	and $\bm \varTheta(k,j) = \Pr(X_t=x_k,X_{t+1}=x_j)$.
	Here, $M$ denotes the number of observable states and $R$ the number of hidden states.
	Then, \eqref{eq:hmm_fac2} can be written compactly as
	\begin{align}\label{eq:MTM}
		\bOmega = \bM\bTheta\bM^\T.
	\end{align}
	The above model	is very similar to the second-order-statistics
	topic model in \eqref{eq:topicec}: both $\M$ and $\bTheta$ are constituted by PMFs, and thus being nonnegative. Similar to the topic modeling example, if one lets $\bm X=\bm \varOmega$, $\W=\M\bTheta$ and $\H=\M$, then the classic NMF model in \eqref{eq:classic_nmf} is recovered.
	

	\subsection{Community Detection}
	
\begin{figure}
\centering
\includegraphics[width=.5\linewidth]{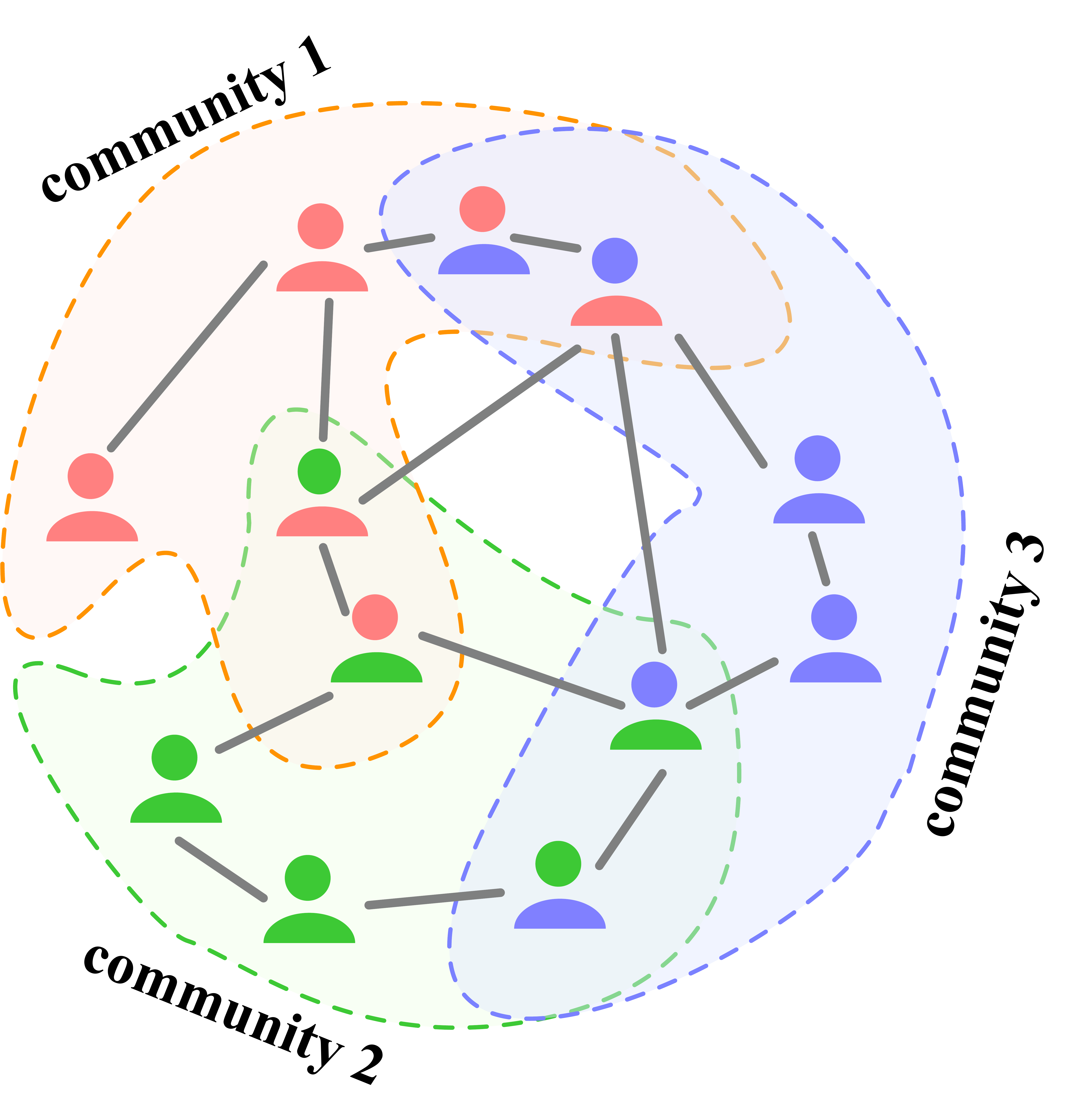}
\caption{{ The problem of overlapped community detection: assigning multiple memberships to different nodes in a network according to their latent attributes}.}
\label{fig:communitydetection}

\end{figure}

	Recent work in \cite{mao2017mixed,panov2017consistent} has related NMF to the celebrated {\it mixed membership stochastic blockmodel} (MMSB) \cite{airoldi2008mixed} that is widely adopted for overlapped community detection---which is a problem of assigning multiple community memberships to nodes (e.g., people) in a network; see Fig.~\ref{fig:communitydetection}.  Overlapped community detection is considered a hard problem since many network datasets are of very large scale and are often recorded in an oversimplified way (e.g., using unweighted undirected adjacency matrices).
	
	The MMSB considers the generative model of a symmetric adjacency matrix $\A\in \{0,1\}^{N\times N}$. Specifically,  $\bm A(i,j)$'s represent the connections between entities in a network---i.e., $\A({i,j})=1$ means that node $i$ and $j$ are connected, while $\A({i,j})=0$ means otherwise.  Assume that every node has a membership indicator vector $\bm \theta_i=[\theta_{1,i},\ldots,\theta_{R,i}]\in\mathbb{R}^{1\times R}$ and there are $R$ underlying communities. The value of $\theta_{r,i}$ indicates the probability that node $i$ is associated with community $r$---and thus $\bm \theta_i$ can naturally represent multiple memberships of node $i$. Apparently, we have
	$\bm \theta_i\geq {\bm 0},\quad \sum_{r=1}^R\theta_{r,i}=1.$
	Define a matrix $\B\in\mathbb{R}^{R\times R}$, where $\B(m,n)\in[0,1]$ represents the probability that communities $m$ and $n$ are connected.
	Under the assumption that nodes are connected with each other through their latent community identities,
	the node-node link-generating process is as follows
	\begin{subequations}
		\begin{align}
			\bm \theta_i&\sim {\rm Dirichlet}(\bm \alpha),\quad{\bm P}= \bm \Theta \bm B \bm \Theta^\T\\
			\A(i,j)&=\A(j,i)\sim {\rm Bernoulli}(\bm P(i,j))
		\end{align}
	\end{subequations}
	where $\bm \Theta\in\mathbb{R}^{M\times R}$ collects all the $\bm \theta_i$'s as its rows, and $\bm \alpha\in\mathbb{R}^R$ is the parameter vector that characterizes the Dirichlet distribution \cite{forbes2011statistical}. In the above, $\bm \theta_i$ and $\bm A(i,j)$ are assumed to be sampled from a Dirichlet distribution and a Bernoulli distribution (with parameter $\bm P(i,j)$) \cite{forbes2011statistical}, respectively.
	Being sampled from the Dirichlet distribution, $\bm \theta_i$ resides in the probability simplex, and thus $\theta_{r,i}$ naturally reflects how much node $i$ is associated with community $r$---e.g., when $R=3$, $\bm \theta_i=[0.5,0.2,0.3]^\T$ means $50\%$, $20\%$, and $30\%$ of node $i$'s participation is associated with communities 1, 2, and 3, respectively.  Using a Bernoulli distribution also makes a lot of sense since plenty of network data is very coarsely recorded---using "0" and "1" to represent absence/presence of a connection, without regard to connection strength.
	Both the model parameters $\bm \Theta$ (node-membership matrix) and $\bm B$ (community-community connection matrix) are of great interest in network analytics.
	The model $\bm P=\bm \Theta\bm B \bm \Theta^\T$ is an NMF model as in the HMM case. However, $\bm P$ is not available in practice. The interesting observation in \cite{panov2017consistent,mao2017mixed} is that ${\rm range}(\bm \Theta)$ can be accurately approximated. Specifically, denote $\widehat{\U}$ as the matrix consisting of the first $R$ principal eigenvectors of $\bm A$.  Under the MMSB model and some regularity conditions, one can show that
	$ \widehat{\bm U}^\T = \bm Z\bm \Theta^\T +\bm N $ holds for a certain nonsingular $\bm Z\in\mathbb{R}^{R\times R}$ \cite{mao2017mixed} and a noise term $\bm N$ with $\|\bm N\|_F$ bounded. Then, by letting $\H=\bm \Theta$ and $\bm W=\bm Z$, one can recover the following noisy matrix factorization model
	\begin{align}\label{eq:cd}
		\X =\W\H^\T + \bm N, 
	\end{align}
	with $\H\geq {\bm 0}$.  
	We should note that the model in \eqref{eq:cd} is not strictly NMF, since $\W$ could contain negative elements. However, \eqref{eq:cd} is usually regarded as a simple extension of NMF, since
	many identifiability results and algorithms for the classic NMF model can be extended to the model in \eqref{eq:cd} as well \cite{fu2017on,fu2015blind,Gillis2012,arora2012practical,kumar2012fast,fu2015self} .
	\color{black}
	
	\bigskip
	
	There are many more applications of NMF which we cannot cover in detail. For example, gene expression learning has been widely recognized as a good domain of application for NMF  \cite{brunet2004metagenes}; 
	NMF can also help in the context of recommender systems \cite{zhang2006learning};
	{ and NMF has proven very effective in handling a variety of blind source separation (BSS) problems \cite{fu2015blind,fevotte2009nonnegative,fu2015power}---see the supplementary materials of this article.}
    Given this diverse palette of applications, a natural (and critical) question to ask is\textit{ what are the appropriate identification criteria and computational tools that we should apply for each of them}?
	{ The answer to this question is strongly dependent on related identifiability issues, as we will see.}
	
	\section{NMF and Model Identifiability}\label{sec:nmfandident}
	All the introduced problems in the previous section have one thing in common: We are concerned with whether we can identify the true latent factors $\H$ and $\W$ from the data $\X$. For example, in topic modeling, identifying $\H=\C$ yields the PMFs that represent the topics.
	To better understand the theoretical limitations of such parameter identification problems under NMF models, in this section, we will formally define identifiability of NMF and introduce some pertinent concepts.
	
	\subsection{Problem Statement}
	The notion of \textit{model uniqueness} or \textit{model identifiability} is not unfamiliar to the signal processing community, since it is one of the basic desiderata in parameter estimation. 
	For NMF, identifiability refers to the ability to identify the data generating factors $\W$ and $\H$ that give rise to $\X=\W\H^\T$.
	A matrix factorization model without constraints on the latent factors is nonunique, since we have
	\begin{equation}\label{eq:Q}
		\X=\W\H^\T=\W\Q(\H\Q^{-\T})^\T
	\end{equation} 
	for any invertible $\Q\in\mathbb{R}^{R\times R}$. Therefore, for whichever $(\W,\H)$ that fits the model $\X=\W\H^\T$, the alternative solution $(\W\Q,\H\Q^{-\T})$ fits it equally well. Intuitively, adding constraints such as nonnegativity on the latent factors may help us achieve identifiability, since constraints can reduce the `volume' of the solution space and remove a lot of possibilities for $\Q$. 
	The study of NMF identifiability, roughly speaking, addresses under what identification criteria and conditions most $\Q$'s can be removed.
	When studying identifiability, we often ignore cases where $\Q$ only results in column reordering and scaling /counter-scaling of $\W$ and $\H$, which is considered unavoidable, but also inconsequential in most applications.
	Bearing this in mind, let us formally define identifiability of nonnegative matrix factorization:
	
	\vspace{-.15cm}
	\begin{Def} (Identifiability)\label{def:ident}
		Consider a data matrix generated from the model $\X=\W_\natural\H_\natural^\T$, where $\W_\natural$ and $\H_\natural$ are the ground-truth factors. Suppose that $\W_\natural$ and $\H_\natural$ satisfy a certain condition. Let $(\W_\star,\H_\star)$ be an optimal solution to an identification criterion or an output from a procedure.
		If, under the aforementioned condition of $\W_\natural$ and $\H_\natural$'s, it holds that
		\begin{equation}\label{eq:ident}
			\W_\star = \W_\natural{\bm \varPi}{\bm D},\quad \H_\star = \H_\natural{\bm \varPi}{\bm D}^{-1},
		\end{equation} where ${\bm \varPi}$ is a permutation matrix and ${\bm D}$ is a full-rank diagonal matrix, then we say that the matrix factorization model is identifiable under that condition\footnote{{ Ideally, it would be much more appealing if the conditions that enable NMF identifiability are defined on $\X$, rather than on $\W_\natural$ and $\H_\natural$, since then the conditions could be easier for checking. However, this is in general very hard, since the latent factors play essential roles for establishing identifiability, as we will see.}}.
	\end{Def}
	
	As an example, consider the most popular NMF identification criterion
	\begin{align}\label{eq:nmf_fit}
		\minimize_{\W\geq{\bm 0},~{\bm H}\geq {\bm 0},}&~\left\| \X - \W\H^\T\right\|_F^2;
	\end{align}
	see ~\cite{donoho2003does,lee1999learning,huang2014non,huang2014putting}.
	Here, the identification criterion is the optimization problem in \eqref{eq:nmf_fit}, and $(\W_\star,\H_\star)$ is an optimal solution to this problem. To establish identifiability of this NMF criterion, one will need to analyze under what conditions \eqref{eq:ident} holds.
	Many NMF approaches can be understood as an optimization criterion, e.g., those in \cite{fu2016robust,lin2014identifiability,fu2015blind,VMAX,fu2017on}, which directly output estimates of $\W_\natural$ and $\H_\natural$. There are, however, methods that employ an estimation/optimization criterion for identifying a different set of variables, and then the latter is used to produce $\W_\natural$ and $\H_\natural$ under some conditions \cite{Gillis2012,VCA,recht2012factoring,fu2015self}---which will become clear later. 
	
	\begin{shaded*}
		\noindent {\bf Should I care about identifiability?} Simply speaking, when one works with an engineering problem that requires parameter estimation, identifiability is essential for guaranteeing sensible results. For example, in the topic mining model \eqref{eq:topicec}, $\W_\natural$'s columns are topics that generate a document. If a chosen NMF algorithm is unable to produce identifiable results, one could have $\W_\natural\Q$ as a solution of the parameter identification process, which is meaningless in interpreting the data since they are still mixtures of the ground-truth topics. An example of text mining is shown in Table~\ref{tab:topics}, where the top-20 key words of 5 mined topics from 1,683 real news articles are shown. The left is from a classic method (i.e., LDA  \cite{blei2003latent}) which has no known identifiability guarantees, and the right is from a method (i.e., Anchor-Free  \cite{fu2018anchor,huafusid2016nips}) that is built to guarantee identifiability under certain mild conditions. One can see that the Anchor-Free method outputs topics that are clean and easily distinguishable---the five groups of words clearly correspond to ``White House Scandal in 1998'', ``GM Strike in Michigan'', ``NASA \& Columbia Space Shuttle'', ``NBA Final in 1997'', and ``School Shooting in Arkansas'', respectively. In the meantime, the topics mined  by the classic method (i.e., LDA \cite{blei2003latent}) are noticeably mixed---particularly, words related to `White House Scandal' are spread across several topics. This illustrates the pivotal role of identifiability in data mining problems.

	\end{shaded*}

	\begin{table}[t!]
		\centering
		\caption{\footnotesize Mined topics by AnchorFree and an anchor word-based competitor from real-word documents. Table reproduced (permission will be sought)  \cite{huafusid2016nips,fu2018anchor}, $\copyright$ IEEE.}
		\resizebox{.99\linewidth}{!}{\huge
			\begin{tabular}{|c|c|c|c|c|c|c|c|c|c|}
				\hline
				\multicolumn{5}{|c|}{LDA \cite{blei2003latent}} & \multicolumn{5}{c|}{AnchorFree \cite{huafusid2016nips}} \\
				\hline
				\textcolor{blue}{lewinsky} & spkr  & school &\textcolor{blue}{clinton} & jordan & lewinsky & gm    & shuttle & bulls & jonesboro \\    starr & news  & gm    &\textcolor{blue}{president} & time  & monica & motors & space & jazz  & arkansas \\
				\textcolor{blue}{president} & people & workers &\textcolor{blue}{house} & game  & starr & plants & columbia & nba   & school \\
				\textcolor{blue}{white} &\textcolor{blue}{monica} & arkansas &\textcolor{blue}{white} & bulls & grand & flint & astronauts & chicago & shooting \\
				\textcolor{blue}{house} & story & people &\textcolor{blue}{clintons} & night & white & workers & nasa  & game  & boys \\
				ms    & voice & children & public & team  & jury  & michigan & crew  & utah  & teacher \\
				grand & reporter & students &\textcolor{blue}{allegations} & left  & house & auto  & experiments & finals & students \\
				lawyers & washington & strike &\textcolor{blue}{presidents} & i''m  & clinton & plant & rats  & jordan & westside \\
				\textcolor{blue}{jury} & dont  & jonesboro &\textcolor{blue}{political} & chicago & counsel & strikes & mission & malone & middle \\
				jones & media & flint &\textcolor{blue}{bill} & jazz  & intern & gms   & nervous & michael & 11year \\
				independent & hes   & boys  &\textcolor{blue}{affair} & malone & independent & strike & brain & series & fire \\
				investigation & thats & day   &\textcolor{blue}{intern} & help  & president & union & aboard & championship & girls \\
				counsel & time  & united & american & sunday & investigation & idled & system & karl  & mitchell \\
				lawyer & believe & union & bennett & home  & affair & assembly & weightlessness & pippen & shootings \\
				\textcolor{blue}{relationship} & theres & shooting & personal & friends & lewinskys & production & earth & basketball & suspects \\
				\textcolor{blue}{office} & talk  & motors & accusations & little & relationship & north & mice  & win   & funerals \\
				\textcolor{blue}{sexual} & york  & plant &\textcolor{blue}{presidential} & michael & sexual & shut  & animals & night & children \\
				\textcolor{blue}{clinton} & sex   & middle &\textcolor{blue}{woman} & series & ken   & talks & fish  & sixth & killed \\
				former & press & plants &\textcolor{blue}{truth} & minutes & former & autoworkers & neurological & games & 13year \\
				\textcolor{blue}{monica} & peterjennings' & outside & former & lead  & starrs & walkouts & seven & title & johnson \\
				\hline
			\end{tabular}}%
			\vspace{-5pt}
			\label{tab:topics}%
		\end{table}

		We would like to remark that in linear algebra, identifiability of factorization models (e.g., tensor factorization) is usually understood as a property of a given model that is independent of identification criteria. However, here we examine identifiability from a signal processing point of view, i.e., associated with a given {\em criterion}. This may seem unusual, but, as it turns out, the choice of estimation/identification criterion is crucial for NMF. 
		In other words, NMF can be identifiable under a suitable identification criterion, but not under another, even under the same generative model---as will be seen shortly. 
		

		
		\subsection{Preliminaries}
		\subsubsection{Geometric Interpretations}
		NMF has an intuitively pleasing underlying geometry.
		The model $\X=\W\H^\T$ can be viewed from different geometric perspectives. 
		In this article, we focus on the geometry of the columns of $\X$; the geometry of the rows of $\X$ follows similarly, by transposition. 
		It is readily seen that
		\[ {\bm x}_\ell \in \cone{\W},~\ell=1,\ldots,N \] 
		where { $ \cone{\W}=\{\bm x\in\mathbb{R}^M|\bm x =\bm W{\bm \theta},{\bm \theta}\geq{\bm 0} \}$} denotes the {\it conic hull} of the columns of $\W$. This geometry holds because of the nonnegativity of $\H$.
		There is another very useful---and powerful---geometric interpretation of the NMF model, which requires an additional assumption. Suppose that $\H$, in addition to being nonnegative, also satisfies a { `row sum-to-one' assumption}, i.e.,
		\begin{equation}\label{eq:hsto}
			\H{\bm 1}={\bm 1},~\H\geq{\bm 0}.
		\end{equation}
		{The assumption is called the {\it row-stochastic} assumption, since every row of $\H$ is constrained to lie in the probability simplex}.
		 The assumption naturally holds in many applications such as hyperspectral unmixing \cite{Ma2013} and image embedding \cite{lee1999learning}, where $\H(\ell,r)$ means the proportion of $\bm w_r$ in $\bm x_\ell$.
		In applications where \eqref{eq:hsto} does not arise from the physical model, it can be `enforced' via pre-processing, i.e., normalizing the columns of $\X$ using the $\ell_1$-norm of the corresponding columns \cite{CANMS,Gillis2012}, \textit{if $\W\geq\bm 0$ holds} and noise is absent; see the next insert.

				\begin{shaded*}
					\noindent
					{\bf Column Normalization to Enforce Row-Stochasticity \cite{Gillis2012}.} Consider $\X=\W\H^\T$ where $\W\geq\bm 0$ and $\H\geq{\bm 0}$. We wish to have a model where the right factor is row-stochastic.
					To this end, consider
					\[       \bar{\bm X}(:,\ell)=\frac{{\bm X}(:,\ell)}{\|{\bm X}(:,\ell)\|_1} .  \]
					The above gives an NMF model as follows:
					\begin{align*}
					\bar{\bm X}(:,\ell)& = \sum_{r=1}^R \underbrace{\frac{{\bm W}(:,r)}{\|{\bm W}(:,r)\|_1} }_{ \bar{\bm W}(:,r)}\underbrace{\frac{\|{\bm W}(:,r)\|_1{\bm H}(\ell,r)}{\|\sum_{r=1}^R{\bm W}(:,r){\bm H}(\ell,r)\|_1}}_{\bar{\bm H}(\ell,r)}\\
					& = \sum_{r=1}^R \bar{\bm W}(:,r)\bar{\bm H}(\ell,r) = \bar{\bm W}\bar{\bm H}(\ell,:)^\T.
					\end{align*}
					Note that identifying $\bar{\bm W}$ recovers $\W$ with a column scaling ambiguity (which is intrinsic anyway), and thus the above normalized model is useful in estimating the original latent factors.
					To show that $\bar{\bm H}({\ell,:}){\bm 1}=1$, note that the following holds when $\W\geq\bm 0$:
					\[ \left\|\sum_{r=1}^R{\bm W}(:,r){\bm H}(\ell,r)\right\|_1 = \sum_{r=1}^R{\bm H}(\ell,r)\|{\bm W}(:,r)\|_1.\]
					The above does not hold if ${\bm W}$ has negative elements. 
				\end{shaded*}

\begin{figure}
		
			\includegraphics[width=0.425\linewidth]{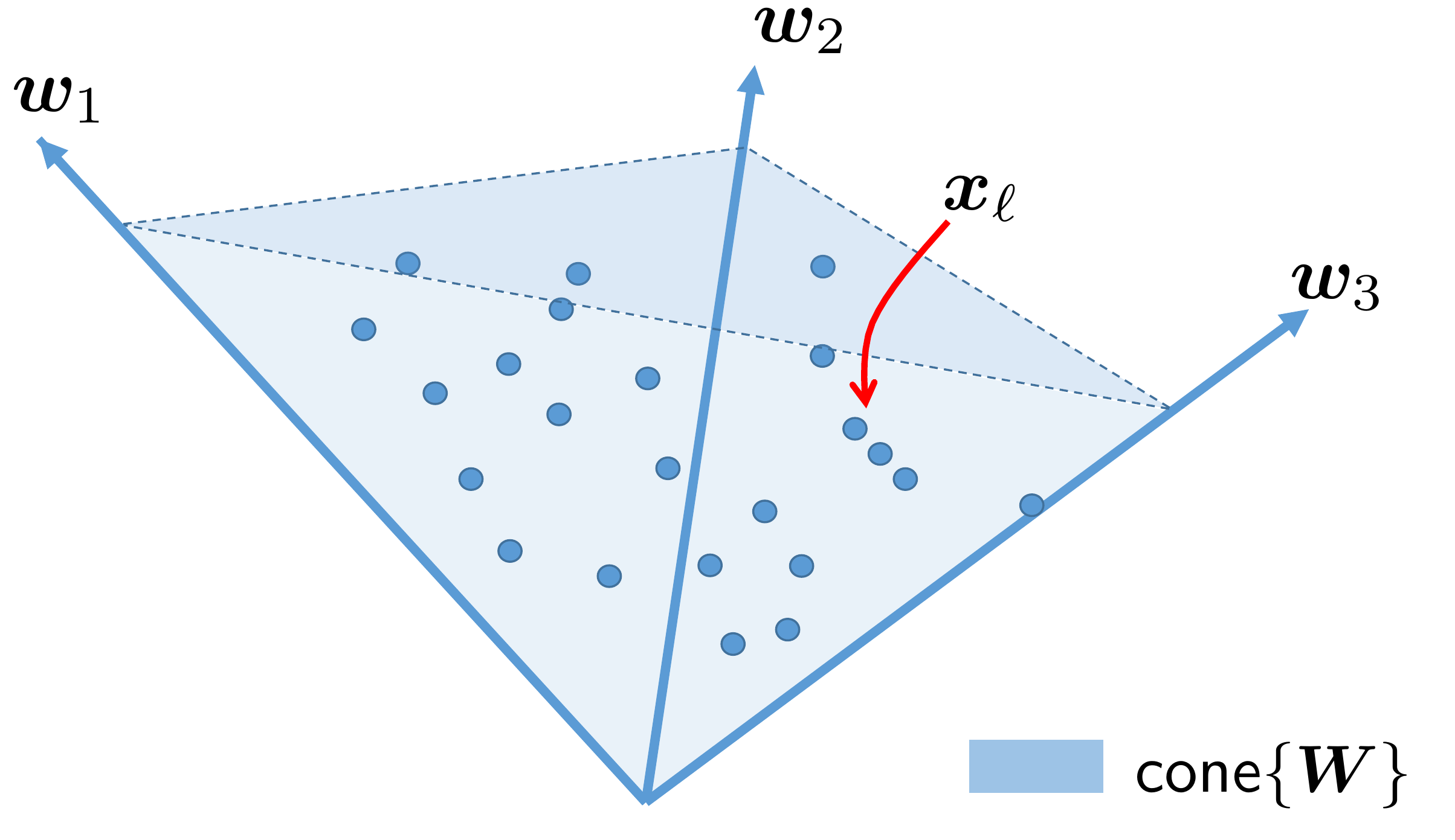}
			\includegraphics[width=0.445\linewidth]{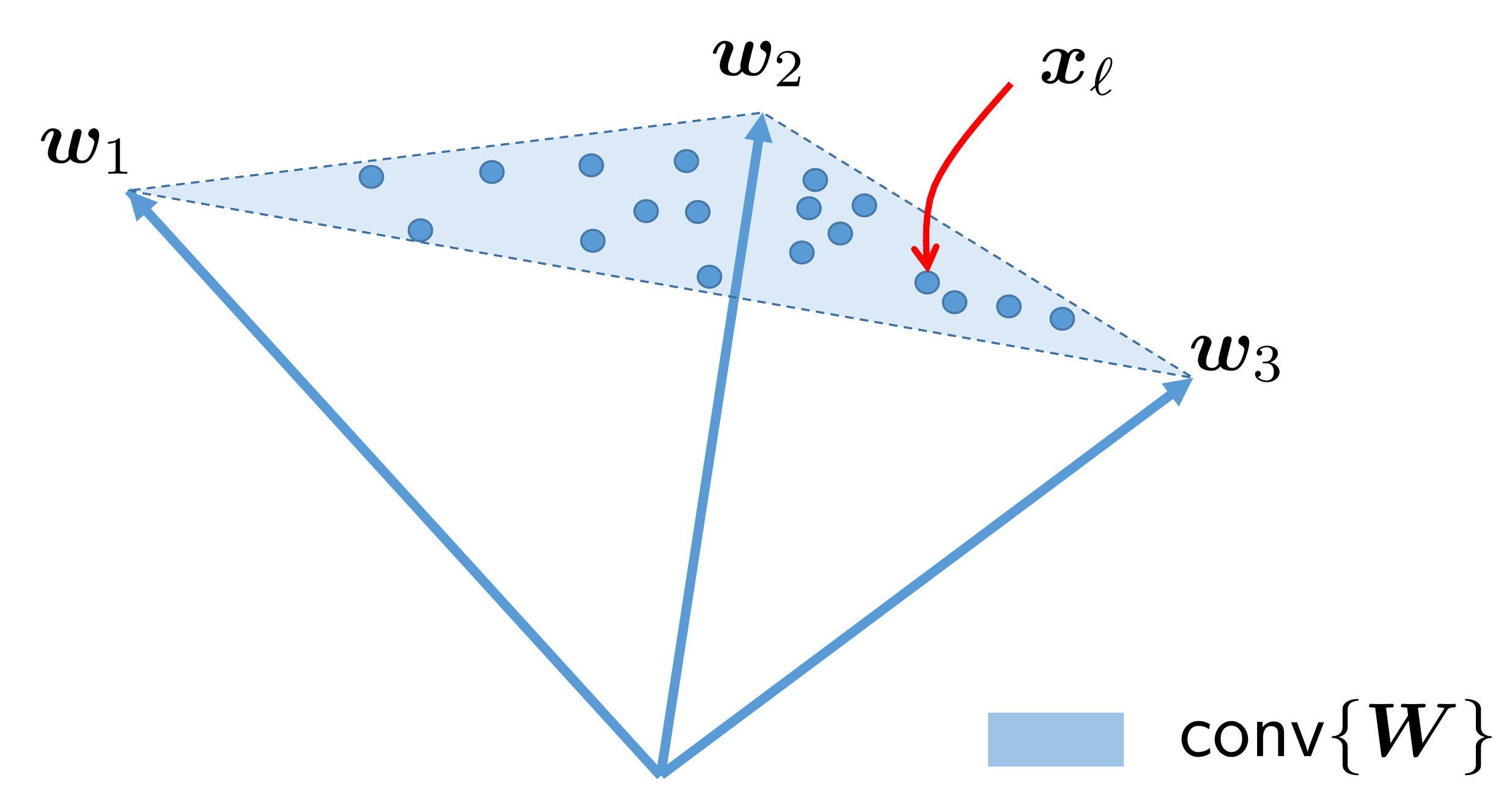}

			\caption{Two versions of the NMF geometry when $R=3$. Left: the original version. Right: the normalized version where $\H$ is row-stochastic. The dots are $\x_\ell$'s.}\label{fig:geometry}
		\end{figure}

\begin{figure}
			\begin{center}
				\includegraphics[width=.55\linewidth]{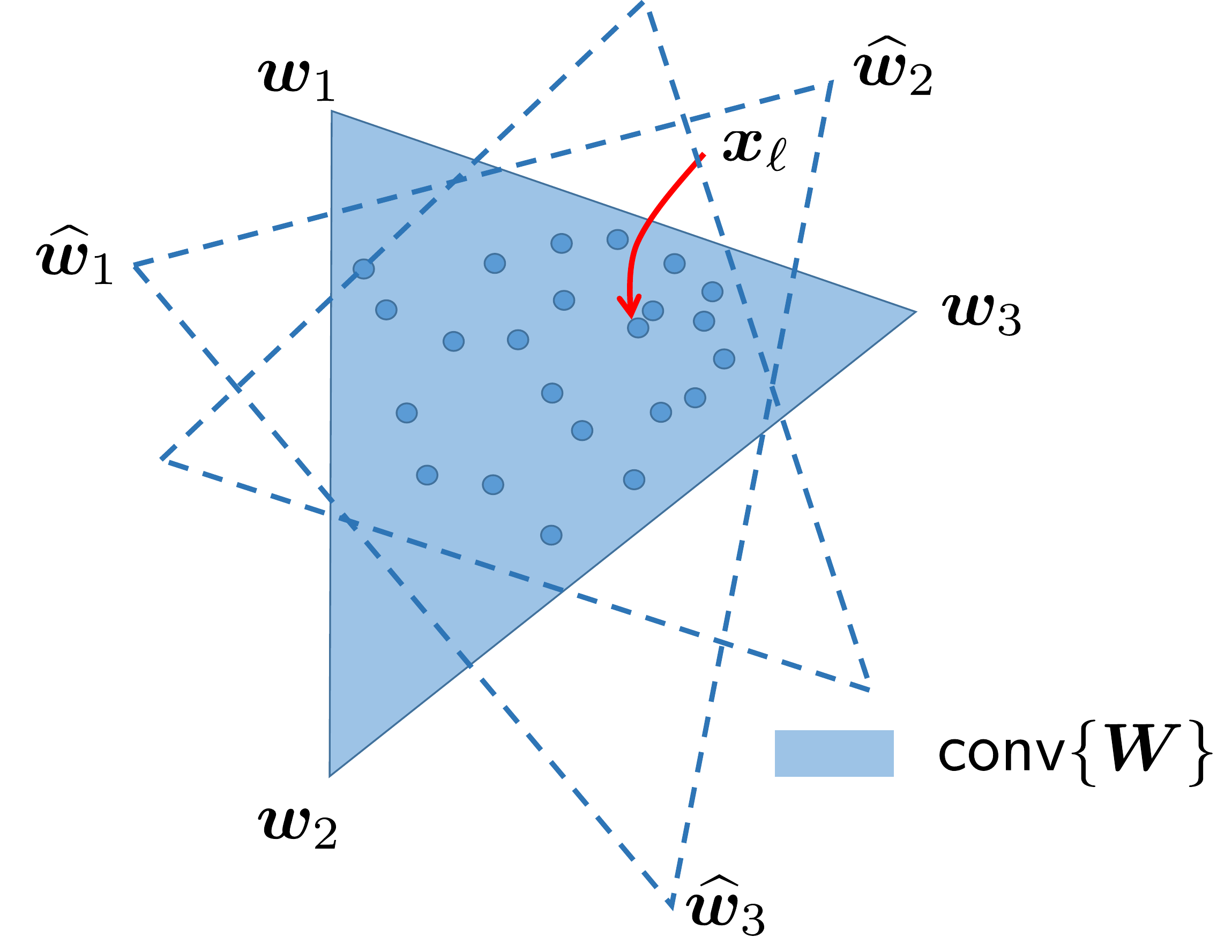}
				\caption{{ Illustration of the ill-posedness of NMF. The dashed lines represent ${\sf conv}\{\widehat{\W}\}$'s that enclose all the $\x_\ell$'s. That is, there are many solutions that satisfy the data model $\X=\W\H^\T$ with $\H\geq \bm 0$ and $\H\bm 1=\bm 1$.}}\label{fig:ill}
			\end{center}
		\end{figure}

		When $\H$ is row-stochastic, we have
		\[ {\bm x}_\ell \in \conv{\W},~\ell=1,\ldots,N, \] 
		where { $ \conv{\W}=\{\bm x\in\mathbb{R}^M|\bm x =\bm W{\bm \theta},{\bm \theta}\geq{\bm 0},~\bm \theta^\T{\bm 1}=1 \}$} denotes the {\it convex hull} of the columns of $\W$. When $\rank(\W)=R$, { it can be shown that} the columns of $\W$ are also the vertices of this convex hull ({ note that the converse is not necessarily true}).
		The convex hull is also called a {\it simplex} if the columns of $\W$ are {\it affinely independent}, i.e., if $\{\bm w_r-\bm w_R\}_{r=1,\ldots,R-1}$ is linearly independent.
		The two geometric interpretations of NMF are shown in Fig.~\ref{fig:geometry}. 
		From the geometries, it is also obvious that NMF is equivalent to recovering a data enclosing conic hull or simplex---which can be very ill-posed as many solutions may exist. { To be more precise, let us take the later case as an example: Assume $\W\geq{\bm 0}$. Then $\x_\ell\in\mathbb{R}^{M}_+$ holds for all $\ell$.
			Any $\widehat{\W}=[\widehat{\bm w}_1,\ldots,\widehat{\bm w}_R]$ that satisfies ${\sf conv}\{\X\}\subseteq{\sf conv}\{\widehat{\W}\}\subseteq \mathbb{R}^{M}_{+}$ also satisfies the data model $\X=\widehat{\W}\widehat{\H}^\T={\W}\H^\T$ for some $\widehat{\H}$, where both $\H$ and $\widehat{\H}$ satisfy \eqref{eq:hsto}, $\W\neq \widehat{\W}$, and $\H\neq \widehat{\H}$---see Fig.~\ref{fig:ill}. }
		
		%
		
		
		\subsubsection{Key Characterizations}
		{Let us take another look at Fig.~\ref{fig:ill}. Assume that $\W\geq \bm 0$ and $\H$ satisfies the row-stochasticity condition in \eqref{eq:hsto}. Intuitively, if the data points $\x_\ell$'s are `well spread' in $\mathbb{R}^{M}_{+}$ such that $\conv{\X}\approx \mathbb{R}^{M}_+$, then it would be hard to find a $\widehat{\bm W}$ such that $\conv{\X}\subseteq{\sf conv}\{\widehat{\W}\}\subseteq \mathbb{R}^{M}_{+}$ holds---and a unique factorization model is then possibly guaranteed.} This suggests that, to understand the identifiability of different NMF approaches, it is important to characterize the `distribution' of vectors within the nonnegative orthant.

		{ In the literature, such characterizations are usually described in terms of the geometric properties of $\H$ (or $\W$, by role symmetry). 
			The scattering of $\H(\ell,:)$'s is directly translated to that of $\x_\ell$'s since they are linked by a full column rank matrix $\W$.}
		The following is arguably the most frequently used characterization:

	\begin{figure}
			\centering
			\includegraphics[width=1\linewidth]{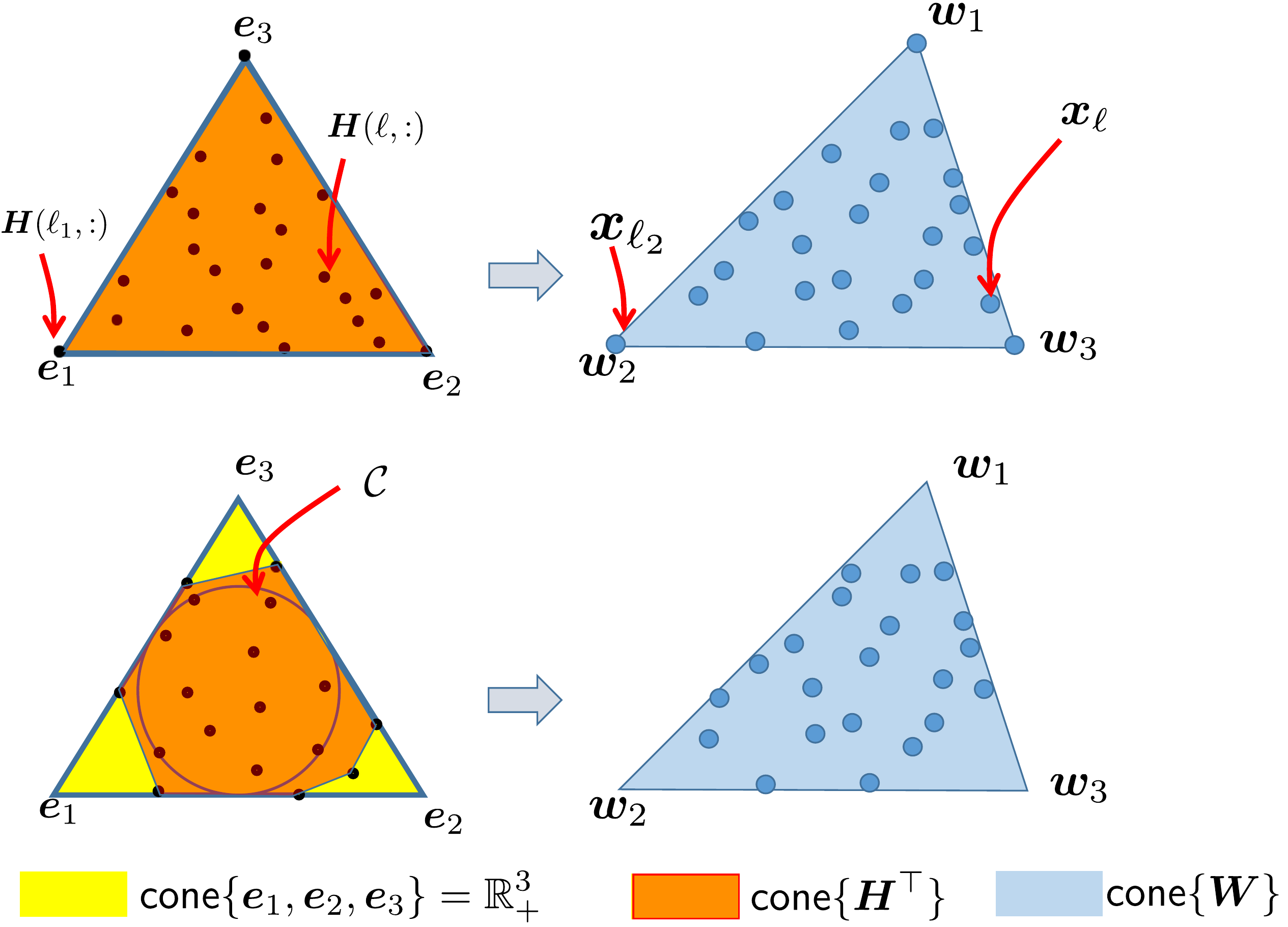}
			\caption{ Illustration of the separability (top) and sufficiently scattered (bottom) conditions where $M=R=3$. The left and right columns are visualized by assuming that the viewer is facing ${\sf cone}\{\bm e_1,\bm e_2,\bm e_3\}$ (towards the origin) and ${\rm cone}\{\W\}$, respectively. The left column is illustrated in the $\H$-domain. The right hand side shows the corresponding views in the data} \label{fig:cond}
	\end{figure}

		\begin{figure}

				\centering
				\includegraphics[width=0.75\linewidth]{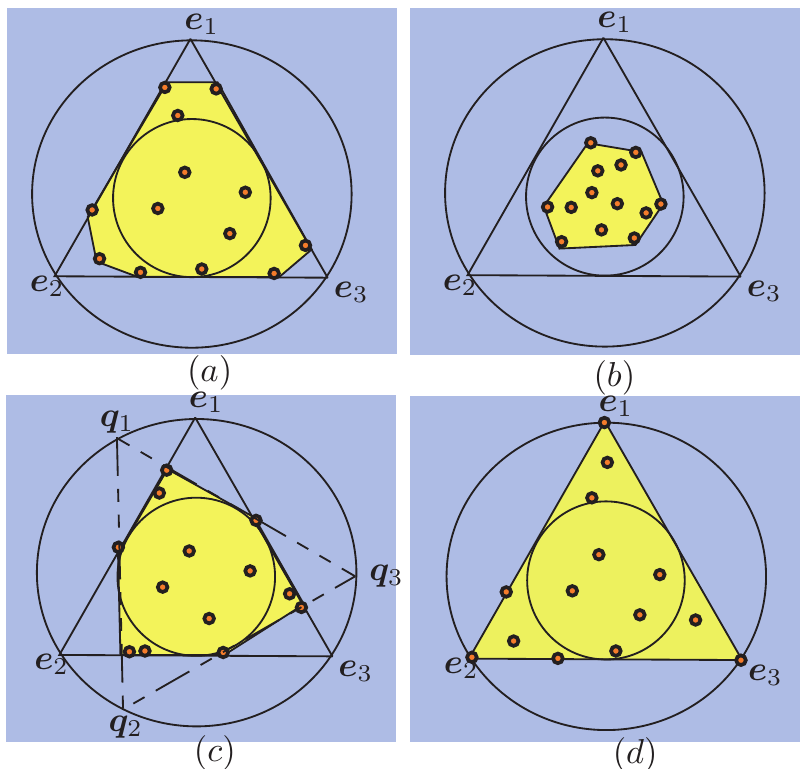}
				\caption{{ The inner circle corresponds to ${\cal C}$, the shadowed polytope corresponds to ${\rm cone}\{\H^\T\}$, the outer circle corresponds to $\{{\bm x}\in\mathbb{R}^R|{\bm 1}^\T{\bm x}\geq \|{\bm x}\|_2\}$, and the red dots correspond to $\H(\ell,:)$'s. (a) $\H$ satisfies the sufficiently scattered condition; (b) $\H$ does not satisfy the sufficiently scattered condition; (c) ${\cal C}\subseteq {\sf cone}\{\H^\T\}$ but the regularity condition is violated; (d) $\H$ satisfies the separability condition. Figure reproduced (permission will be sought) from \cite{fu2015blind}, $\copyright$ IEEE.}}
				\label{fig:reg}
		\end{figure}

		\begin{Def}[Separability] \label{def:sep}
			A nonnegative matrix $\bm H \in\mathbb{R}^{N\times R}$ is said to satisfy the separability condition if
			\begin{equation}\label{eq:sep}
				\cone{\H^\T} =\cone{\bm e_1,\ldots,\bm e_R}= \mathbb{R}_+^R.  
			\end{equation}
		\end{Def}

		Note that in Definition~\ref{def:sep}, the way we state separability is somewhat different from that in the literature \cite{donoho2003does,Gillis2012,laurberg2008theorems,huang2014non,fu2015self}; our intention with using  Definition~\ref{def:sep} is to provide geometric interpretation. In the literature, $\H$ is said to satisfy the separability condition if for every $r=1,...,R$, there exists a row index $\ell_r$ such that 
	\[\bm H({\ell_r},:) = \alpha_r\bm{e}_r^\T,\] where $\alpha_r>0$ is a scalar and $\bm{e}_r$ is the $r$th coordinate vector in $\mathbb{R}^R$. When \eqref{eq:hsto} is satisfied, $\alpha_r=1$ for $r=1,\ldots,R$---which is exactly what we summarized in Definition~\ref{def:sep}. 
		\color{black}
	
The separability condition\footnote{Another remark is that, in the literature, often times separability is defined for $\X$; i.e., when $\H$ satisfies Definition~\ref{def:sep}, it is said that $\X$ is separable. In this tutorial, we do not use this notion of separability. Instead, we unify the geometric characterizations of the NMF model using the structure of the latent factors as in Definitions~\ref{def:sep}-\ref{def:suf}.} has been used to come up with effective NMF algorithms in many fields---but it is considered a relatively restrictive condition.
Essentially, it means that the extreme rays of the nonnegative orthant (i.e., $\alpha_r\bm e_r$ for $\alpha_r>0$) are `touched' by some rows of $\bm H$ and thus the conic hull of $\bm H^\T$ { equals to the entire nonnegative orthant; see Fig.~\ref{fig:cond} (top). Correspondingly, in the data domain ({ corresponding to the right column of Fig.~\ref{fig:cond}}), one can see that there exists $\x_{\ell_r}=\alpha_r\bm w_r$ for each $r=1,\ldots, R$, i.e., the $\x_\ell$'s are actually `touching the corners' in $\cone{\W}$---we have $\cone{\X}=\cone{\W}$ under separability.}
Another condition that can effectively model the spread of the nonnegative vectors but is much more relaxed \cite{huang2014non,fu2015blind,fu2017on} (see also \cite{lin2014identifiability,lin2017maximum}) is as follows:

\begin{Def}[Sufficiently Scattered]\label{def:suf}
	A {nonnegative matrix $\bm H\in\mathbb{R}^{N\times R}$} is said to be sufficiently scattered if the following two conditions are satisfied:
	\begin{enumerate}
		\item the rows of $\H$ are spread enough so that
		\begin{equation}\label{eq:suff}
		{\cal C} \subseteq \cone{\bm H^\T}  
		\end{equation}
		where ${\cal C}$ is a second-order cone defined as
		$\mathcal{C} = \{\x \in\mathbb{R}^R~|~ \x^\T\mathbf{1} \geq \sqrt{R-1}\|\x\|_2\}.$
		\item ${\sf cone}\{\H^\T\}\subseteq {\sf cone}\{\Q\}$ does not hold for any orthonormal $\Q$ except the permutation matrices.
	\end{enumerate}
\end{Def}

\begin{figure}
		\centering
		\includegraphics[width=.8\linewidth]{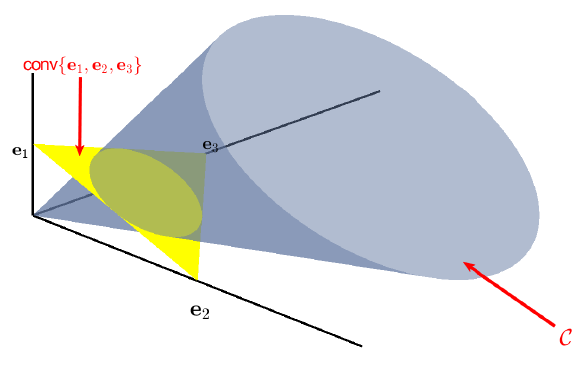}
		\caption{A closer look at the second-order cone ${\cal C}$ for $R=3$ from another angle. A \textit{second-order cone} \cite{CVX} is sometimes called the ice cream cone due to its shape in the three-dimensional space.}
		\label{fig:coneplot}
\vspace{-.5cm}
\end{figure}

To understand the sufficiently scattered condition, one key is to understand the second-order cone ${\cal C}$.
This second-order cone is tangent to every facet of the nonnegative orthant (see Figs.~\ref{fig:reg}--\ref{fig:coneplot}). 
Hence, if $\H$ satisfies the sufficiently scattered condition, it means that the rows of $\H$ are spread enough in the nonnegative orthant (at least every facet of the nonnegative orthant has to be `touched' by some rows of $\H$; see Fig.~\ref{fig:reg} (a)).  
The second condition has a couple of equivalent forms \cite{huang2014non,fu2015blind} and seems a bit technical at first glance. Geometrically, it means that ${\sf cone}\{\H^\T\}$ needs to be slightly `larger' than ${\cal C}$, not just tangentially containing it; see \cite{fu2015blind,fu2016robust} and the illustration in Fig.~\ref{fig:reg} (c) for a case that violates the regularity condition.

\color{black}

However, the rows of $\bm H$ need not be so `extremely scattered' that its conic hull `covers' the entire nonnegative orthant, but only `well scattered' to contain the second-order cone ${\cal C}$---which is much more relaxed than the separability condition. Fig.~\ref{fig:reg} gives more illustrations on different scenarios for ${\sf cone}\{\H^\T\}$. A key difference in the ambient data geometries resulting from Definitions~\ref{def:sep}-\ref{def:suf} is that under the sufficiently scattered condition, there need not be data points touching the $\bm w_{r}$'s; see Fig.~\ref{fig:cond} (bottom right).

\color{black}

Before we move to the next section, one important remark is that we will use $\W_\natural$ and $\H_\natural$ in the sequel to denote the true generating factors that give rise to the data $\X=\W_\natural\H_\natural^\T$, to distinguish them from the optimization variables $\W$ and $\H$ in the NMF criteria.
We will assume the ground-truth generating factors $\W_\natural$ and $\H_\natural$ to be full-rank throughout this article, since most NMF methods work under this condition (although there exist interesting exceptions; see \cite{gillis2014successive}). 	
	
\section{Separability-Based Methods for NMF}\label{sec:sep}
In this section, we will introduce some representative approaches that explicitly make use of the separability condition (cf. Definition 2) and have identifiability guarantees.
From a terminology point of view, the separability condition was first coined by Donoho \textit{et al.} in 2004 \cite{donoho2003does}. However, this condition was used for NMF in the context of hyperspectral unmixing well before 2004. In HU, researchers noticed that there are instances in which the remotely sensed hyperspectral images have many pixels containing only a single material; i.e., if noise is absent, there exist induces $\ell_r$ for $r=1,\ldots,R$ such that $\bm x_{\ell_r}=\W_\natural(:,r)$ (or equivalently,  $\H_\natural(\ell_r,r)=1$ and $\H_\natural(\ell_r,j)=0$ for $j\neq r$). The existence of such `pure pixels' has been exploited to come up with quite effective hyperpsectral unmixing algorithms \cite{VCA,VMAX,MC01}. 

To understand the reason why separability helps,
we assume that $\H_\natural{\bm 1}={\bm 1}$ and $\H_\natural\geq\bm 0$, i.e., $\H_\natural$ is row-stochastic. 
Under separability, there is { an index set $\varLambda =\{\ell_1,\ell_2,\ldots,\ell_R\}$ that collects the indices satisfying $\H_\natural(\ell_r,:)=\bm e_r^\T$ for $r=1,\ldots,R$.}
As we have seen in Fig.~\ref{fig:cond} { (top)}, the vertices of the convex hull are now `touched' by some data samples (i.e., columns of $\X$). In other words, we have $\X(:,\varLambda)=\W_\natural$ in the noiseless case. Instead of finding a simplex that encloses all the data columns, now the problem boils down to finding data columns which span a convex hull (or conic hull) that encloses all the other data columns. If $\varLambda$ can be successfully identified, one can simply retrieve the latent factors via $\widehat{\W}=\X(:,\varLambda)$ and $\widehat{\H}^\T=\widehat{\bm W}^\dag \X$ (or $\widehat{\H}=\arg\min_{\H\geq{\bm 0}}~\|\X-\widehat{\W}\H^\T\|^2_F$).




\subsection{Convex Formulations}
\noindent
$\bullet$ {\bf Basic Insight}~Under separability, the critical step towards identifying $\W_\natural$ is to design an efficient algorithm to `pick out' the vertices of the convex hull of the data (or extreme rays if $\H_\natural{\bm 1}=\bm 1$ is not assumed) from $\x_1,\ldots,\x_N$. It turns out that this can be done via a variety of convex formulations. The idea behind is based on a simple fact: The vertices (resp. extreme rays) of a convex hull (resp. conic hull) cannot be represented by any other vectors that reside inside the convex hull ({ resp.} conic hull) using their convex combinations ({ resp.} nonnegative combinations). For example, if noise is absent, { all $ \x_\ell$'s are distinct}, Eq.~\eqref{eq:hsto} (row-stochasticity of $\H_\natural$) is satisfied, and $\ell \in \varLambda$, then we always have
\begin{equation}\label{eq:cvx_1}
	\left(\min_{{\bm \theta}\geq{\bm 0},{\bm 1}^\T{\bm \theta}=1}~\|\x_{\ell}-\X_{-\ell}{\bm \theta}\|_q\right) > {0} ,  
\end{equation}
where the notation $\X_{-\ell}$ means that $\X_{-\ell} = [~ \x_1,\ldots,\x_{\ell-1}, \x_{\ell+1}, \ldots \x_N ~]$ (i.e., the column $\x_\ell$ is removed from $\X$),
$q \geq 1$,
and $\| \cdot \|_q$ denotes the (convex) $q$-norm. When $\ell\notin\varLambda$, we always have
$\min_{{\bm \theta}\geq{\bm 0},{\bm 1}^\T{\bm \theta}=1}~\|\x_{\ell}-\X_{-\ell}{\bm \theta}\|_q={0}$ since $\x_\ell \in \conv{\X(:,\varLambda)}$.

A relatively naive implementation of separability-based NMF is to repeatedly solve Problem~\eqref{eq:cvx_1} for each $\ell\in\{1,\ldots,N\}$ and compare the objective values, which is the basic idea in \cite{arora2012learning,naanaa2005blind}. It can be shown that, even if there is noise, \eqref{eq:cvx_1} can be used to estimate  $\widehat{\bm W}$ and there is some guarantee on the noise robustness of the estimate \cite{arora2012learning}. However, this formulation needs to solve $N$ convex programs, each of which has $N-1$ decision variables, which can be quite inefficient when $N$ is large. 

\noindent
$\bullet$ {\bf Self-dictionary Convex Formulations}~To avoid solving a large number of convex programs, a class of convex optimization-based NMF methods were proposed \cite{Esser2012,recht2012factoring,fu2015self,fu2015robust,gillis2018fast}. The main idea of these methods is the same as that in \eqref{eq:cvx_1}, i.e., utilizing the self-expressiveness of 
$\X$ by $\X(:,\varLambda)$ under the separability assumption.
By assuming row-stochasticity of $\H_\natural$, the formulations in this line of work can be cast in the framework of block-sparse optimization:
\begin{subequations}\label{eq:rowzero}
	\begin{align}
		\minimize_{{\bm C}}&~\|\C\|_{{\rm row}-0}\\
		{\rm subject~to}&~\X=\X\C\\
		&~{\bm C}\geq{\bm 0},~ {\bm 1}^\T{\bm C}={\bm 1}^\T,
	\end{align}
\end{subequations}
where $\|\bm Y\|_{{\rm row}-0}$ counts the number of non-zero rows in $\bm Y$. To understand the formulation, consider, without loss of generality, a case where $\X=[\W_\natural,\X']$, i.e., $\varLambda =\{1,\ldots,R\}$. Then, one can see that
\[        \C^\star =\begin{bmatrix}
\H_\natural^\T\\
{\bm 0}
\end{bmatrix},     \]
is an optimal solution. Intuitively, the reasons are that 1) $\X\C^\star=\W_\natural\H_\natural^\T$ and 2) no other nonzero rows of $\C^\star$ can be removed, since $\X(:,\varLambda)$ can only be represented by $\X(:,\varLambda){\bm H}_\natural(\varLambda,:)^\T$.
One can easily identify $\varLambda$ via inspecting the indices of the nonzero rows of $\C^\star$.

The formulation in \eqref{eq:rowzero} is nonconvex due to the combinatorial nature of the `row zero-norm'. Nevertheless, it can be handled by relaxing $\|\cdot\|_{{\rm row}-0}$ to a convex function, e.g., $\|\bm Y\|_{q,1}=\sum_{m=1}^M \|\bm Y(m,:)\|_q$ where $\bm Y\in\mathbb{R}^{M\times N}$ and $q\geq 1$ \cite{fu2015power}. In addition, it was proven in \cite{fu2015robust} that when there is noise, i.e., $\x_\ell=\W_\natural\H_\natural(\ell,:)^\T+\bm v_\ell$, the modified versions such as
\begin{subequations}\label{eq:Linf}
	\begin{align}
		\minimize_{~{\bm C}\geq{\bm 0},~ {\bm 1}^\T{\bm C}={\bm 1}^\T,}&~\|\C\|_{\infty,1}\\
		{\rm subject~to}&~\|\x_\ell-\X\bm c_\ell\|_q\leq \epsilon,~\forall~\ell
	\end{align}
\end{subequations}
\color{black}
where  { $q=1$ \cite{gillis2013robustness,recht2012factoring} or $q=2$ \cite{fu2015self,fu2015robust},}  $\bm c_\ell$ is the $\ell$th column of $\C$, and $\epsilon$ is a pre-specified parameters, can guarantee correct identification of $\varLambda$, provided that the noise is not too significant and some more assumptions ({ e.g., there are no repeated columns in the noiseless data $\W\H^\T$}) are satisfied.

As we have seen, the self-dictionary sparse optimization methods are appealing in terms of vertex identifiability and noise robustness. 
They can also facilitate the derivation of distributed algorithms, if the $\X(m,:)$'s are collected by different agents \cite{fu2015power}.
The downside, however, is complexity. All these self-dictionary formulations lift the number of decision variables to $N^2$, which is not affordable when $N$ is large. 
In \cite{recht2012factoring}, a linear program-based algorithm was employed to handle a self-dictionary criterion, with judiciously designed updates. In \cite{gillis2018fast}, a first-order fast method was proposed. However, the orders of the computational and memory complexities are still high. { One workaround is to use some simple methods, such as the greedy methods to be introduced in the next subsection, to identify a small subset of $\x_\ell$ that are candidates for the vertices, and then apply the self-dictionary methods to refine the results \cite{ammanouil2014blind}.}

\bigskip

{
There are other convex formulations under separability. For example, recently, a new method has been proposed to identify the vertices of ${\sf conv}\{\W_\natural\}$ via  finding a minimum-volume ellipsoid that is centered at the origin and encloses the data columns $\bm x_1,\ldots,\bm x_N$ and their `mirrors' $-\bm x_1,\ldots,-\bm x_N$ \cite{mizutani2014ellipsoidal,gillis2015semidefinite}. 
This method has a small number of primal optimization variables (i.e., $R^2$ primal variables) and thus could potentially lead to more economical algorithms relative to the self-dictionary formulations. The downside is that the method requires accurate knowledge of $R$, which the self-dictionary methods do not need---and this is an additional advantage of the self-dictionary methods \cite{fu2015self}. See more details of ellipsoid volume minimization-based NMF in the supplementary materials.
}

\subsection{Greedy Algorithms}
Another very important class of methods that can provably identify $\varLambda$ is greedy algorithms---i.e., the class of algorithms that identify $\ell_1,\ldots,\ell_R$ one by one. 
Here we describe a particularly simple algorithm in this context.
Assuming that $\X=\W_\natural\H_\natural^\T$ and $\H_\natural$ satisfies { the row-stochasticity condition in} \eqref{eq:hsto},
the algorithm discovers the first vertex of the target convex hull by 
\begin{equation}\label{eq:spa}
	\hat{\ell}_1 =\arg\max_{\ell \in\{1,\ldots,N\}}~\|\x_\ell\|_q,  
\end{equation}   
where $1<q\leq\infty$.
The proof is simple: given that $\H_\natural$ satisfies \eqref{eq:hsto}, we have
\begin{align*}
	\|\x_\ell\|_q &= \left\|\W_\natural\H_\natural(\ell,:)^\T\right\|_q\\& \leq \sum_{r=1}^R \H_\natural(\ell,r)\left\|\W_{\natural}(:,r)\right\|_q \leq \max_{r\in\{1,\ldots,R\}}~\|\W_{\natural}(:,r)\|_q,
\end{align*}
{ where the maximum is attained if and only if $\H_\natural(:,\ell)$ is a unit vector}.
Note that the first inequality is by the triangle inequality and the second uses the fact that $\H_\natural$ satisfies \eqref{eq:hsto}. After finding the first column index that belongs to $\varLambda$, the algorithm projects all the data vectors $\x_\ell$ onto the orthogonal complement of $\X(:,\hat{\ell}_1)$, i.e., forming \begin{equation} \label{eq:SPA_proj}
{\x_\ell:=\bm P^\perp_{\X(:,{\hat{\ell}_1})}\x_\ell, \quad \text{ for $\ell=1,\ldots,N$.} }
\end{equation} Then, if one repeats \eqref{eq:spa}, the already found vertex will not appear again and the next vertex will be identified. Repeating \eqref{eq:SPA_proj} for $R$ times will enable us to identify all the vertices.

The described algorithm has many names, e.g., successive simplex volume maximization (SVMAX) \cite{VMAX} and self-dictionary simultaneous matching pursuit (SD-SOMP) \cite{fu2015self}, since it was re-discovered from different perspectives ({see the supplementary materials for more detailed discussion}). The algorithm has also many closely related variants, such as the vertex component analysis (VCA) algorithm \cite{VCA} and FastAnchor \cite{arora2012practical}, which share the same spirit.
The most commonly used name for this algorithm in signal processing and machine learning is {\it successive projection algorithm} (SPA), which first appeared in \cite{MC01}. { Bearing a similar structure as the Gram-Schmidt procedure}, SPA has very lightweight updates: both the $q$-norm comparison and the orthogonal projection can be carried out easily. It was shown by Gillis and Vavasis in \cite{Gillis2012} that the algorithm is robust to noise under most convex norms (except $q=1$). It was further shown that using $q=2$ exhibits the best noise-robustness---which is consistent with what has been most frequently used in many fields \cite{MC01,VMAX,fu2015blind,arora2012practical}.
The downside of SPA is that it is prone to error propagation, as with all the other greedy algorithms.

\bigskip

{ A remark is that most separability-based methods work without assuming $\W_\natural$ to be nonnegative, if $\H_\natural$ is row-stochastic. 
This allows us to apply the algorithms to handle problems like community detection under the model in \eqref{eq:cd}.
However, when $\H_\natural$ does not satisfy \eqref{eq:hsto}, then row-stochasticity has to be enforced through column normalization of $\X$, which does not work without the assumption $\W_\natural\geq {\bm 0}$ \cite{Gillis2012}; also see the previous insert on column normalization. }

\section{Separability-Free Methods for NMF}
Separability-based NMF approaches are usually associated with elegant formulations and tractable algorithms. They are also well understood---the performance under noisy scenarios has been studied and characterized. On the other hand, the success of these algorithms hinges on the separability condition, which makes them `fragile' in terms of model robustness. Separability usually has plausible physical interpretations in applications. { As discussed before, in hyperspectral imaging, separability means that there are spectral pixels that only contain a single material, i.e., pure pixels.
Moreover, in text mining, separability means that every topic has some characteristic words, which other topics do not use. These are legitimate assumptions to some extent.} 
However, there is also a (sometimes considerable) risk that separability does not hold in practice. If separability is (grossly) violated, can we still guarantee identifiability of $\W_\natural$ and $\H_\natural$? We will review a series of important recent developments that address this question.

\subsection{Plain NMF Revisited} 
The arguably most popular NMF criterion is the fitting based formulation defined in \eqref{eq:nmf_fit}, i.e.,  \[ \minimize_{\W\geq{\bm 0},~{\bm H}\geq {\bm 0},}
\left\| \X - \W\H^\T\right\|_F^2.\]
{ For convenience, we will call \eqref{eq:nmf_fit} the {\em plain NMF} criterion in the sequel.}
To study the identifiability of the criterion in \eqref{eq:nmf_fit}, it suffices to study the identifiability of the following feasibility problem:
\begin{subequations}\label{eq:feas}
	\begin{align}
		{\rm find}~&\W,\H\\
		{\rm subject~to}~&\X=\W\H^\T \label{eq:x=wh}\\
		&\W\geq {\bm 0},~\H\geq {\bm 0},
	\end{align}
\end{subequations}
since \eqref{eq:nmf_fit} and \eqref{eq:feas} share the same optimal solutions if noise is absent.
In \cite{donoho2003does}, Donoho \textit{et al.} derived the first result on the identifiability of \eqref{eq:nmf_fit} and \eqref{eq:feas}.  In 2008, Laurberg \textit{et al.} \cite{laurberg2008theorems} came up with a rather similar sufficient condition. 
In a nutshell, the sufficient conditions in \cite{donoho2003does} and \cite{laurberg2008theorems} can be summarized as follows: 1) the $\H_\natural$ matrix satisfies the separability condition; and 2) there is a certain zero pattern in $\W_\natural$. { For example, the condition in \cite{donoho2003does} requires that every column of $\W_{\natural}$ has a zero element, and the zero elements from different columns appear in different rows. Geometrically, this assumes that there are $\W_{\natural}(m,:)$'s `touching' every facet of $\mathbb{R}_+^R$---see a detailed summary in \cite{huang2014non}}. 
Huang \textit{et al.} \cite{huang2014non} derived another interesting sufficient condition for NMF identifiability in 2014:

\vspace{.2cm}
\noindent
\boxed{
	\begin{minipage}{.99\linewidth}
		\textit{If both $\W_\natural$ and $\H_\natural$ are sufficiently scattered (cf. Definition 3), then any optimal solution $(\W_\star,\H_\star)$ to \eqref{eq:nmf_fit} (and the feasibility problem in \eqref{eq:feas}) must satisfy $\W_\star =\W_\natural\bm \varPi\bm D$ and $\H_\star = \H_\natural\bm \varPi\bm D^{-1}$, { where ${\bm \varPi}$ is a permutation matrix and ${\bm D}$ is a full-rank diagonal matrix}.}
	\end{minipage}
}
\vspace{.2cm}

\begin{figure}
	\centering
	\includegraphics[width=1\linewidth]{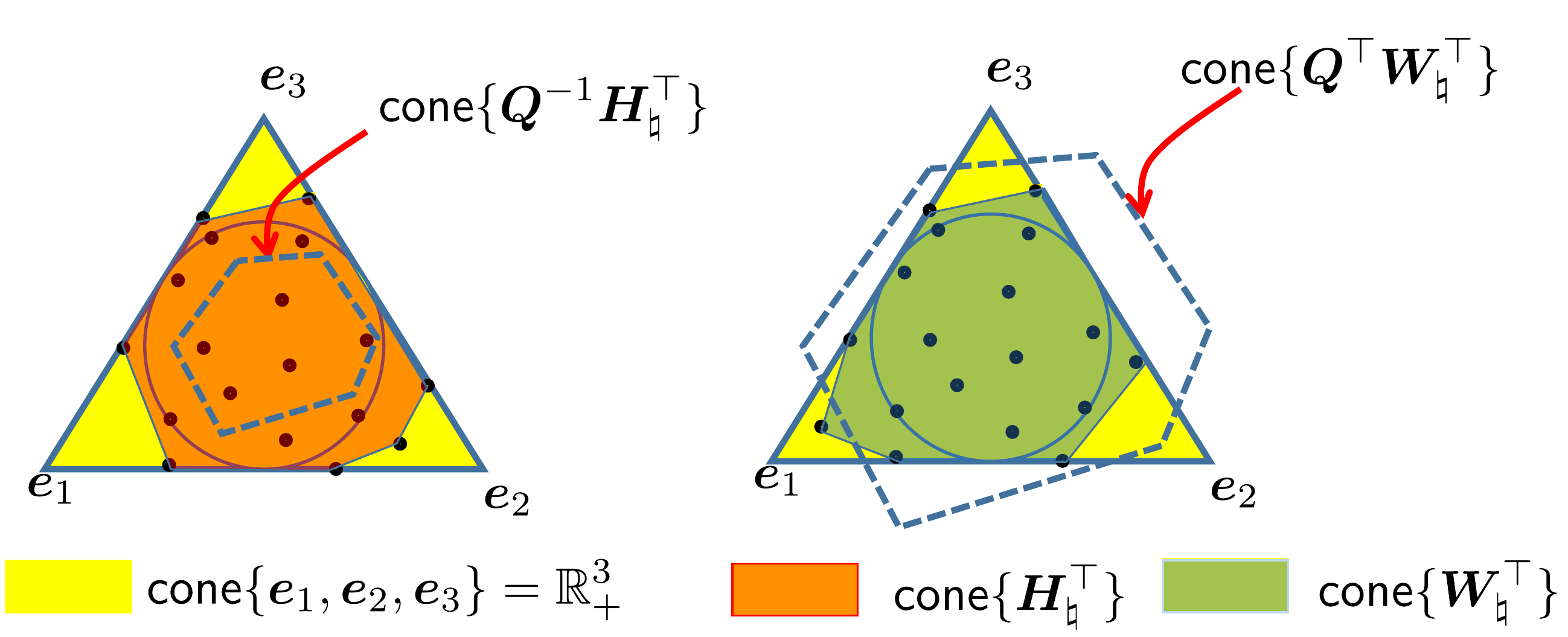}
	\caption{Intuition behind the NMF identifiability conditions in \cite{huang2014non,donoho2003does,laurberg2008theorems}.}\label{fig:plain}
	\vspace{-.85cm}
\end{figure}

\noindent
The insights behind the three sufficient conditions in \cite{donoho2003does,laurberg2008theorems,huang2014non} can be understood in a unified way by looking at Problem~\eqref{eq:feas}.
Specifically, let us assume that there exists a $\Q$ such that $\X=\W_\natural\Q(\H_\natural\Q^{-\T})^\T$ holds.
Intuitively, $\Q^{-\T}$ is an operator that can rotate, enlarge, or shrink the red color-shaded region in Fig.~\ref{fig:plain} (left) (note that this shaded region represents ${\sf conv}\{\H_{\natural}^\T\}$ here). If $\H_\natural$ is separable or sufficiently scattered and $\H_\natural\Q^{-\T}$ is constrained to be nonnegative, one can see that $\Q^{-\T}$ cannot rotate or enlarge the area since this will result in negative elements of $\H_\natural\Q^{-\T}$. In fact, it cannot shrink the area either---since then $\Q$ will enlarge a corresponding area determined by  ${\sf cone}\{\W_\natural^\T\}$ (cf. Fig.~\ref{fig:plain} (right)). If $\W_\natural$ is sufficiently scattered (Huang \textit{et al.}'s condition) or if there are some rows in $\W_\natural$ touching the boundary of the nonnegative orthant (Donoho \textit{et al.} and Laurberg {\it et al.}'s conditions), then $\W_\natural\Q$ will be infeasible. 
Hence, $\Q$ can only be a permutation and column-scaling matrix.

Although Huang \textit{et al.}'s identifiability condition does not, strictly speaking, subsume the conditions in \cite{donoho2003does} and \cite{laurberg2008theorems}, the former presents a significant departure from the separability condition which \cite{donoho2003does} and \cite{laurberg2008theorems} rely upon.  
The region covered by ${\cal C}$ rapidly shrinks at a geometric rate when $R$ grows and thus the sufficiently scattered condition can be expected to be satisfied in many cases---see \cite{huang2018hmm} and the next insert. The sufficiently scattered condition {\it per se} is also a very important discovery, which has not only helped establish identifiability of the plain NMF criterion, but has also been generalized and connected to many more other important NMF criteria, which will be introduced next.

\begin{figure}
		\centering
		\includegraphics[width=.8\linewidth]{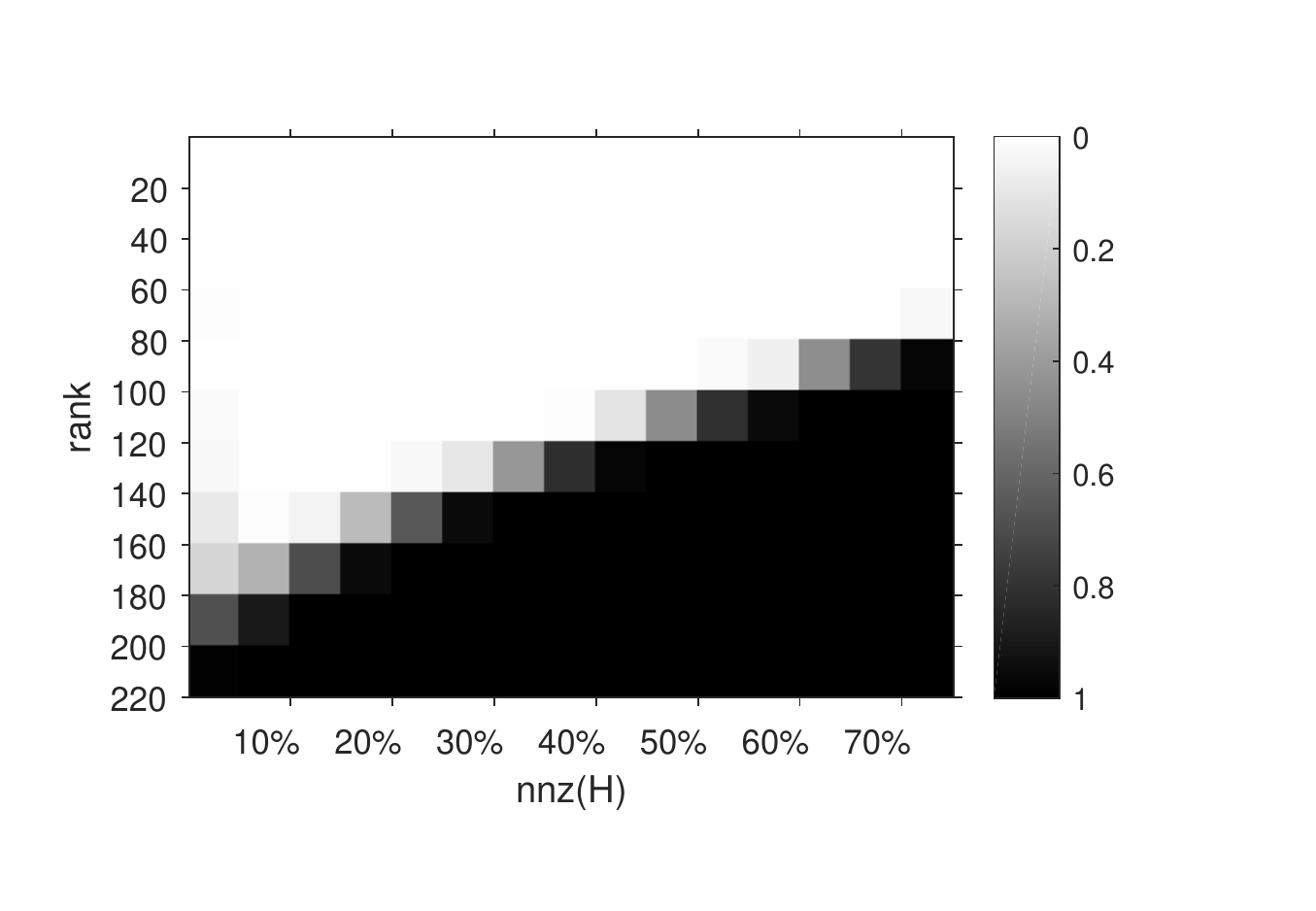}
		\caption{{ Transition plot of $\H\in\mathbb{R}^{500\times R}$; in the grayscale bar, ``1'' means that $\bm H$ does not satisfy the sufficiently scattered condition; the zero elements appear uniformly at random in different locations of the matrix.}}
		\label{fig:trans}
\end{figure}

\begin{shaded*}	
	\noindent
	{\bf Is the sufficiently scattered condition easy to satisfy?}
	One possibly annoying fact is that the sufficiently scattered condition is NP-hard to check \cite{huang2014non}. One way to empirically check the condition is as follows. Consider the following optimization problem:
	\begin{align}\label{eq:check}
		\maximize_{\x}~~ & \|\x\|_2^2	\\ 
		\text{subject to }~~ & \H\x \geq 0,~~ \x^\T{\bf 1} = 1. \nonumber
	\end{align}
	It can be shown that $\H$ is sufficiently scattered if and only if the maximum of the above is attained at $\x\in\{\bm e_1,\ldots,\bm e_R\}$, since the constraint set represents the dual cone ({ the dual cone of a cone ${\cal K}$ is defined as ${\cal K}^\ast =\{\bm y|\bm y^\T\x\geq \bm 0,~\forall \x\in{\cal K}\}$})---and the dual cone is restricted inside $\|\bm x\|_2\leq 1$ and touches the second-order cone $\|\bm x\|_2\leq 1$ at the canonical vectors only, if $\H$ is sufficiently scattered \cite{huang2014non}.  
	
	Problem~\eqref{eq:check} is a nonconvex problem, but can be approximated by successive linearization of the objective.  We have conducted an experiment using randomly generated nonnegative $\H$ with $N=500$ under various { ranks} and density levels. The elements of $\H$ are drawn from the i.i.d. exponential distribution (with the parameter being 1), and then a certain amount of selected entries  (which are chosen uniformly at random across all the entries)  are set to zeros according to a pre-specified density level. 
	Fig.~\ref{fig:trans} shows the probabilities that the generated $\H$ is {\it not} sufficiently scattered (in the grayscale bar, black means all the generated $\bm H$'s do not satisfy the sufficiently scattered condition, and white means the opposite).  One can see that for every fixed ${\rm nnz}(\bm H)$ (recall that ${\rm nnz}(\H)$ counts the nonzero elements of $\H$), there is a very sharp transition happening somewhere, and $R$ is directly related to where the transition happens. 
	
	In fact, as empirically observed in \cite{huang2014non}, for such random $\bm H$, if every column has $R-1$ zero elements, then the sufficiently scattered condition is satisfied with very high probability---which is a fairly mild condition. 

	

%
	\color{black}
\end{shaded*}

\color{black}



%

%

\subsection{Simplex Volume Minimization (VolMin)} 
An important issue that was noticed in the literature of NMF is that a {\it necessary condition} of the plain NMF identifiability is that both $\W_\natural$ and $\H_\natural$ contain some zero elements \cite{huang2014non}. In some applications, this seemingly mild condition can be rather damaging for model identification. For example, in hyperspectral unmixing, the columns of $\W_\natural$ model the spectral signatures of different materials. Many spectral signatures do not contain any zero elements, and dense $\W_\natural$'s frequently arise. 
Under such cases, can we still guarantee identifiability of $\W_\natural$ and $\H_\natural$? The answer is affirmative---and it turns out the main idea behind was proposed almost 30 years ago \cite{craig1994minimum}.
%

\begin{figure}
		\centering
		\includegraphics[width=.8\linewidth]{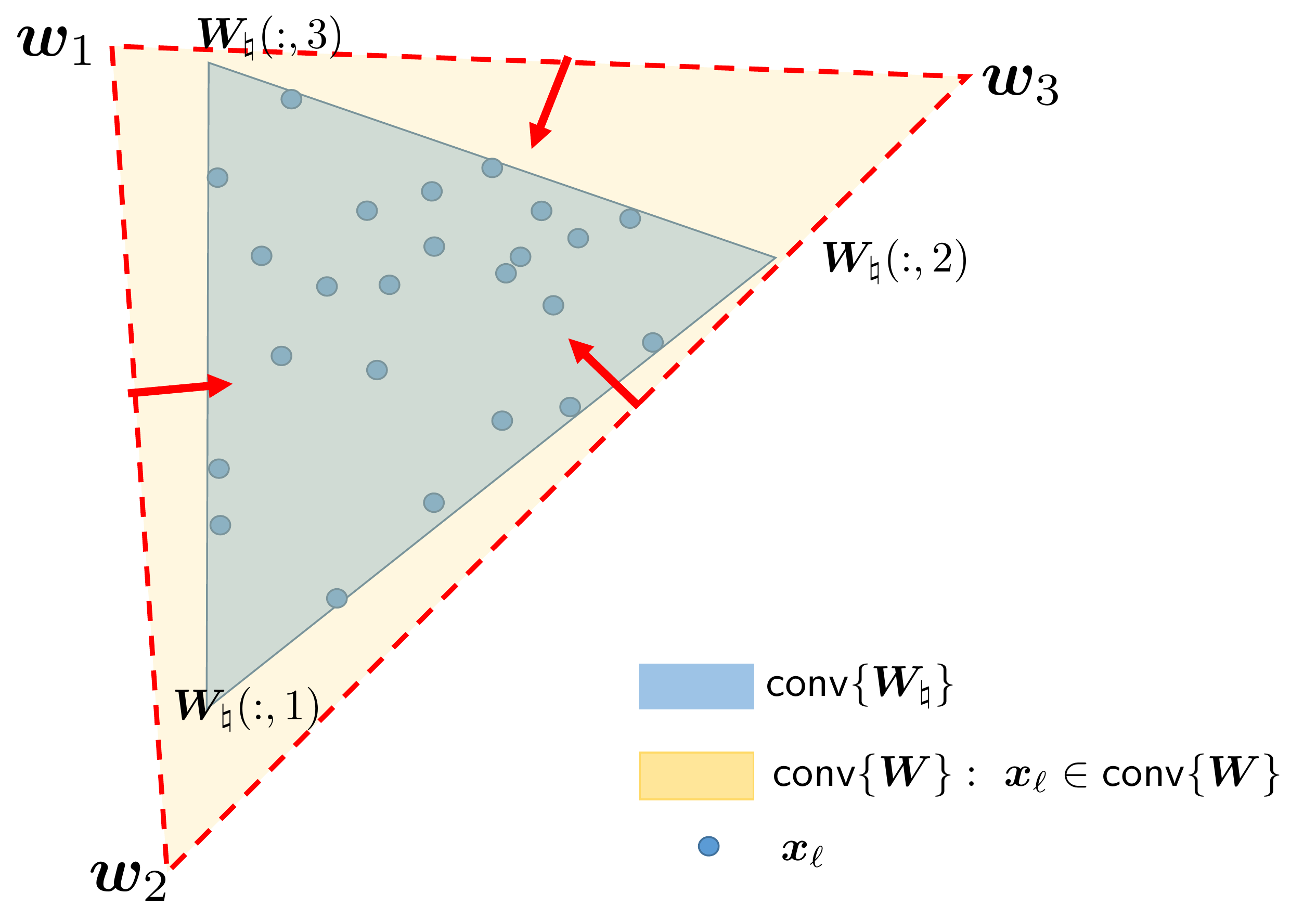}
		\caption{{ Intuition of simplex volume minimization based NMF. Shrinking a data-enclosing simplex to have minimum volume will recover the ground truth ${\sf conv}\{\W_\natural\}$ if the data is sufficiently spread in ${\sf conv}\{\W_\natural\}$.}}
		\label{fig:volminnew}
\end{figure}

To handle the matrix factorization problem without assuming $\W_\natural$ to contain any zeros, let us again assume row-stochasticity of $\H_\natural$, i.e., $\H_\natural{\bm 1}={\bm 1}$, $\H_\natural\geq {\bm 0}$. This way, $\x_\ell\in \conv{\W_\natural}$ holds for $\ell=1,\ldots,N$, as mentioned before. In 1994, Craig proposed an geometrically intriguing way to recover $\W_\natural(:,1),\ldots,\W_\natural(:,R)$ \cite{craig1994minimum}. The so-called {\it Craig's belief} is as follows: {\it if the $\x_\ell$'s are sufficiently spread in  $\conv{\W_\natural}$, then finding the minimum-volume enclosing simplex of $\{\x_1,\ldots,\x_N\}$ identifies $\W_\natural$} (cf. Fig.~\ref{fig:volminnew}).
Craig's paper has provided an elegant geometric viewpoint for tackling NMF. One can see that there are very few assumptions on $\W_{\natural}$ in Craig's belief.
Craig did not provide a mathematical programming form for the stated problem, i.e., finding the minimum-volume data-enclosing simplex. 
In 2015, Fu \textit{et al.} \cite{fu2015blind} formulated this volume-minimization (VolMin) problem as follows:
\begin{subequations}\label{eq:svolmin}
	\begin{align}
		\minimize_{\W,\H}&~\det(\W^\T\W)\\
		{\rm subject~to}&~\X=\W\H^\T\\
		&~\H\geq{\bm 0},~{\bm 1}^\T\H^\T={\bm 1}^\T,
	\end{align}
\end{subequations}
where $\det(\W^\T\W)$ is a surrogate of the volume of $\conv{\bm W}$ and the constraints mean that $\x_\ell \in \conv{\W}$ for every $\ell\in\{1,\ldots,N\}$ (also see other similar formulations in \cite{li2008minimum,bioucas2009variable,MVES,zhou2011minimum}). 

Craig's belief was a conjecture, which had been supported by extensive empirical evidence over the years. However, there had been a lack of theoretical justification. 
In 2015, two parallel works \cite{fu2015blind,lin2014identifiability} pushed forward the understanding of VolMin identifiability substantially. Fu \textit{et al.} showed in \cite{fu2015blind} the following:

\vspace{.2cm}
\noindent
\boxed{
	\begin{minipage}{.98\linewidth}
		{\it If ${\rm rank}(\W_\natural)={\rm rank}(\H_\natural)=R$ and $\H_\natural$ satisfies the sufficiently scattered condition (Definition 3) and \eqref{eq:hsto}, then any optimal solution of Problem~\eqref{eq:svolmin} is $\W_\star = \W_\natural\bm \Pi, \H_\star =\H_\natural\bm \Pi^\T$, { where $\bm \Pi$ is a permutation matrix}.}
	\end{minipage}
}
\vspace{.2cm}


In retrospect, it is not surprising that the identifiability of VolMin relates to the scattering of the rows of $\H_\natural$---since the scattering of $\H_\natural$ translates to the spread of $\x_1,\ldots,\x_N$ in the convex hull spanned by $\W_\natural(:,1),\ldots,\W_\natural(:,R)$, and this is what enables the success of volume minimization. Remarkably, VolMin sets $\W_\natural$ free---it merely requires ${\rm rank}(\W_\natural)=R$ to be satisfied, yet it allows $\W_\natural$ to have negative elements and/or be completely dense. As we have discussed, this can potentially benefit a lot of applications for which the plain NMF cannot offer identifiability. The price to pay is that the optimization problem could be even more cumbersome to handle compared to the plain NMF, due to the determinant term for measuring volume.

\color{black}

\subsection{Towards More Relaxed Conditions}
Simplex volume minimization has significant advantages over plain NMF and separability-based NMF, 
as we have previously seen that the known sufficient identifiability condition of simplex volume minimization is much better or much more relaxed than those of plain NMF and separability-based NMF.
But there is a small caveat: the identifiability is established under the assumption that $\H_\natural$ is row-stochastic. This, in some applications, is not a very restrictive assumption. For example, in hyperspectral imaging, $\H_\natural(\ell,:)$ corresponds to the so-called abundance vector whose elements represent the proportions of the materials contained in pixel $\ell$, and thus is naturally nonnegative and sum-to-one. 
On the other hand, this assumption is not without loss of generality: one cannot assume $\H_\natural\bm 1=\bm 1$ for all the factorization models $\X=\W_\natural\H_\natural^\T$. As mentioned, if $\W_\natural$ and $\H_\natural$ are nonnegative, the sum-to-one condition on the rows of $\H_\natural$ can be enforced by data column normalization \cite{Gillis2012}. However, when $\W_\natural$ contains negative elements, normalization does not work (cf. the insert in Sec.~\ref{sec:nmfandident}). 

Very recently, Fu \textit{et al.} \cite{fu2017on} proposed a simple fix to the above issue. Specifically, the following matrix factorization criterion is proposed:
\begin{subequations}\label{eq:spl}
	\begin{align}
		\minimize_{\W,\H}&~\det\left(\W^\T\W\right)\\
		{\rm subject~to}&~\X=\W\H^\T\\
		&~\H\geq{\bm 0},~{\bm 1}^\T\H=\rho{\bm 1}^\T,
	\end{align}
\end{subequations}
where $\rho>0$ is any positive real number.
The above criterion looks very much like the VolMin criterion in \eqref{eq:svolmin}, but with the constraint on $\H$ changed---the sum-to-one constraint is now on the {\it columns} of $\H$ (if one lets $\rho=1$), which is without loss of generality since column scaling of $\W_\natural$ and counter column scaling of $\H_\natural$ are intrinsic degrees of freedom in every matrix factorization model. In other words, no column normalization is needed for assuming $\bm 1^\T\H=\rho\bm 1^\T$, and this assumption can always be made \textit{without loss of generality}: { $\H\geq{\bm 0}$ and ${\bm 1}^\T\H=\rho{\bm 1}^\T$ imply that $\| \bm H(:,r) \|_1=\rho$ for all $r$, which means that we wish to fix the column scaling of the right latent factor.} It is proven in \cite{fu2017on} that:

\vspace{.2cm}
\noindent
\boxed{
	\begin{minipage}{.98\linewidth}
		{\it If ${\rm rank}(\W_\natural)={\rm rank}(\H_\natural)=R$ and $\H_\natural$ is sufficiently scattered (Definition 3), then the optimal solution of Problem~\eqref{eq:spl} is $\W_\star = \W_\natural\bm \Pi\bm D, \H_\star =\H_\natural\bm \Pi^\T\bm D^{-1}$, { where $\bm \Pi$ and ${\bm D}$ are permutation and full-rank scaling matrices as before}.}
	\end{minipage}
}
\vspace{.2cm}

\noindent
This generalizes the result of VolMin identifiability in a significant way: the matrix factorization model $\X =\W_\natural\H_\natural^\T$ is identifiable given any full column rank $\W_\natural$ and a sufficiently scattered nonnegative $\H_\natural$---which is by far the mildest identifiability condition for nonnegative matrix factorization.

\begin{shaded*}
	\noindent
	{\bf Insights Behind Criterion \eqref{eq:spl}}
	The criterion is inspired by the intuition that we demonstrated in Fig.~\ref{fig:plain}. There may exist a nontrivial $\Q$ such that
	\begin{equation}\label{eq:feasi}
		\begin{aligned}
			&\X=\W_\natural\Q\Q^{-1}\H_\natural^\T,\\
			&\H_\natural\Q^{-\T}\geq{\bm 0},~{\bm 1}^\T\H_\natural\Q^{-\T}=\rho{\bm 1}^\T,
		\end{aligned}
	\end{equation}
	Such a $\Q$ must be `shrinking' the red region in Fig.~\ref{fig:plain} (left), which makes the rows of $\H=\H_\natural\Q^{-\T}$ closer to each other in terms of the Euclidean distance in the nonnegative orthant; otherwise, $\H$ is infeasible if $\H_\natural$ is sufficiently scattered. To prevent this from happening, we can maximize $\det(\H\H^\T)$---which, intuitively speaking, maximizes the area of the red region covered by ${\sf cone}\{\H^\T\}$ in Fig.~\ref{fig:plain} (left). Note that maximizing $\det(\H\H^\T)$ is equivalent to minimizing $\det(\W^\T\W)$ under the equality constraint $\X=\W\H^\T$, which leads to the criterion in \eqref{eq:spl}. In fact, one can rigorously show that
	\[|\det(\Q)|\geq \rho\]
	for any $\Q$ satisfying \eqref{eq:feasi} if $\H_\natural$ is sufficiently scattered and the column sums are $\rho$, and the equality holds if and only if $\Q$ is a permutation matrix \cite{fu2017on,huafusid2016nips}. This leads to 
	$ \det(\W^\T\W)=|\det(\Q)|^2\det(\W_\natural^\T\W_\natural)\geq \rho^2\det(\W_\natural^\T\W_\natural),$
	where the lower bound is attained if and only if $\Q$ is a permutation matrix. This is the key for showing the soundness of the criterion in \eqref{eq:spl}.
\end{shaded*}

\color{black}

\begin{shaded*}
	\noindent
	{\bf How important is choosing the right factorization model?}
	Here, we present a toy example to demonstrate how identifiability of the factorization tools affects the performance under different data models.
	We generate $\X=\W_\natural\H_\natural^\T$ with different types of latent factors. Following the simulation setup in \cite{fu2017on}, three cases of $\W_\natural$ are generated: case 1 has elements of $\W_\natural$ following the uniform distribution between 0 and 1 with a density level $s={\rm nnz}(\W_\natural)/(MR)=0.65$, case 2 the uniform distribution between 0 and 1 with $s=1$, and case 3 the i.i.d. standard Gaussian distribution. For all three cases, the elements of $\H_\natural$ follow the uniform distribution between 0 and 1 with a density level $s=0.65$. 
	The first two cases have both $\W_\natural$ and $\H_\natural$ being nonnegative, while $\W_\natural$ has many negative values in the third case. We put under test separability-based NMF (SPA \cite{Gillis2012}), plain NMF (HALS \cite{cichocki2009fast}), volume-minimization (VolMin) \cite{bioucas2009variable}, and the determinant-based criterion in \eqref{eq:spl} (ALP \cite{fu2017on}), respectively, under $R=3,5,10,15$. The \textit{mean squared error} (MSE) of the estimated $\widehat{\H}$ (as defined in \cite{fu2017on,fu2016robust}) by different approaches can be seen in Table~\ref{tab:toy_example}. 
	
	Several important observations that reflect the importance of identifiability are in order: First, for the separability-based algorithm, when $R$ is small and the separability condition is relatively easy to be satisfied by a moderately sparse $\H_\natural$, it works very well for case 1 and case 2. It does not work well for case 3 since the adopted algorithm, i.e., SPA, needs the row-stochasticity assumption of $\H_\natural$, which cannot be enforced when $\W_\natural$ contains negative elements. Second, the plain NMF works well for case 1, since both $\W_\natural$ and $\H_\natural$ are very likely sufficiently scattered, but it fails for the second and third cases since the $\W_\natural$'s are dense there, which violates a \textit{necessary condition} for plain NMF identifiability. Third, VolMin exhibits high estimation accuracy for the first two cases since its identifiability holds for any full column-rank $\W_\natural$. But it fails in the third case because it cannot enforce the row-stochasticity of $\H_\natural$ there (note that we have applied the column normalization trick before applying VolMin and SPA---which does not work if $\W_\natural$ is not nonnegative). Fourth, the determinant based method can successfully identify $\H_\natural$ in all three cases, which echoes our comment that this criterion needs by far the mildest condition for identifiability. Table~\ref{tab:toy_example} illustrates the pivotal role of identifiability in NMF problems. 
	
	We should finish with a remark that it is not always { recommended} to use the approaches that offer the strongest identifiability guarantees (e.g., VolMin and the criterion in \eqref{eq:spl}), since these pose harder optimization problems in practice { and normally require more runtime; also see Table~\ref{tab:toy_example}}. This will be explained in the next section.
\end{shaded*}

\begin{table}[htbp]
	\centering
	\caption{The estimated MSE of $\widehat{\H}$ (after fixing permutation and scaling ambiguities) and average runtime given by different approaches under different data models.}
	\resizebox{1\linewidth}{!}{\huge
		\begin{tabular}{|c|cccc|rrrr|}
			\hline
			\multirow{2}[4]{*}{Case} & \multicolumn{4}{c|}{Separability-Based \cite{Gillis2012}} & \multicolumn{4}{c|}{Plain NMF \cite{huang2014non,cichocki2009fast} } \\
			\cline{2-9}          & $R=3$   & $R=5$   & $R=10$  & $R=15$  & \multicolumn{1}{c}{$R=3$} & \multicolumn{1}{c}{$R=5$} & \multicolumn{1}{c}{$R=10$} & \multicolumn{1}{c|}{$R=15$} \\
    \hline
Case 1 & 1.11E-32 & 0.0008 & 0.1037 & 0.2475 & \multicolumn{1}{r}{1.31E-05} & \multicolumn{1}{r}{4.89E-05} & 0.0005 & 0.0014 \\
Case 2 & 1.08E-32 & 0.0002 & 0.0424 & 0.0881 & 0.0077 & 0.0149 & 0.0411 & 0.1322 \\
Case 3 & 0.4714 & 0.7242 & 0.7280 & 0.7979 & 0.3489 & 0.2787 & 0.2628 & 0.3260 \\
runtime (sec.) & 0.0006 & 0.0005 & 0.0008 & 0.0010 & 0.0550 & 0.0872 & 0.2915 & 0.5622 \\
\hline
			\multirow{2}[4]{*}{Case} & \multicolumn{4}{c|}{VolMin \cite{fu2015blind,bioucas2009variable} }   & \multicolumn{4}{c|}{Determinant-Based Criterion \eqref{eq:spl} \cite{fu2017on}} \\
			\cline{2-9}          & $R=3$   & $R=5$   & $R=10$  & $R=15$  & \multicolumn{1}{c}{$R=3$} & \multicolumn{1}{c}{$R=5$} & \multicolumn{1}{c}{$R=10$} & \multicolumn{1}{c|}{$R=15$} \\
    \hline
Case 1 & 7.64E-06 & 7.33E-08 & 2.89E-08 & 1.47E-04 & \multicolumn{1}{r}{1.76E-17} & \multicolumn{1}{r}{1.07E-17} & \multicolumn{1}{r}{1.91E-17} & \multicolumn{1}{r|}{2.43E-17}\\
Case 2 & 2.78E-05 & 7.78E-10 & 1.17E-08 & 5.76E-04 & \multicolumn{1}{r}{2.02E-17} & \multicolumn{1}{r}{1.54E-17} & \multicolumn{1}{r}{1.96E-17} & \multicolumn{1}{r|}{2.02E-17} \\
Case 3 & 0.6566 & 1.0035 & 1.2103 & 1.2541 & \multicolumn{1}{r}{2.08E-17} & \multicolumn{1}{r}{1.90E-17} & \multicolumn{1}{r}{1.84E-17} & \multicolumn{1}{r|}{1.86E-17} \\
runtime (sec.) & 0.0128 & 0.0124 & 0.0224 & 0.0341 & 2.6780 & 4.3743 & 9.2779 & 13.9156 \\
\hline
\end{tabular}}%
		\label{tab:toy_example}%
	\end{table}%
	\section{Algorithms}
	Identifiability, as a theoretical subject, has significant implications on how well an NMF criterion or procedure will perform in real-world applications.
	On the other hand, identifiability is not equivalent to \textit{solvability} or \textit{tractability}. Upon a closer look at the NMF methods described in the above sections,
	one will realize that many separability-based methods lead to polynomial-time algorithms,
	but separability-free methods such as determinant minimization require us to solve nonconvex optimization problems.
	In fact, both the plain NMF and determinant minimization problems are known to fall in the NP-hard problem class \cite{vavasis2009complexity,packer2002np}. Hence, how to effectively handle an NMF criterion is often times an art, which must also take into consideration computational and memory complexities, regularization, and (local) convergence guarantees simultaneously---all of which contribute to good performance in practice.
	
	In this section, we will review the main ideas behind the separability-free algorithms, since separability-based approaches are either associated with tractable convex formulations or greedy procedures---which are already quite well-understood.
	
	\subsection{Fitting-Based NMF}
	\noindent
	The computational aspects of plain NMF is very well studied. Let us denote \[f(\W,\H)=\left\|\X-\W\H^\T\right\|_F^2+r_1(\W)+r_2(\H),\] where $r_1(\cdot)$ and $r_2(\cdot)$ are certain regularizers that take into consideration prior knowledge, e.g., sparsity. In some cases, $\|\X-\W\H^\T\|_F^2$ is replaced by $L(\X||\W\H^\T)$ where $L(\cdot||\cdot)$ denotes distance measures that { are} non-Euclidean and is suitable for certain types of data \cite{fevotte2009nonnegative} (e.g., the Kullback-Leibler (KL)-divergence for count data). The general problem of interest is
	\begin{align}\label{eq:general_fit}
		\minimize_{\W\geq{\bm 0},\H\geq{\bm 0}}~f(\W,\H).
	\end{align}
	This naturally leads to the following block coordinate descent (BCD) scheme:
	\begin{subequations}\label{eq:exactBCD}
		\begin{align}
			\W^{(t+1)} &\leftarrow \arg\min_{\W\geq{\bm 0}}~f\left(\W,\H^{(t)}\right) \label{eq:wupdate},\\
			\H^{(t+1)}&\leftarrow \arg\min_{\H\geq{\bm 0}}~f\left(\W^{(t+1)},\H\right)  \label{eq:hupdate},
		\end{align}
	\end{subequations}
	where $t$ is the iteration index.
	Note that Problems~\eqref{eq:wupdate}-\eqref{eq:hupdate} are both convex problems under a variety of distance measures (e.g., Euclidean distance, KL-divergence, and $\beta$-divergence)  { and convex regularizations (e.g., $\|\cdot\|_1$ and $\|\cdot\|_F^2$). Therefore, the subproblems can be solved by a large variety of off-the-shelf convex optimization algorithms}.
	Many such BCD algorithms were summarized in \cite{zhou2014nonnegative,huang2014putting}.
	One of the recently developed algorithms, namely, the AO-ADMM algorithm \cite{huang2016flexible} employs ADMM for solving the subproblems \eqref{eq:wupdate}-\eqref{eq:hupdate}. The salient feature of ADMM is that it can easily handle a large variety of regularizations and constraints simultaneously via introducing slack variables judiciously and leveraging easily solvable subproblems.

	Note that the subproblems need not be solved exactly. The popular multiplicative update (MU) and a recently developed algorithm \cite{xu2013block} both update $\W$ and $\H$ via solving local approximations of Problems~\eqref{eq:wupdate} and \eqref{eq:hupdate}. 
	There is an interesting trade-off between using exact and inexact solutions of \eqref{eq:wupdate} and \eqref{eq:hupdate}.
	Generally speaking, exact BCD uses fewer iterations to obtain a good solution, but the per-iteration complexity is higher relative to the inexact ones.
	Inexact updating uses many more iterations in general, but some interesting and pragmatic tricks can help substantially reduce the number of iterations, e.g., the combination of Nesterov's extrapolation and alternating gradient projection in \cite{xu2013block}.
	We refer the readers to \cite{zhou2014nonnegative,huang2014putting,xu2013block,gillis2014and} for detailed survey of plain NMF algorithms.

	\subsection{Optimizing Determinant Related Criteria}
    Next, we turn our attention to the determinant minimization problems in \eqref{eq:svolmin} and \eqref{eq:spl}. Determinant minimization is quite a hard nonconvex problem. The remote sensing community has spent much effort in this direction (to implement \textit{Craig's belief})  \cite{li2008minimum,bioucas2009variable,MVES,miao2007endmember}.

	\noindent
	$\bullet$ {\textbf{Successive Convex Approximation (SCA)}} The work in \cite{li2008minimum,bioucas2009variable,MVES} considered a case where $\W_\natural$ is a square matrix and $\H_\natural$ satisfies \eqref{eq:hsto}. The methods can be summarized in a unified way: Consider a square and full rank $\W$. It has a unique inverse $\Z=\W^{-1}$.
	Hence, the problem in \eqref{eq:svolmin} can be recast as
	\begin{subequations}\label{eq:qmax}
		\begin{align}
			\maximize_{\bm Z}&~|\det({\bm Z})|\\
			{\rm subject~to}&~{\bm 1}^\T \bm Z\X={\bm 1^\T}, ~\Z\X\geq {\bm 0},
		\end{align}
	\end{subequations}
	{ since $\det(\W^\T\W)=|\det(\W)|^2$ when $M=R$, and minimizing $|\det(\W)|$ is equivalent to maximizing $|\det(\Z)|$}. Problem~\eqref{eq:qmax} is still nonconvex, but now the constraints are all convex sets. From this point, the methods in \cite{MVES,fu2017on,huafusid2016nips} and the methods in \cite{bioucas2009variable,li2008minimum} take two different routes to handle the above (and its variants).

	The first route is successive convex approximation \cite{li2008minimum,bioucas2009variable}. First, it is noted that the objective can be changed to 
	minimizing $-\log|\det(\Z)|$ which does not change the optimal solution { but the $\log$ function keeps $\log|\det(\Z)|$ `safely' inside the differentiable domain---as it penalizes nonsingular $\Q$ very heavily. This way, the gradient of the cost function can be used for optimization}.
	Then, by right-multiplying $\X^\dag = \X^\T(\X\X^\T)^{-1}$ to both sides of the equality constraint of Problem~\eqref{eq:qmax}, we have $\bm 1^\T\Z =\a$, where $\a = \bm 1^\T \X^\dag$.
	The resulting optimization problem may be re-expressed as
	\begin{subequations}\label{eq:qsucc}
		\begin{align}
			\minimize_{\bm Q}&~-\log|\det{\bm Z}|\\
			{\rm subject~to}&~{\bm 1}^\T \bm Z={\a}, ~\Z\X\geq {\bm 0},
		\end{align}
	\end{subequations}
	The cost function can be approximated by the following quadratic function locally around $\Z=\Z^{(t)}$:
	\begin{align*}
		\widehat{f}\left(\Z;\Z^{(t)}\right)&=f\left(\Z^{(t)}\right)+ \left<\nabla_{\Z} f\left({\Z^{(t)}}\right),\Z-\Z^{(t)}\right> \\
		&+ \frac{\gamma}{2}\left\|\Z-\Z^{(t)}\right\|_F^2 
	\end{align*}
	where $f(\Z)$ denotes the objective in \eqref{eq:qsucc}, and the $\gamma$ is a parameter that is associated with the step size. 
	Given $\widehat{f}(\Z;\Z^{(t)})$, the subproblem in iteration $t$ is
	\begin{align}\label{eq:ttttttt}
		\Z^{(t+1)}\leftarrow &\arg\min_{{\bm 1}^\T \bm Z={\a}, ~\Z\X\geq {\bm 0}}~\widehat{f}\left(\Z;\Z^{(t)}\right)
	\end{align}
	which is a convex quadratic program. 
	Using a general-purpose second-order optimization algorithm (e.g., the Newton method) to solve Problem~\eqref{eq:ttttttt} needs $O(R^6\sqrt{RN})$ flops \cite{li2008minimum}, since there are $R^2$ optimization variables and $RN$ inequality constraints. However, there are special structures to be exploited, and using a customized ADMM algorithm to solve it brings down the per-iteration complexity to $O(R^2N)$ flops \cite{bioucas2009variable,fu2018anchor}.
	
	\noindent
	$\bullet$ \textbf{Alternating Linear Programming (ALP)} The second route is alternating linear programming \cite{MVES,huafusid2016nips}. The major idea is to optimize w.r.t. the $j$th row of $\Z$ while fixing all the other rows in $\Z$.
	Note that $\det(\Z) = \bm q_j^\T \bm r_j$ where { ${\bm r}_j=[\bm r_j(1),\ldots,\bm r_j(R)  ]^\T$}, $\bm r_j(k) = (-1)^{j+k}\det({\cal Z}_{j,k})$ { by the cofactor expansion of determinant}, and { ${\cal Z}_{j,k}\in\mathbb{R}^{(R-1)\times (R-1)}$ is a submatrix of $\Z$ that is formed by nullifying the $j$th row and $k$th column of $\Z$}.
	Hence, by considering $\maximize |\det(\Z)|$ subject to the constraints in \eqref{eq:qmax}, the resulting update is
	\begin{subequations}\label{eq:qbcd}
		\begin{align}
			\bm z_j^{(t+1)}\leftarrow\arg\max_{\bm z_j}&~|\bm r_j^\T\bm z_j|\\
			{\rm subject~to}&~{\bm 1}^\T {\bm z}_j={\bm b}_j, ~{\bm z}_j^\T\X\geq {\bm 0},
		\end{align}
	\end{subequations}
	where $\bm b_j = \bm 1^\T -\sum_{m\neq j}(\bm z_m^{(t)})^\T\X$. The objective is still nonconvex, but can be solved by comparing the solutions of two linear programs with objective functions being $\bm r_j^\T\bm z_j$ and $-\bm r_j^\T\bm z_j$, respectively.
	The idea of alternating linear programming can be applied to the modified criterion in \eqref{eq:spl} \cite{fu2017on,huafusid2016nips} as well.
	In terms of complexity, the algorithm costs $O(R^2N)$ flops for every iteration, even if we use the interior-point algorithm to solve \eqref{eq:qbcd}, due to the small scale of the linear programs; see more discussion in \cite{huafusid2016nips} and an acceleration strategy in \cite{fu2018anchor}.
	
	\noindent
	$\bullet$ \textbf{Regularized Fitting} 
	Another way
	to tackle the determinant minimization problems is to consider a regularized variant:
	\begin{subequations}\label{eq:regfit}
		\begin{align}
			\minimize_{\W,\H}&~\left\|\X-\W\H^\T\right\|_F^2 + \lambda\cdot g(\W)\\
			{\rm subject~to}&~\W\in{\cal {W}},~\H\in{\cal H},
		\end{align}
	\end{subequations}
	where $g(\cdot)$ is an optimization surrogate of $\det(\W^\T\W)$ and ${\cal W}$ and ${\cal H}$ are constraint sets for $\W$ and $\H$, respectively.
	{ The formulation in \eqref{eq:regfit} can easily incorporate prior information on $\W$ and $\bm H$ (via adding different constraints), and its use of the fitting term $\left\|\X-\W\H^\T\right\|_F^2$ in the objective function makes the formulation exhibit some form of noise robustness.
	} 
	Some commonly used constraints are
	${\cal W}=\{\W|\W\geq\bm 0\},~{\cal H}=\{\H|\H{\bm 1}={\bm 1}, \H\geq\bm 0\},$
	where $\cal H$ is naturally inherited from the VolMin criterion and $\cal W$ takes into consideration of the prior knowledge of $\W$.
	The problem in \eqref{eq:regfit} is essentially a regularized version of plain NMF, and can also be handled via two block coordinate descent w.r.t. $\W$ and $\H$, respectively.
	However, the subproblems can be much harder than those in plain NMF---depending on the regularization term $g(\cdot)$ that is employed.
	For example, in \cite{zhou2011minimum,fu2016robust}, \[g(\W)=\det(\W^\T\W)\] was used. Consequently, the $\W$-subproblem is no longer convex. A commonly used trick to handle such a nonconvex $\W$-subproblem is to use projected gradient descent combined with backtracking line search (to ensure decrease of the cost function) \cite{miao2007endmember,fu2016robust}.
	To circumvent line search which could be time consuming, in \cite{fu2016robust}, it was proposed to use a surrogate for $\det(\W^\T\W)$, namely,
	\begin{equation}\label{eq:logdetreg}
		g(\W)=\log\det(\W^\T\W+\epsilon\bm I).
	\end{equation}
	for some pre-specified small constant $\epsilon>0$. The function
	 $\log\det(\W^\T\W+\epsilon\bm I)$ still promotes small determinant of $\W^\T\W$ since $\log(\cdot)$ is monotonic.
	The term $\epsilon\bm I$ is to prevent { $g(\W)$} from going to $-\infty$. { Such a regularization term entails one to design a  {\it block successive upperbound minimization} (BSUM) algorithm \cite{razaviyayn2013unified} to handle the regularized fitting problem, using simple updates of $\W$ and $\bm H$ without involving any line search procedure  \cite{fu2016robust}.}
	There are various ways of solving the $\W$- and $\H$-problems in the BSUM framework, e.g., ADMM~\cite{huang2016flexible,fu2016robusticassp} and (accelerated) proximal gradient~\cite{fu2016robust,xu2013block}. All these first-order optimization algorithms cost $O(R^2(M+N))$ flops for one iteration.

	
	
	We should mention that many other different approximations for simplex volume or determinant of $\W^\T\W$ exist \cite{berman2004ice,miao2007endmember,li2012collaborative}. For example, in \cite{berman2004ice}, Berman {\it et al.} employs an even simpler regularization, which is \[g(\W)={\rm trace}(\bm F\W\W^\T),\] where $\bm F=R\bm I-\bm 1\bm 1^\T$. Essentially, this regularization approximates the volume of $\conv{{\bm w}_1,\ldots,{\bm w}_R}$ using the sum of squared distances between all the pairs of ${\bm w}_i$ and ${\bm w}_j$ (since $g(\W)\propto \sum_{i=1}^R\sum_{j=i+1}^R\|\bm w_i-\bm w_j\|_2^2$)---{ which intuitively relates to the volume of ${\sf conv}\{\W\}$}. This makes the $\W$-subproblem convex and easy to handle. Nevertheless, this approximation of simplex volume is considered rough and usually using determinant or log-determinant outputs better results according to empirical evidences \cite{fu2016robust}.

\section{More Discussions on Applications}
In this section, with everything that we have learned, we will take another look at the applications to discuss how to choose appropriate factorization tools for them.
\subsection{Case Study: Topic Modeling}
Recall that $\X$ is a word-document matrix, $\X(m,n)$ is the term-frequency of the $m$th word in document $n$, $\W_\natural$ is the word-topic matrix, and $\H_\natural$ is the document-topic matrix. It makes sense to assume that there are a few characteristic words for each topic, which other topics do not use. 
In the literature, such characteristic words are called {\it anchor words}, whose existence is the same as the separability condition holding for $\W_\natural$. Note that if the anchor word assumption holds for the model $\X=\C_\natural\bm \varSigma\bm D_\natural^\T$ (where $\W_\natural = \bm C_\natural$ and $\H_\natural = \bm D_\natural \bm \varSigma_\natural$), it also holds for the second-order model $\bm P=\C_\natural \E_\natural\C_\natural^\T$ since the $\bm P$ matrix carries the word-topic matrix with it.
Therefore, many separability-based NMF algorithms were applied to $\bm P$ or $\X$ for topic mining \cite{recht2012factoring,arora2012practical,kumar2012fast}.
In some cases, e.g., when the dictionary size is small, the topics may not have anchor words captured in the dictionary. 
Under such scenarios, it is `safer' to apply the computationally more challenging yet identifiability-wise more robust methods such as the introduced determinant based method in \eqref{eq:spl} and its variants \cite{huafusid2016nips,fu2018anchor}.

\begin{shaded*}
	\noindent
	{\bf Community Detection}
	{ Note that the discussion here for anchor words-based topic modeling also holds for the community detection problem in \eqref{eq:cd}. The only difference is that the anchor-word assumption is replaced by the `pure-node' assumption, where a pure node is a node who has a single community membership; i.e., there exists $\bm \theta_{\ell_r}=\bm e_r^\T$ for $r=1,\ldots,R$ \cite{mao2017mixed,panov2017consistent}. 
	Under the pure-node assumption, $\H$ in \eqref{eq:cd} satisfies the separability condition, and thus the algorithms that we introduced in Sec.~\ref{sec:sep} can be directly applied.
	Also note that in the community detection model \eqref{eq:cd}, $\W$ is not necessarily nonnegative. Nevertheless, the separability-based and determinant-based algorithms are not affected in terms of identifiability, as we have learned from the previous sections.}
\end{shaded*}

Another point that is worth mentioning is that to estimate $\C_\natural$ from the second-order model $\bm P=\C_\natural\E_\natural\C_\natural^\T$, it is desired to directly work on the symmetric {\it tri-factorization} model, instead of treating $\bm P$ as an asymmetric factorization model (i.e., by letting $\W_\natural=\C_\natural \bm E_\natural$ and $\bm H_\natural =\C_\natural$)---since the former is more natural. It was proved in the literature \cite{huafusid2016nips,fu2018anchor} that the criterion
 \begin{subequations}\label{eq:cec }
 	\begin{align}
 	\minimize_{{\bm C},\E}&~|\det\left(\E\right)|\\
 	{\rm subject~to}&~\X =\C\E\C^\T \label{eq:lift}\\
 	&~\bm 1^\T\C=\bm 1^\T,~\C\geq\bm 0.
 	\end{align}
 \end{subequations}
 can identify $\C_\natural$ and $\bm E_\natural$ with trivial ambiguities---{\it if $\C_\natural$ is sufficiently scattered}. Note that the column-stochasticity constraint on $\C$ in \eqref{eq:cec } is added because the topics are modeled as PMFs. 
 In terms of computation, such a {\it tri-factorization} model is seemingly harder to handle relative to the bi-factorization models. Nevertheless, some good { heuristic} algorithms exist \cite{huafusid2016nips,fu2018anchor,huang2018hmm}; see the supplementary materials on tri-factorization and symmetric NMF for more information.		
 
Table~\ref{tab:new} shows the performance of two anchor-based algorithms (i.e., SPA \cite{Gillis2012} and XRAY \cite{kumar2012fast}) and an anchor-free algorithm \cite{huafusid2016nips,fu2018anchor}, when applied to 319 real documents. One can see that SPA only identifies a single topic (all the columns of $\widehat{\W}$ correspond to a sexual misconduct case in the military). XRAY identifies two topics, i.e., the sexual misconduct story and the Cavalese cable car disaster which happened in Italy, 1998. The AnchorFree algorithm \cite{huafusid2016nips,fu2018anchor} that employs the criterion in \eqref{eq:cec } identifies three different topics. Note that the topic that is clearly related to the New York City's traffic and taxi drivers was never discovered by the separability-based methods. This example shows the power of the sufficiently scattered condition-based algorithms in topic mining.

\begin{table}[htbp]
  \centering
  \caption{Prominent 3 topics found from 319 documents in the TDT2 dataset ({ dictionary} size $= 4,864$) by different algorithms.}
  \resizebox{1\linewidth}{!}{
    \begin{tabular}{|c|c|c|c|c|c|c|c|c|}
    \hline
    \multicolumn{3}{|c|}{SPA \cite{Gillis2012}} & \multicolumn{3}{c|}{XRAY \cite{kumar2012fast} } & \multicolumn{3}{c|}{AchorFree \cite{huafusid2016nips} } \\
    \hline
    mckinney & mckinney & mckinney & mckinney & mckinney & cable & mckinney & cable & mayor \\
    sergeant & sergeant & sergeant & sergeant & sergeant & plane & sergeant & marine & giuliani \\
    gene  & gene  & gene  & gene  & gene  & italian & gene  & italy & city \\
    martial & army  & martial & martial & martial & italy & martial & italian & drivers \\
    sexual & major & armys & major & sexual & car   & sexual & plane & york \\
    major & sexual & sexual & sexual & misconduct & marine & misconduct & accident & mayors \\
    court & rank  & misconduct & court & armys & accident & armys & car   & giulianis \\
    misconduct & obstruction & major & army  & enlisted & jet   & major & jet   & yorkers \\
    mckinneys & martial & court & misconduct & major & flying & enlisted & crew  & citys \\
    armys & misconduct & enlisted & accusers & court & ski   & court & ski   & rudolph \\
    enlisted & justice & counts & armys & mckinneys & crew  & army  & flying & street \\
    army  & enlisted & topenlisted & mckinneys & army  & 20    & mckinneys & 20    & cab \\
    accusers & reprimanded & army  & enlisted & soldier & aviano & soldier & pilot & taxi \\
    women & sentence & 19    & women & charges & pilot & women & aviano & pedestrians \\
    six   & armys & former & six   & women & gondola & charges & gondola & civility \\
    charges & conviction & soldier & former & former & military & former & military & police \\
    soldier & court & mckinneys & charges & six   & base  & six   & base  & cabbies \\
    former & charges & jurors & soldier & 19    & severed & accusers & severed & manhattan \\
    testified & jury  & women & sexually & testified & altitude & jury  & resort & traffic \\
    sexually & mckinneys & charges & harassing & jury  & killed & counts & altitude & vendors \\
    \hline
    \end{tabular}}%
  \label{tab:new}%
\end{table}%

\color{black}

\subsection{Case Study: HMM Identification}
In the HMM model, $\M(i,r)={\rm Pr}[Y_t=y_i|X_t=x_r]$ which is the emission probability of $Y_t=y_i$ given that the latent state $X_t=x_r$.
In this case, the separability condition translates to that \textit{for every latent state, there is an observable state that is emitted by this latent state with an overwhelming probability.} This is not impossible, but it heavily hinges on the application that one is working with. 
For more general cases where the separability condition is hard to argue, using a determinant based method ({ such as that in \eqref{eq:cec }}) is always `safer' in terms of identifiability.
%

A practical estimation criterion for HMM identification is as follows:
\begin{align}\label{eq:hua}
\minimize_{{\bm M},{\bm \varTheta}}&~{\sf KL}\left({\bm \varOmega} ||\M{\bm \varTheta}\M^\T\right) + \lambda |\det ({\bm \varTheta})|\\
{\rm subject~to}&~{\bm 1}^\T\M = {\bm 1}^\T,~\M\geq {\bm 0},~{\bm 1}^\T{\bm \varTheta}{\bm 1}=1,~{\bm \varTheta}\geq {\bm 0},\nonumber
\end{align}
where the constraints are due to the fact the columns of $\M$ and the matrix $\bm \varTheta$ are both respectable PMFs, { and ${\sf KL}\left({\bm \varOmega} ||\M{\bm \varTheta}\M^\T\right) $ measures the KL-divergence between ${\bm \varOmega}$ and $\M{\bm \varTheta}\M^\T$. The formulation in \eqref{eq:hua} can be understood as a reformulation of \eqref{eq:cec } by lifting constraint \eqref{eq:lift} to the objective function under the KL-divergence based distance measure.}
Using the regularized fitting criterion instead of a hard constraint ${\bm \varOmega} =\M{\bm \varTheta}\M^\T$ is because there is noise in $\bm \varOmega$---since it is usually estimated via sample averaging. The data fitting part employs the KL-divergence as a distance measure since $\bm \varOmega$ is an estimate of a joint PMF, and $\M{\bm \varTheta}\M^\T$ is supposed to be a parameterization of the true PMF---and KL divergence is the most natural measure for the distance between PMFs.
{ The term $|\det({\bm \varTheta})|$ is critical, since it offers identifiability support.}
Optimizing the Problem~\eqref{eq:hua} is highly nontrivial, since it is a symmetric model under the KL divergence, but can be approached by carefully approximating the objective using local optimization surrogates; see \cite{huang2018hmm} for details.

\begin{figure}[t]
	\centering
	\includegraphics[width=.88\linewidth]{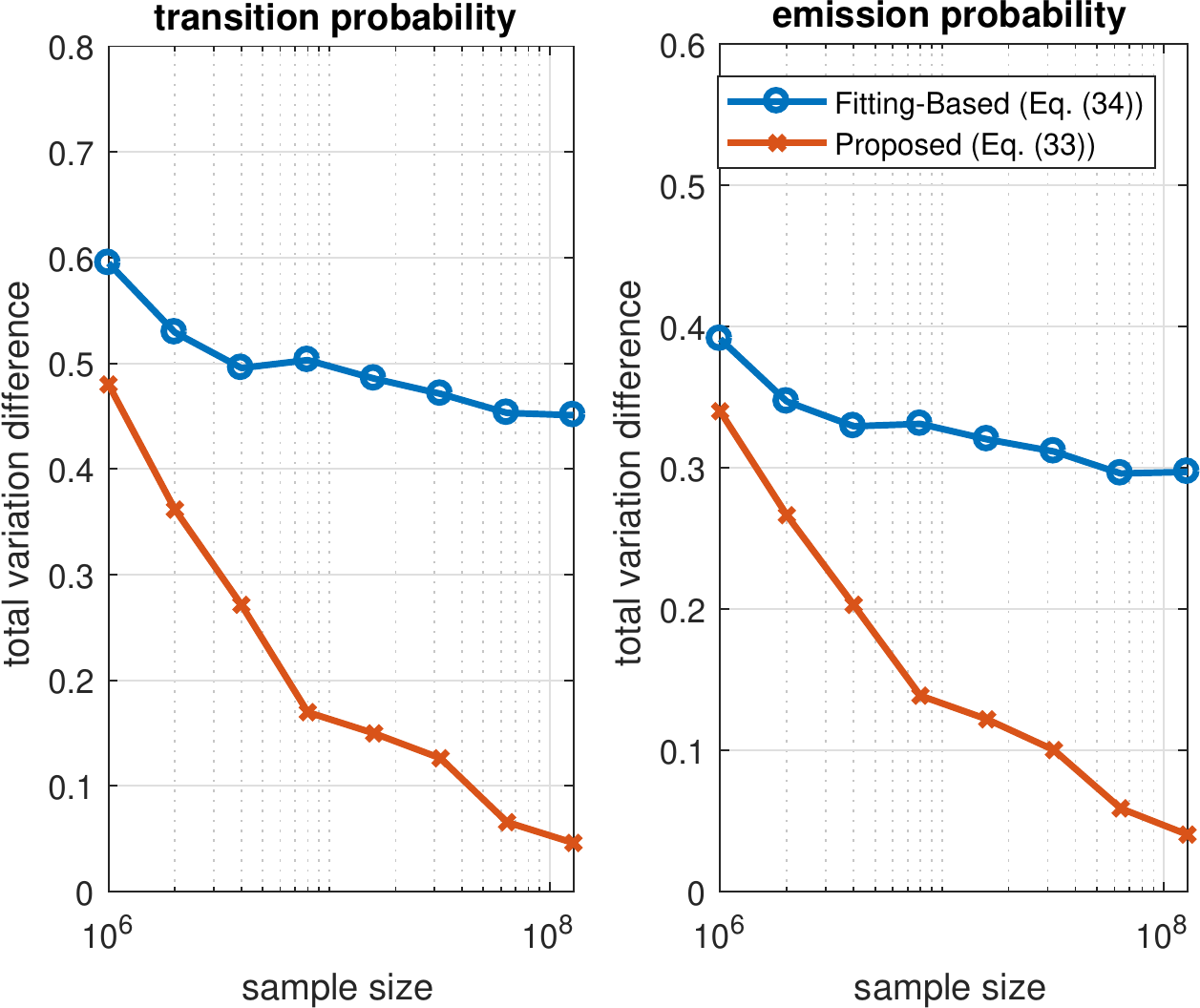}
	\caption{{ The total variation difference (defined as $\nicefrac{1}{2R}\|\bM_\natural-\widehat{\bM}\|_1$ and   $\nicefrac{1}{2R}\|\bTheta_\natural-\widehat{\bTheta}\|_1$, respectively) between the ground truth and algorithm-estimated transition probability (left) and emission probability (right). The result is averaged over 10 random problem instances. In this example, we have $M=100$ observable states and $R=20$ latent states. Both the transition and the emission probability matrices are generated randomly following an exponential distribution, and $50\%$ elements of the emission matrix are set to be zeros---so that the sufficiently scattered condition holds for $\M$.}}
	\label{fig:sim_exp1}
	\vspace{-10pt}
\end{figure}

In the literature, the HMM identification problem was formulated as follows \cite{lakshminarayanan2010non}:
\begin{align}\label{eq:raich}
	\minimize_{{\bm M},{\bm \varTheta}}&~\left\|{\bm \varOmega} -\M{\bm \varTheta}\M^\T\right\|_F^2\\
	{\rm subject~to}&~{\bm 1}^\T\M = {\bm 1}^T,~\M\geq {\bm 0},~{\bm 1}^\T{\bm \varTheta}{\bm 1}=1,~{\bm \varTheta}\geq {\bm 0}. \nonumber
\end{align}
The key difference between \eqref{eq:hua} and  \eqref{eq:raich} is that the latter does not have identifiability guarantees of the latent parameters. The impact in practice is quite significant---see the experiment results in Fig.~\ref{fig:sim_exp1}, where both formulations are applied to estimating the transition and emission probabilities from synthetic HMM data. One can see that the identifiability-guaranteed approach largely outperforms the other approach. 
This example well demonstrates the impact of identifiability in practice---under the same data model, the performance can be quite different with or without identifiability support.

	\color{black}
	\section{Take-home Points, Open Questions, Concluding Remarks}
	In this feature article, we have reviewed many recent developments in identifiable nonnegative matrix factorization, including models, identifiability theory and methods, and timely applications in signal processing and machine learning. Some take-home points are summarized as follows:
	
	\noindent
	$\bullet$ Different applications may have different characteristics, and thus choosing a right NMF tool for the application at hand is essential. For example, plain NMF is considered not suitable for hyperspectral unmixing, { since the $\W_\natural$ matrices there are in general dense, which violates the {\it necessary condition} for plain NMF identifiability.}
	
	\noindent
	$\bullet$ Separability-based methods have many attractive features, such as identifiability, solvability, provable noise robustness, and the existence of lightweight greedy algorithms. In applications where separability is likely to hold (e.g., {community detection in the presence of `pure nodes' and hyperspectral unmixing with pure pixels}), it is a highly recommended approach to try first.
	
	\noindent
	$\bullet$ Sufficiently scattered condition-based approaches are very powerful and need a minimum amount of model assumptions. These approaches ensure identifiability under mild conditions and often outperform separability-based algorithms in applications like topic mining and HMM identification, where the separability condition is likely violated. The caveat is that the optimization problems arising in such approaches are usually hard, and judicious design that is robust to noise and outliers is often needed.
	
	There are also some very important open questions in the field:
	
	\noindent
	$\bullet$ The plain NMF,  VolMin, and Problem~\eqref{eq:spl} are NP-hard problems. But in practice we often observe very accurate solutions (up to machine accuracy), at least when noise is absent. The conjecture is that with some additional assumptions, the problems can be shown to be solvable with high probability---while now the understanding to this aspect is still limited. If solvability can be established under some conditions of practical interest, then, combining with identifiability, NMF's power as a learning tool will be lifted to another level.
	
	\noindent
	$\bullet$ Noise robustness of the NMF models is not entirely well understood so far. This is particularly true for separability-free cases. The Cram\'er-Rao bound evaluation in \cite{huang2014putting} may help identify effective algorithms, but this approach is still a heuristic.
	Worst-case analysis is desired since it helps understand the limitations of the adopted NMF methods for any problem instance.
	
	\noindent
	$\bullet$ Necessary conditions for NMF identifiability is not very well understood, but necessary conditions are often useful in saving practitioners' effort for trying NMF on some hopeless cases. It is conjectured that the sufficiently scattered condition is also necessary for VolMin, which is consistent with numerical evidence. But the proof is elusive.

	\bibliographystyle{IEEEtran}
	

\begin{thebibliography}{10}
\providecommand{\url}[1]{#1}
\csname url@samestyle\endcsname
\providecommand{\newblock}{\relax}
\providecommand{\bibinfo}[2]{#2}
\providecommand{\BIBentrySTDinterwordspacing}{\spaceskip=0pt\relax}
\providecommand{\BIBentryALTinterwordstretchfactor}{4}
\providecommand{\BIBentryALTinterwordspacing}{\spaceskip=\fontdimen2\font plus
\BIBentryALTinterwordstretchfactor\fontdimen3\font minus
  \fontdimen4\font\relax}
\providecommand{\BIBforeignlanguage}[2]{{%
\expandafter\ifx\csname l@#1\endcsname\relax
\typeout{** WARNING: IEEEtran.bst: No hyphenation pattern has been}%
\typeout{** loaded for the language `#1'. Using the pattern for}%
\typeout{** the default language instead.}%
\else
\language=\csname l@#1\endcsname
\fi
#2}}
\providecommand{\BIBdecl}{\relax}
\BIBdecl

\bibitem{GHGolub1996}
G.~H. Golub and C.~F.~V. Loan., \emph{Matrix Computations}.\hskip 1em plus
  0.5em minus 0.4em\relax The Johns Hopkins University Press, 1996.

\bibitem{Comon1994}
P.~Common, ``Independent component analysis, a new concept?'' \emph{Signal
  Processing}, vol.~36, no.~3, pp. 287 -- 314, 1994.

\bibitem{chen1984nonnegative}
J.-C. Chen, ``The nonnegative rank factorizations of nonnegative matrices,''
  \emph{Linear algebra and its applications}, vol.~62, pp. 207--217, 1984.

\bibitem{craig1994minimum}
M.~D. Craig, ``Minimum-volume transforms for remotely sensed data,'' \emph{IEEE
  Trans. Geosci. Remote Sens.}, vol.~32, no.~3, pp. 542--552, 1994.

\bibitem{paatero1994positive}
P.~Paatero and U.~Tapper, ``Positive matrix factorization: A non-negative
  factor model with optimal utilization of error estimates of data values,''
  \emph{Environmetrics}, vol.~5, no.~2, pp. 111--126, 1994.

\bibitem{lee1999learning}
D.~Lee and H.~Seung, ``Learning the parts of objects by non-negative matrix
  factorization,'' \emph{Nature}, vol. 401, no. 6755, pp. 788--791, 1999.

\bibitem{arora2012practical}
S.~Arora, R.~Ge, Y.~Halpern, D.~Mimno, A.~Moitra, D.~Sontag, Y.~Wu, and M.~Zhu,
  ``A practical algorithm for topic modeling with provable guarantees,'' in
  \emph{International Conference on Machine Learning (ICML)}, 2013.

\bibitem{huafusid2016nips}
K.~Huang, X.~Fu, and N.~D. Sidiropoulos, ``Anchor-free correlated topic
  modeling: Identifiability and algorithm,'' in \emph{Advances in Neural
  Information Processing Systems}, 2016.

\bibitem{huang2014non}
K.~Huang, N.~Sidiropoulos, and A.~Swami, ``Non-negative matrix factorization
  revisited: Uniqueness and algorithm for symmetric decomposition,'' \emph{IEEE
  Trans. Signal Process.}, vol.~62, no.~1, pp. 211--224, 2014.

\bibitem{mao2017mixed}
X.~Mao, P.~Sarkar, and D.~Chakrabarti, ``On mixed memberships and symmetric
  nonnegative matrix factorizations,'' in \emph{International Conference on
  Machine Learning}, 2017, pp. 2324--2333.

\bibitem{donoho2003does}
D.~Donoho and V.~Stodden, ``When does non-negative matrix factorization give a
  correct decomposition into parts?'' in \emph{NIPS}, vol.~16, 2003.

\bibitem{laurberg2008theorems}
H.~Laurberg, M.~G. Christensen, M.~D. Plumbley, L.~K. Hansen, and S.~Jensen,
  ``Theorems on positive data: On the uniqueness of {NMF},''
  \emph{Computational Intelligence and Neuroscience}, vol. 2008, 2008.

\bibitem{CANMS}
T.-H. Chan, W.-K. Ma, C.-Y. Chi, and Y.~Wang, ``A convex analysis framework for
  blind separation of non-negative sources,'' \emph{IEEE Trans. Signal
  Process.}, vol.~56, no.~10, pp. 5120 --5134, oct. 2008.

\bibitem{fu2016robust}
X.~Fu, K.~Huang, B.~Yang, W.-K. Ma, and N.~Sidiropoulos, ``Robust
  volume-minimization based matrix factorization for remote sensing and
  document clustering,'' \emph{IEEE Trans. Signal Process.}, vol.~64, no.~23,
  pp. 6254--6268, 2016.

\bibitem{fu2017on}
X.~Fu, K.~Huang, and N.~D. Sidiropoulos, ``On identifiability of nonnegative
  matrix factorization,'' \emph{IEEE Signal Process. Lett.}, vol.~25, no.~3,
  pp. 328--332, 2018.

\bibitem{lin2014identifiability}
C.-H. Lin, W.-K. Ma, W.-C. Li, C.-Y. Chi, and A.~Ambikapathi, ``Identifiability
  of the simplex volume minimization criterion for blind hyperspectral
  unmixing: The no-pure-pixel case,'' \emph{IEEE Trans. Geosci. Remote Sens.},
  vol.~53, no.~10, pp. 5530--5546, Oct 2015.

\bibitem{fu2015blind}
X.~Fu, W.-K. Ma, K.~Huang, and N.~D. Sidiropoulos, ``Blind separation of
  quasi-stationary sources: Exploiting convex geometry in covariance domain,''
  \emph{IEEE Trans. Signal Process.}, vol.~63, no.~9, pp. 2306--2320, May 2015.

\bibitem{fevotte2009nonnegative}
C.~F{\'e}votte, N.~Bertin, and J.-L. Durrieu, ``Nonnegative matrix
  factorization with the itakura-saito divergence: With application to music
  analysis,'' \emph{Neural computation}, vol.~21, no.~3, pp. 793--830, 2009.

\bibitem{fu2015power}
X.~Fu, W.-K. Ma, and N.~Sidiropoulos, ``Power spectra separation via structured
  matrix factorization,'' \emph{IEEE Trans. Signal Process.}, vol.~64, no.~17,
  pp. 4592--4605, 2016.

\bibitem{lakshminarayanan2010non}
B.~Lakshminarayanan and R.~Raich, ``Non-negative matrix factorization for
  parameter estimation in hidden markov models,'' in \emph{Proc. IEEE MLSP
  2010}.\hskip 1em plus 0.5em minus 0.4em\relax IEEE, 2010, pp. 89--94.

\bibitem{huang2018hmm}
K.~Huang, X.~Fu, and N.~D. Sidiropoulos, ``Learning hidden markov models using
  pairwise co-occurrences with applications to topic modeling,'' \emph{Proc.
  ICML 2018}, 2018.

\bibitem{zhou2014nonnegative}
G.~Zhou, A.~Cichocki, Q.~Zhao, and S.~Xie, ``Nonnegative matrix and tensor
  factorizations: An algorithmic perspective,'' \emph{IEEE Signal Processing
  Magazine}, vol.~31, no.~3, pp. 54--65, 2014.

\bibitem{gillis2014and}
N.~Gillis, ``The why and how of nonnegative matrix factorization,''
  \emph{Regularization, Optimization, Kernels, and Support Vector Machines},
  vol.~12, p. 257, 2014.

\bibitem{gillis2017introduction}
------, ``Introduction to nonnegative matrix factorization,'' \emph{arXiv
  preprint arXiv:1703.00663}, 2017.

\bibitem{wang2013nonnegative}
Y.-X. Wang and Y.-J. Zhang, ``Nonnegative matrix factorization: A comprehensive
  review,'' \emph{IEEE Transactions on Knowledge and Data Engineering},
  vol.~25, no.~6, pp. 1336--1353, 2013.

\bibitem{Ma2013}
W.-K. Ma, J.~Bioucas-Dias, T.-H. Chan, N.~Gillis, P.~Gader, A.~Plaza,
  A.~Ambikapathi, and C.-Y. Chi, ``A signal processing perspective on
  hyperspectral unmixing,'' \emph{IEEE Signal Process. Mag.}, vol.~31, no.~1,
  pp. 67--81, Jan 2014.

\bibitem{huang2014putting}
K.~Huang and N.~Sidiropoulos, ``Putting nonnegative matrix factorization to the
  test: a tutorial derivation of pertinent {Cramer-Rao} bounds and performance
  benchmarking,'' \emph{IEEE Signal Process. Mag.}, vol.~31, no.~3, pp. 76--86,
  2014.

\bibitem{imbrie1964vector}
J.~Imbrie and T.~H. Van~Andel, ``Vector analysis of heavy-mineral data,''
  \emph{Geological Society of America Bulletin}, vol.~75, no.~11, pp.
  1131--1156, 1964.

\bibitem{lee2001algorithms}
D.~D. Lee and H.~S. Seung, ``Algorithms for non-negative matrix
  factorization,'' in \emph{Advances in neural information processing systems},
  2001, pp. 556--562.

\bibitem{recht2012factoring}
B.~Recht, C.~Re, J.~Tropp, and V.~Bittorf, ``Factoring nonnegative matrices
  with linear programs,'' in \emph{Advances in Neural Information Processing
  Systems}, 2012, pp. 1214--1222.

\bibitem{arora2012learning}
S.~Arora, R.~Ge, and A.~Moitra, ``Learning topic models--going beyond {SVD},''
  in \emph{IEEE Symposium on Foundations of Computer Science (FOCS)}.\hskip 1em
  plus 0.5em minus 0.4em\relax IEEE, 2012, pp. 1--10.

\bibitem{cai2009probabilistic}
D.~Cai, X.~Wang, and X.~He, ``Probabilistic dyadic data analysis with local and
  global consistency,'' in \emph{Proceedings of the 26th Annual International
  Conference on Machine Learning (ICML'09)}, 2009, pp. 105--112.

\bibitem{blei2003latent}
D.~M. Blei, A.~Y. Ng, and M.~I. Jordan, ``Latent {Dirichlet} allocation,''
  \emph{Journal of Machine Learning Research}, vol.~3, pp. 993--1022, 2003.

\bibitem{panov2017consistent}
M.~Panov, K.~Slavnov, and R.~Ushakov, ``Consistent estimation of mixed
  memberships with successive projections,'' in \emph{International Workshop on
  Complex Networks and their Applications}.\hskip 1em plus 0.5em minus
  0.4em\relax Springer, 2017, pp. 53--64.

\bibitem{airoldi2008mixed}
E.~M. Airoldi, D.~M. Blei, S.~E. Fienberg, and E.~P. Xing, ``Mixed membership
  stochastic blockmodels,'' \emph{Journal of Machine Learning Research},
  vol.~9, no. Sep, pp. 1981--2014, 2008.

\bibitem{forbes2011statistical}
C.~Forbes, M.~Evans, N.~Hastings, and B.~Peacock, \emph{Statistical
  distributions}.\hskip 1em plus 0.5em minus 0.4em\relax John Wiley \& Sons,
  2011.

\bibitem{Gillis2012}
N.~Gillis and S.~Vavasis, ``Fast and robust recursive algorithms for separable
  nonnegative matrix factorization,'' \emph{IEEE Trans. Pattern Anal. Mach.
  Intell.}, vol.~36, no.~4, pp. 698--714, April 2014.

\bibitem{kumar2012fast}
A.~Kumar, V.~Sindhwani, and P.~Kambadur, ``Fast conical hull algorithms for
  near-separable non-negative matrix factorization,'' pp. 231--239, 2013.

\bibitem{fu2015self}
X.~Fu, W.-K. Ma, T.-H. Chan, and J.~M. Bioucas-Dias, ``Self-dictionary sparse
  regression for hyperspectral unmixing: Greedy pursuit and pure pixel search
  are related,'' \emph{IEEE J. Sel. Topics Signal Process.}, vol.~9, no.~6, pp.
  1128--1141, Sep. 2015.

\bibitem{brunet2004metagenes}
J.-P. Brunet, P.~Tamayo, T.~R. Golub, and J.~P. Mesirov, ``Metagenes and
  molecular pattern discovery using matrix factorization,'' \emph{Proceedings
  of the national academy of sciences}, vol. 101, no.~12, pp. 4164--4169, 2004.

\bibitem{zhang2006learning}
S.~Zhang, W.~Wang, J.~Ford, and F.~Makedon, ``Learning from incomplete ratings
  using non-negative matrix factorization,'' in \emph{Proceedings of the 2006
  SIAM International Conference on Data Mining}.\hskip 1em plus 0.5em minus
  0.4em\relax SIAM, 2006, pp. 549--553.

\bibitem{VMAX}
T.-H. Chan, W.-K. Ma, A.~Ambikapathi, and C.-Y. Chi, ``A simplex volume
  maximization framework for hyperspectral endmember extraction,'' \emph{IEEE
  Trans. Geosci. Remote Sens.}, vol.~49, no.~11, pp. 4177 --4193, Nov. 2011.

\bibitem{VCA}
J.~Nascimento and J.~Bioucas-Dias, ``Vertex component analysis: A fast
  algorithm to unmix hyperspectral data,'' \emph{IEEE Trans. Geosci. Remote
  Sens.}, vol.~43, no.~4, pp. 898--910, 2005.

\bibitem{fu2018anchor}
X.~Fu, K.~Huang, N.~D. Sidiropoulos, Q.~Shi, and M.~Hong, ``Anchor-free
  correlated topic modeling,'' \emph{IEEE Trans. Patt. Anal. Machine Intell.},
  vol. to appear, {DOI:} 10.1109/TPAMI.2018.2827377, 2018.

\bibitem{lin2017maximum}
C.-H. Lin, R.~Wu, W.-K. Ma, C.-Y. Chi, and Y.~Wang, ``Maximum volume inscribed
  ellipsoid: A new simplex-structured matrix factorization framework via facet
  enumeration and convex optimization,'' \emph{arXiv preprint
  arXiv:1708.02883}, 2017.

\bibitem{CVX}
S.~Boyd and L.~Vandenberghe, \emph{Convex Optimization}.\hskip 1em plus 0.5em
  minus 0.4em\relax Cambriadge Press, 2004.

\bibitem{gillis2014successive}
N.~Gillis, ``Successive nonnegative projection algorithm for robust nonnegative
  blind source separation,'' \emph{SIAM Journal on Imaging Sciences}, vol.~7,
  no.~2, pp. 1420--1450, 2014.

\bibitem{MC01}
U.~Ara\'ujo, B.~Saldanha, R.~Galv\~ao, T.~Yoneyama, H.~Chame, and V.~Visani,
  ``The successive projections algorithm for variable selection in
  spectroscopic multicomponent analysis,'' \emph{Chemometrics and Intelligent
  Laboratory Systems}, vol.~57, no.~2, pp. 65--73, 2001.

\bibitem{naanaa2005blind}
W.~Naanaa and J.-M. Nuzillard, ``Blind source separation of positive and
  partially correlated data,'' \emph{Signal Processing}, vol.~85, no.~9, pp.
  1711--1722, 2005.

\bibitem{Esser2012}
E.~Esser, M.~Moller, S.~Osher, G.~Sapiro, and J.~Xin, ``A convex model for
  nonnegative matrix factorization and dimensionality reduction on physical
  space,'' \emph{IEEE Trans. Image Process.}, vol.~21, no.~7, pp. 3239 --3252,
  July 2012.

\bibitem{fu2015robust}
X.~Fu and W.-K. Ma, ``Robustness analysis of structured matrix factorization
  via self-dictionary mixed-norm optimization,'' \emph{IEEE Signal Process.
  Lett.}, vol.~23, no.~1, pp. 60--64, 2016.

\bibitem{gillis2018fast}
N.~Gillis and R.~Luce, ``A fast gradient method for nonnegative sparse
  regression with self-dictionary,'' \emph{IEEE Transactions on Image
  Processing}, vol.~27, no.~1, pp. 24--37, 2018.

\bibitem{gillis2013robustness}
N.~Gillis, ``Robustness analysis of hottopixx, a linear programming model for
  factoring nonnegative matrices,'' \emph{SIAM Journal on Matrix Analysis and
  Applications}, vol.~34, no.~3, pp. 1189--1212, 2013.

\bibitem{ammanouil2014blind}
R.~Ammanouil, A.~Ferrari, C.~Richard, and D.~Mary, ``Blind and fully
  constrained unmixing of hyperspectral images,'' \emph{IEEE Trans. Image
  Process.}, vol.~23, no.~12, pp. 5510--5518, Dec. 2014.

\bibitem{mizutani2014ellipsoidal}
T.~Mizutani, ``Ellipsoidal rounding for nonnegative matrix factorization under
  noisy separability,'' \emph{The Journal of Machine Learning Research},
  vol.~15, no.~1, pp. 1011--1039, 2014.

\bibitem{gillis2015semidefinite}
N.~Gillis and S.~A. Vavasis, ``Semidefinite programming based preconditioning
  for more robust near-separable nonnegative matrix factorization,'' \emph{SIAM
  Journal on Optimization}, vol.~25, no.~1, pp. 677--698, 2015.

\bibitem{li2008minimum}
J.~Li and J.~M. Bioucas-Dias, ``Minimum volume simplex analysis: a fast
  algorithm to unmix hyperspectral data,'' in \emph{Proc. IEEE IGARSS 2008},
  vol.~3, 2008, pp. III--250.

\bibitem{bioucas2009variable}
J.~M. Bioucas-Dias, ``A variable splitting augmented lagrangian approach to
  linear spectral unmixing,'' in \emph{Proc. IEEE WHISPERS'09}, 2009, pp. 1--4.

\bibitem{MVES}
T.-H. Chan, C.-Y. Chi, Y.-M. Huang, and W.-K. Ma, ``A convex analysis-based
  minimum-volume enclosing simplex algorithm for hyperspectral unmixing,''
  \emph{IEEE Trans. Signal Process.}, vol.~57, no.~11, pp. 4418 --4432, Nov.
  2009.

\bibitem{zhou2011minimum}
G.~Zhou, S.~Xie, Z.~Yang, J.-M. Yang, and Z.~He, ``Minimum-volume-constrained
  nonnegative matrix factorization: Enhanced ability of learning parts,''
  \emph{IEEE Trans. Neural Netw.}, vol.~22, no.~10, pp. 1626--1637, 2011.

\bibitem{cichocki2009fast}
A.~Cichocki and A.-H. Phan, ``Fast local algorithms for large scale nonnegative
  matrix and tensor factorizations,'' \emph{IEICE transactions on fundamentals
  of electronics, communications and computer sciences}, vol.~92, no.~3, pp.
  708--721, 2009.

\bibitem{vavasis2009complexity}
S.~A. Vavasis, ``On the complexity of nonnegative matrix factorization,''
  \emph{SIAM Journal on Optimization}, vol.~20, no.~3, pp. 1364--1377, 2009.

\bibitem{packer2002np}
A.~Packer, ``Np-hardness of largest contained and smallest containing simplices
  for v-and h-polytopes,'' \emph{Discrete and Computational Geometry}, vol.~28,
  no.~3, pp. 349--377, 2002.

\bibitem{huang2016flexible}
K.~Huang, N.~D. Sidiropoulos, and A.~P. Liavas, ``A flexible and efficient
  algorithmic framework for constrained matrix and tensor factorization,''
  \emph{IEEE Transactions on Signal Processing}, vol.~64, no.~19, pp.
  5052--5065, 2016.

\bibitem{xu2013block}
Y.~Xu and W.~Yin, ``A block coordinate descent method for regularized
  multiconvex optimization with applications to nonnegative tensor
  factorization and completion,'' \emph{SIAM Journal on imaging sciences},
  vol.~6, no.~3, pp. 1758--1789, 2013.

\bibitem{miao2007endmember}
L.~Miao and H.~Qi, ``Endmember extraction from highly mixed data using minimum
  volume constrained nonnegative matrix factorization,'' \emph{IEEE Trans.
  Geosci. Remote Sens.}, vol.~45, no.~3, pp. 765--777, 2007.

\bibitem{razaviyayn2013unified}
M.~Razaviyayn, M.~Hong, and Z.-Q. Luo, ``A unified convergence analysis of
  block successive minimization methods for nonsmooth optimization,''
  \emph{SIAM Journal on Optimization}, vol.~23, no.~2, pp. 1126--1153, 2013.

\bibitem{fu2016robusticassp}
X.~Fu, W.-K. Ma, K.~Huang, and N.~D. Sidiropoulos, ``Robust volume
  minimization-based matrix factorization via alternating optimization.'' in
  \emph{ICASSP}, 2016, pp. 2534--2538.

\bibitem{berman2004ice}
M.~Berman, H.~Kiiveri, R.~Lagerstrom, A.~Ernst, R.~Dunne, and J.~Huntington,
  ``{ICE}: A statistical approach to identifying endmembers in hyperspectral
  images,'' \emph{IEEE Trans. Geosci. Remote Sens.}, vol.~42, no.~10, pp.
  2085--2095, 2004.

\bibitem{li2012collaborative}
J.~Li, J.~Bioucas-Dias, and A.~Plaza, ``Collaborative nonnegative matrix
  factorization for remotely sensed hyperspectral unmixing,'' in \emph{Proc.
  IGARSS 2012}, 2012, pp. 3078--3081.

\bibitem{Reilly2005}
K.~Rahbar and J.~Reilly, ``A frequency domain method for blind source
  separation of convolutive audio mixtures,'' \emph{IEEE Speech Audio
  Process.}, vol.~13, no.~5, pp. 832 -- 844, Sep. 2005.

\bibitem{sidiropoulos2017tensor}
N.~D. Sidiropoulos, L.~De~Lathauwer, X.~Fu, K.~Huang, E.~E. Papalexakis, and
  C.~Faloutsos, ``Tensor decomposition for signal processing and machine
  learning,'' \emph{IEEE Transactions on Signal Processing}, vol.~65, no.~13,
  pp. 3551--3582.

\bibitem{ACDC}
A.~Yeredor, ``Non-orthogonal joint diagonalization in the least-squares sense
  with application in blind source separation,'' \emph{IEEE Trans. Signal
  Process.}, vol.~50, no.~7, pp. 1545 --1553, Jul. 2002.

\bibitem{Nion2010}
D.~Nion, K.~N. Mokios, N.~D. Sidiropoulos, and A.~Potamianos, ``Batch and
  adaptive {PARAFAC}-based blind separation of convolutive speech mixtures,''
  \emph{IEEE Audio, Speech, Language Process.}, vol.~18, no.~6, pp. 1193
  --1207, Aug. 2010.

\bibitem{ziehe2004ffdiag}
A.~Ziehe, P.~Laskov, G.~Nolte, and K.-R. M\"{u}ller, ``A fast algorithm for
  joint diagonalization with non-orthogonal transformations and its application
  to blind source separation,'' \emph{Journal of Machine Learning Research},
  pp. 777--800, 2004.

\bibitem{vandaele2016efficient}
A.~Vandaele, N.~Gillis, Q.~Lei, K.~Zhong, and I.~Dhillon, ``Efficient and
  non-convex coordinate descent for symmetric nonnegative matrix
  factorization,'' \emph{IEEE Tran. Signal Process.}, vol.~64, no.~21, pp.
  5571--5584, 2016.

\bibitem{shi2017inexact}
Q.~Shi, H.~Sun, S.~Lu, M.~Hong, and M.~Razaviyayn, ``Inexact block coordinate
  descent methods for symmetric nonnegative matrix factorization,'' \emph{IEEE
  Tran. Signal Process.}, vol.~65, no.~22, pp. 5995--6008, 2017.

\end{thebibliography}
	

%
	\appendices
	
	\clearpage
	
	\section{Ellipsoid Volume Minimization for Separability-Based NMF}
	In \cite{mizutani2014ellipsoidal}, Mizutani introduced the {\it ellipsoid-volume minimization-based NMF}, whose intuition is shown in Fig.~\ref{fig:e-volmin} using a case where $M=R=2$. This method finds a minimum-volume ellipsoid that is centered at the origin and encloses the data columns $\bm x_1,\ldots,\bm x_N$ and their `mirrors' $-\bm x_1,\ldots,-\bm x_N$. Then, under the separability condition, the minimum-volume enclosing ellipsoid `touches' the data cloud at $ \pm \bm W_\natural(:,r)$ for $r=1,\ldots,R$. The identification criterion is formulated as follows:
	\begin{figure}[h!]
		\centering
		\includegraphics[width=0.45\linewidth]{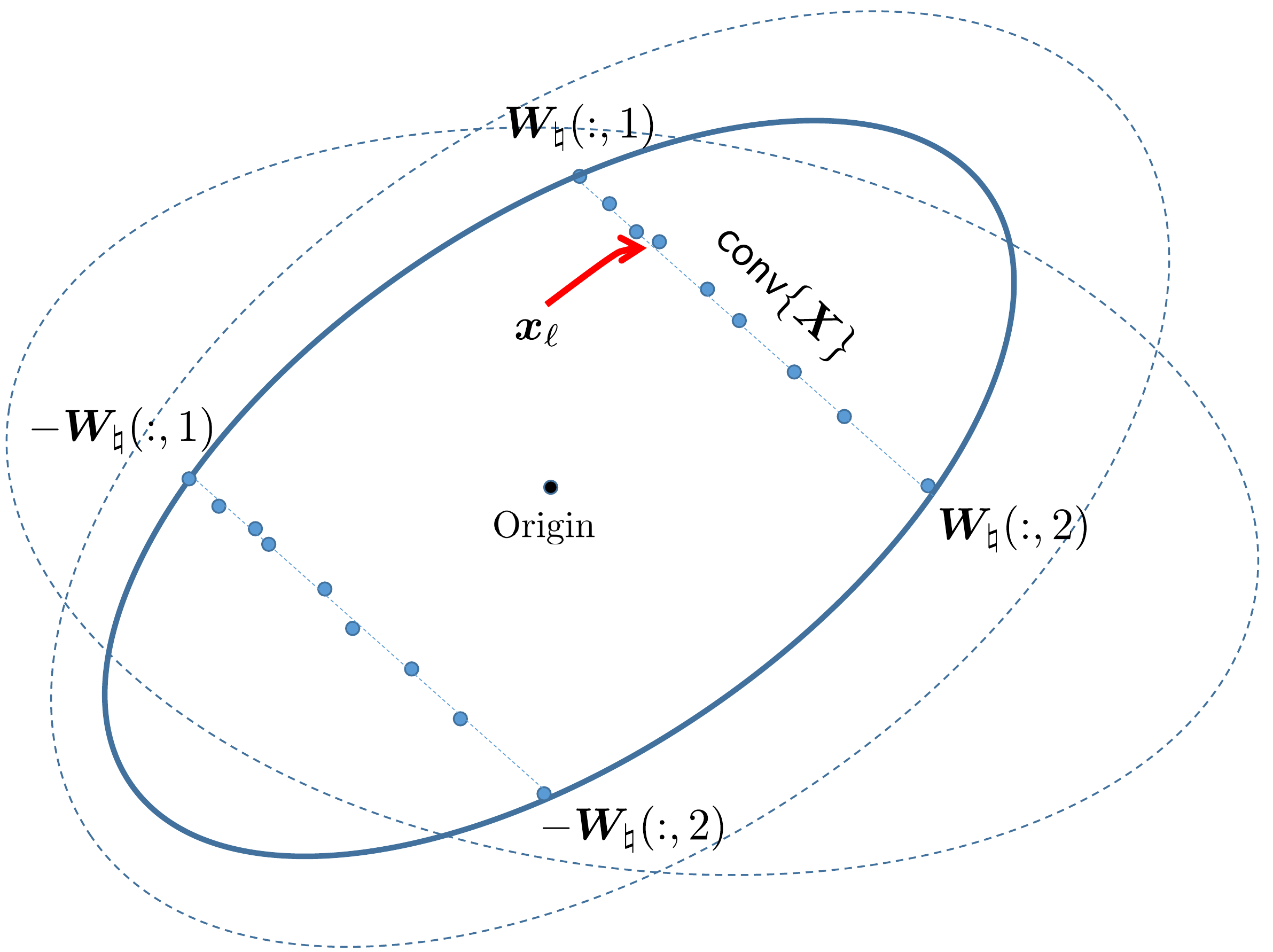}
		\caption{Geometric illustration of the ER method when $M=R=2$. Note that $\H_\natural \bm 1 =\bm 1$ is assumed and that the solid circle represent the minimum-volume ellipsoid that is sought.}
		\label{fig:e-volmin}
	\end{figure}
	\begin{subequations}\label{eq:e-volmin}
		\begin{align}
		\minimize_{{\bm L}}&~-\log\det({\bm L})\\
		{\rm subject~to}&~{\bm L}\succ {\bm 0}\\
		&~{\bm x}_\ell^\T\bm L {\bm x}_\ell \leq 1, ~\forall \ell \label{eq:elip}
		\end{align}
	\end{subequations}
	Here, $\x^\T\bm L\x \leq 1$ represents an ellipsoid in the $R$-dimensional space, the constraint \eqref{eq:elip} means that all the $\x_\ell$'s are enclosed in the ellipsoid parametrized by $\bm L$, and
	$-\log\det(\bm L)$ measures the volume of the ellipsoid.
	Note that Problem~\eqref{eq:e-volmin} is very well studied in convex geometry and thus many off-the-shelf algorithms can be directly applied for NMF.

	By assuming that $\H_\natural$ is row-stochastic and $M=R$, it was shown that the active points that satisfy ${\bm x}_\ell^\T{\bm L}_\star {\bm x}_\ell=1$ are those $\bm x_{\ell_r}={\bm W}_\natural(:,{\ell_r})$ for $r=1,\ldots, R$, where ${\bm L}_\star$ is the optimal solution to Problem~\eqref{eq:e-volmin}---which is consistent with our observation in Fig.~\ref{fig:e-volmin}. This algorithm, referred to as the {\it ellipsoid rounding} (ER) algorithm, is also robust to noise. One possible downside of ER is that it only works with square $\W_\natural$, which means that when $M>R$, a dimensionality reduction procedure has to be employed, resulting in a two-stage approach, which could be prone to error propagation. Working with the reduced-dimension $\W_\natural$ also loses the opportunity to incorporate prior information of the original $\W_\natural$, e.g., sparsity and nonnegativity, which are crucial for performance enhancement in noisy cases. 
	Dimensionality reduction also requires an accurate knowledge of $R$, which the self-dictionary methods do not need---in fact, the self-dictionary methods can even help estimate $R$ in noisy cases \cite{fu2015self}.
	The upshot of ER is that it only has $R^2$ optimization variables, which is considerably smaller than $N^2$ in many real-world applications (e.g., in topic modeling $R$ is the number of topics and $N$ is usually the number of documents or number of words in a dictionary), which may potentially lead to memory-wise more economical algorithms.

	\section{SPA, Greedy Pursuit, and Simplex Volume Maximization}
	
\begin{figure}
		\centering
		\includegraphics[width=0.65\linewidth]{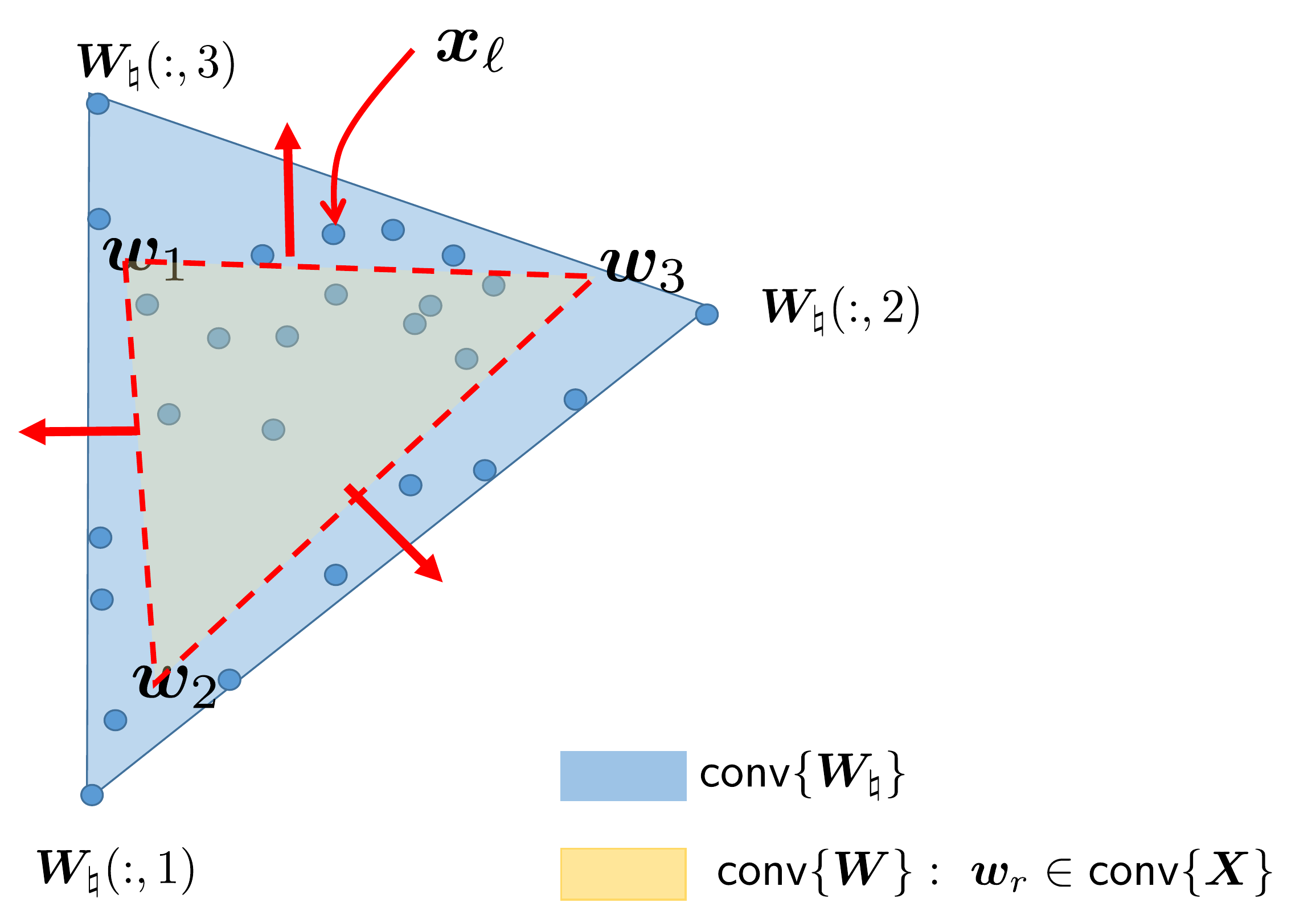}
		\caption{{ Intuition behind simplex volume maximization: the ground-truth convex hull $\conv{\W_\natural}$ is found via finding the maximum-volume simplex spanned by $\W$ which is enclosed in the data convex hull $\conv{\X}$.}}
		\label{fig:simpvolmax}
	\end{figure}

	Besides the pure algebraic derivation of SPA, SPA can be derived from many other ways. Consider the self-dictionary sparse optimization problem:
	\begin{align*}
	\minimize_{{\bm C}}&~\|\C\|_{{\rm row}-0}\\
	{\rm subject~to}&~\X=\X\C\\
	&~{\bm C}\geq{\bm 0},~ {\bm 1}^\T{\bm C}={\bm 1}^\T,
	\end{align*}
	One way to handle the problem is to apply the so-called simultaneous orthogonal matching pursuit (SOMP) algorithm in compressive sensing; i.e., one can find out an `active atom' in the dictionary $\X$ by the following:
	\begin{align}\label{eq:somp}
	\hat{\ell}_1 =\max_{\ell=1,\ldots,N} \|\X^\T\x_\ell\|_q ,    
	\end{align}
	since an active atom ($\x_{\ell_r}$ for $\ell_r\in\varLambda$) in the dictionary is likely to exhibit large correlation with the data $\X$. By setting $q=\infty$, one can see that
	\begin{equation*}
	\begin{aligned}
	\max_{\ell=1,\ldots,N} \|\X^\T\x_\ell\|_\infty &= \max_{\ell\in\{1,\ldots,N\}} \max_{n\in\{1,\ldots,N\}}|\x_n^\T\x_\ell|\\
	&  = \max_{\ell\in\{1,\ldots,N\}} \|\x_\ell\|^2_2,  
	\end{aligned}
	\end{equation*}
	where the last equality is by the Cauchy-Shwartz inequality, and the above is exactly the first step of SPA. In fact, SPA and SOMP are completely equivalent when $q=\infty$ in \eqref{eq:somp} \cite{fu2015self}. Another way to derive SPA is to consider the so-called simplex volume maximization criterion; i.e., the vertices of $\conv{\X}$ can be found by finding a maximum-volume simplex that is inside the data cloud, if separability is satisfied. The intuition behind the formulation is shown in Fig.~\ref{fig:simpvolmax}, which is sometimes referred to as the {\it Winter's belief} \cite{VMAX} in the remote sensing literature.
	The optimization problem can be expressed as
	\begin{align*}
	\maximize_{\W}&~\det(\W^\T\W)\\
	{\rm subject~to}&~\W=\X\bm C,~\bm C\geq{\bm 0},{\bm 1}^\T\bm C=\bm 1^\T,
	\end{align*}
	where the objective is a measure of the simplex spanned by the columns of $\W$ and the constraints means that $\W\in\conv{\X}$. 
	The above optimization problem can be solved by a simple greedy algorithm---which turns out to be exactly SPA;
	see derivations in \cite{VMAX,Ma2013,arora2012practical}.
	
	\section{More Applications: Blind Separation of Non-Stationary and Stationary Sources}
	Blind source separation (BSS) is a classic signal processing problem. A simple instantaneous mixture case of BSS admits the following signal model:
	\begin{equation}\label{eq:ybss}
	\bm y[t] =\A\bm s[t],~t=0,1,2,\ldots 
	\end{equation}  
	where $ y_{m}[t]$ (the $m$th element of the vector $\bm y[t]\in\mathbb{C}^M$) denotes the $m$th received signal (by sensor $m$) at time $t$, $s_{r}[t]$ (the $r$th element of the vector $\bm s[t]\in\mathbb{C}^R$) denotes the $r$th source at time $t$ and $\A\in\mathbb{C}^{M\times R}$ is the mixing matrix. 
	Here we assume the signals and the mixing system are both complex-valued since BSS are commonly applied in the complex domain (e.g., the frequency domain in convolutive speech mixture separation \cite{Reilly2005} and communication signal separation \cite{fu2015power}).
	With the observations $\bm y[t]$ for $t=1,\ldots, T$, the goal of BSS is to estimate $\A$ and $\bm s[t]$ for $t=1,\ldots,T$. 
	BSS finds many applications in speech and audio processing, array processing, and communications.
	
	
	\subsubsection{Speech Separation} To recast the BSS problem as an NMF problem, consider a case where $s_{r}[t]$ is a zero-mean piecewise stationary (or quasi-stationary---meaning that the source is stationary within a short time interval but the statistics vary from interval to interval), which is a commonly used model for speech sources. Consequently, one can approximate the following local correlation matrix (e.g., via sample averaging):
	\[ \bm R_\ell = \mathbb{E}\{ \bm y[t]\bm y[t]^H \}=\A \bm D_n \A^H,~t\in \text{interval $\ell$}, \]
	and $\bm D_n=\mathbb{E}\{ \bm s[t]\bm s[t]^H \}$ is the local { covariance of the sources (assuming zero-mean sources)} within interval $n$. Note that $\bm D_\ell$ is diagonal if the sources are zero-mean and are uncorrelated, and $\bm D_\ell(r,r)$ is the {\it local average power} of source $r$ within interval $n$---which is nonnegative.
	Then, by forming a matrix $\X=[\x_1,\ldots,\x_N]$ where $\x_\ell={\rm vec}(\bm R_\ell)$, we have 
	\[ \X =\W\H^\T, \]
	where $\W= \A^\ast \odot \A$ and $\H(\ell,:)^\T = {\rm diag}(\bm D_\ell)$, where ${\rm diag}(\bm Z)$ means taking the diagonal of $\bm Z$ and making it a column vector (as in {\sc Matlab}) and $\odot$ is the Khatri-Rao product \cite{sidiropoulos2017tensor}. Here, one can see that $\H$ is a nonnegative factor, and the factorization model above is considered { NMF in a broader sense, as discussed before}. 
	Factoring $\X$ to $\W$ and $\H$ identifies the mixing system $\A$\footnote{This is because $\W(:,r)=\A(:,r)^\ast\otimes \A(:,r)$ where $\otimes$ denotes the Kronecker product \cite{sidiropoulos2017tensor} and thus $\A(:,r)$ can be estimated from $\W(:,r)$ via eigendecomposition \cite{fu2015blind}.}, which is usually a critical stepping stone towards source separation.

	\begin{figure}
		\centering
		\includegraphics[width=0.75\linewidth]{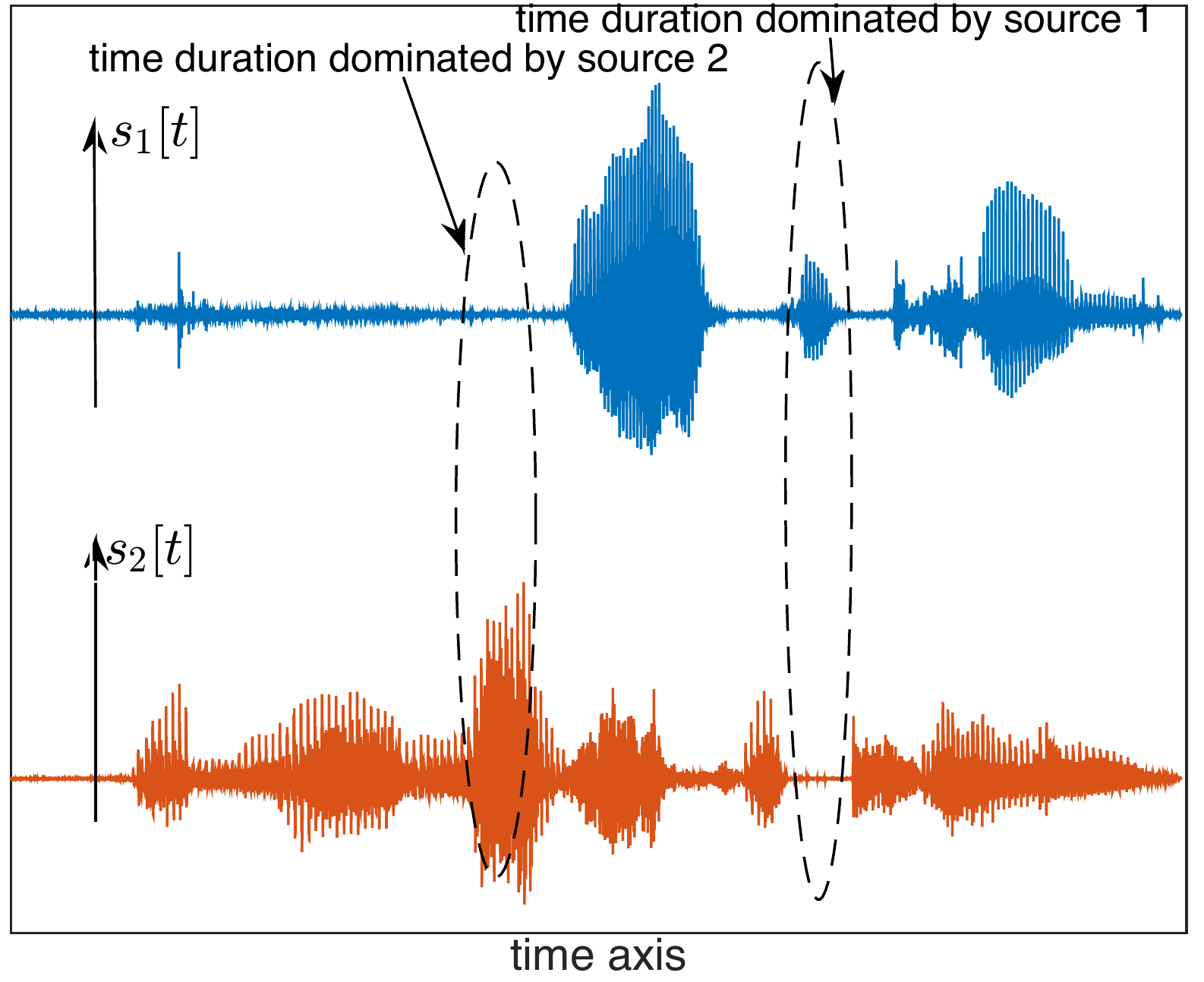}
		\caption{The local dominance phenomenon in speech separation, which gives rise to the spearability condition if the problem is recast as an NMF problem as described.}
		\label{fig:twospeech}
	\end{figure}

\begin{figure}
		\begin{minipage}{.45\linewidth}
			\centering
			\includegraphics[width=1\linewidth]{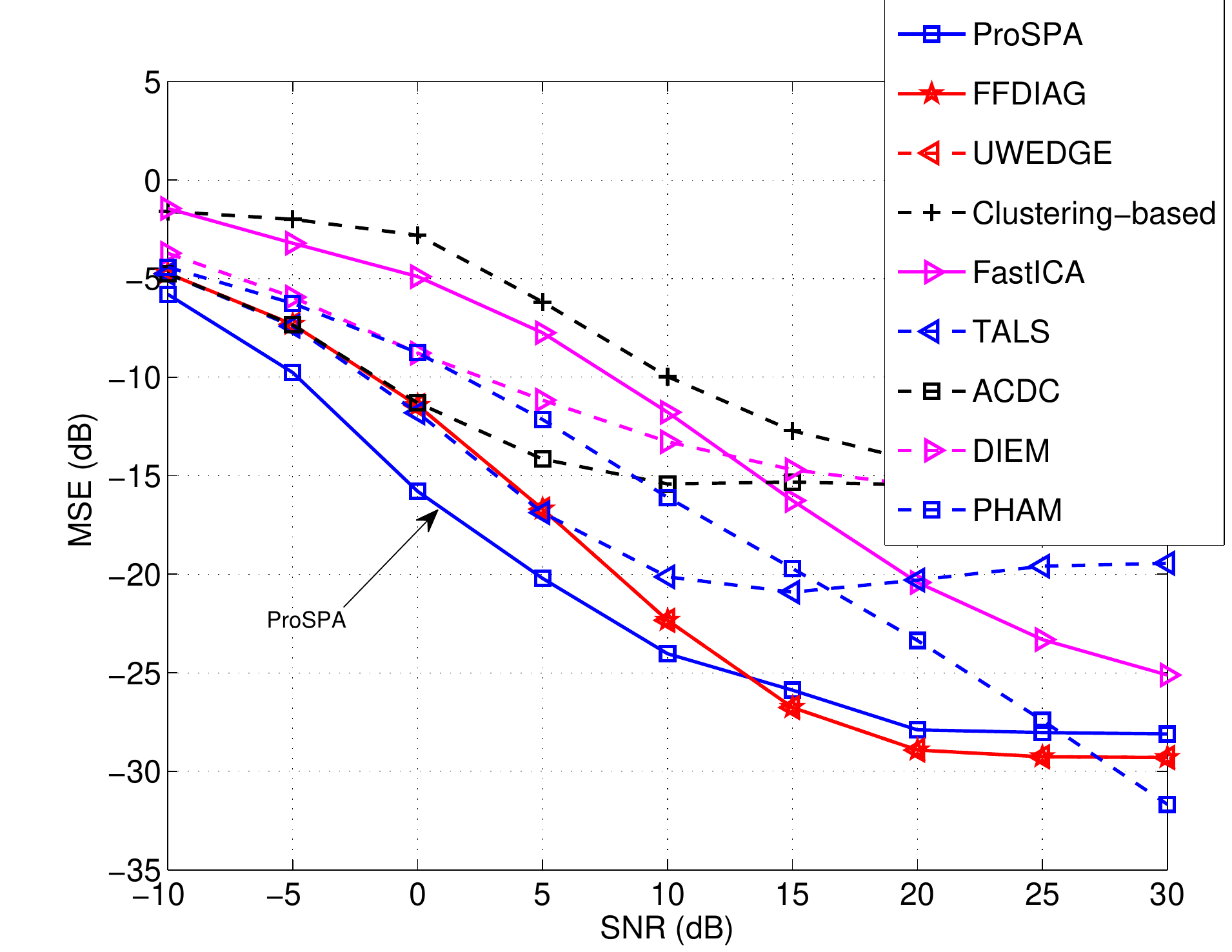}
		\end{minipage}
		\begin{minipage}{.45\linewidth}
			\centering
			\includegraphics[width=1\linewidth]{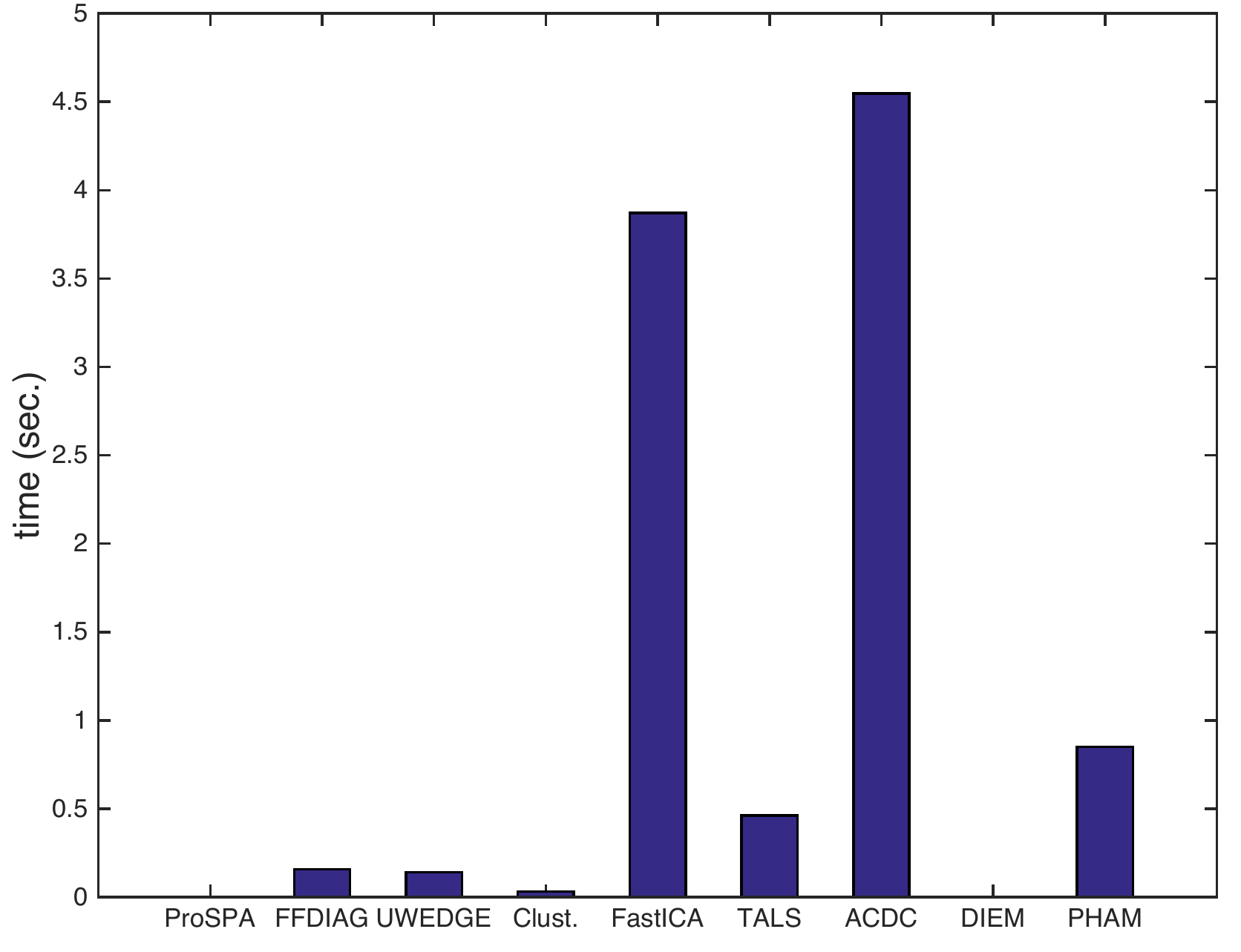}
		\end{minipage}
		\caption{{ The MSE (measured at SNR=10dB) of the estimated $\A$ and average runtime performance in the speech separation problem, respectively. Here, $5$ sources with a 6-second duration are used; 6 sensors are employed.  `{\sf ProSPA}' \cite{fu2015blind} is an NMF based approach under the local correlation formulation. All the baselines are competitive source separation methods. {\sf ProSPA} exhibits 50 times faster execution time compared to the competitor that has similar accuracy (FFDIAG) \cite{ziehe2004ffdiag}.
				More details can be found in \cite{fu2015blind}.} }
		\label{fig:allalgosnrdefense}
	\end{figure}

	Although many sophisticated algorithms exist for source separation (e.g., independent component analysis (ICA) \cite{Comon1994}, joint diagonalization \cite{ACDC}, and tensor decomposition \cite{Nion2010}), using NMF techniques (especially separability-based NMF) can further boost the performance---in particular, the runtime performance.
	In speech separation, separability approximately holds for $\H_\natural$ if there exists short time durations where only one speaker is speaking. This makes sense since there are a lot of pauses and stops during utterances; see Fig.~\ref{fig:twospeech}. SPA and its variants can be applied to identify $\W_\natural$ (and thus $\A$). Since SPA is a Gram-Schmidt-like algorithm, it helps unmix the speech signal in nearly real time (cf. the `{\sf ProSPA}' algorithm in Fig.~\ref{fig:allalgosnrdefense}, which is based on SPA and judicious preprocessing). When there are many speakers talking simultaneously, using volume minimization initialized by ProSPA gives more promising results \cite{fu2015blind}, since now the separability condition is likely violated.

	\subsubsection{Power Spectra Separation} 
	In some wireless communication systems such as cognitive radio, it is desired to establish spectral awareness---i.e., to build up the knowledge about the spectral (un)occupancy so that transmission opportunities can be identified.
	
	Consider the received signals at the receivers of a cognitive radio system, which can be represented as in \eqref{eq:ybss}. Note that now the sources are transmitted signals from some primary or licensed users, and are usually assumed to be zero-mean and {\it wide-sense stationary}. The goal is to identify the power spectrum of each individual source, so that interference avoidance strategies can be designed for the cognitive transmitters.
	In such cases, we can use the following idea to construct an NMF model. We first compute the auto-correlation of $\bm y_{m}[t]$, which is
	\[  \bm r[\tau] = \mathbb{E}\left[\bm y^\ast [t]\circledast \bm y[t-\tau] \right]=\left(\A^\ast \circledast \A\right) \bm q[\tau],~\forall \tau, \]
	if the sources are spatially uncorrelated, where $\circledast$ is the Hadamard product, $\bm q[\tau]\in\mathbb{C}^R$ and $q_r[\tau]=\mathbb{E}[s^\ast_{r}[t]s_{r}[t-\tau]]$ is the autocorrelation of the $r$th source.
	Taking Fourier transform of each element in $\bm r[\tau]$, we have
	\[  \bm z(\omega) =   \left(\A^\ast \circledast \A\right)  \bm p(\omega),   ~\forall \omega\]
	where ${\bm p}(\omega)\in\mathbb{R}^R$ contains the power spectral density (PSD) of the $R$ source signals at frequency $\omega$. By putting the above in a matrix $\X$ whose columns correspond to different (discretized) frequencies, we have a model $\X=\W\H^\T$ where $\W= \left(\A^\ast \circledast \A\right) \geq {\bm 0}$ and $\H(:,r)\geq {\bm 0}$ the PSD of the $r$th source sampled at $N$ frequencies. 

\begin{figure}
		\centering
		\includegraphics[width=0.95\linewidth]{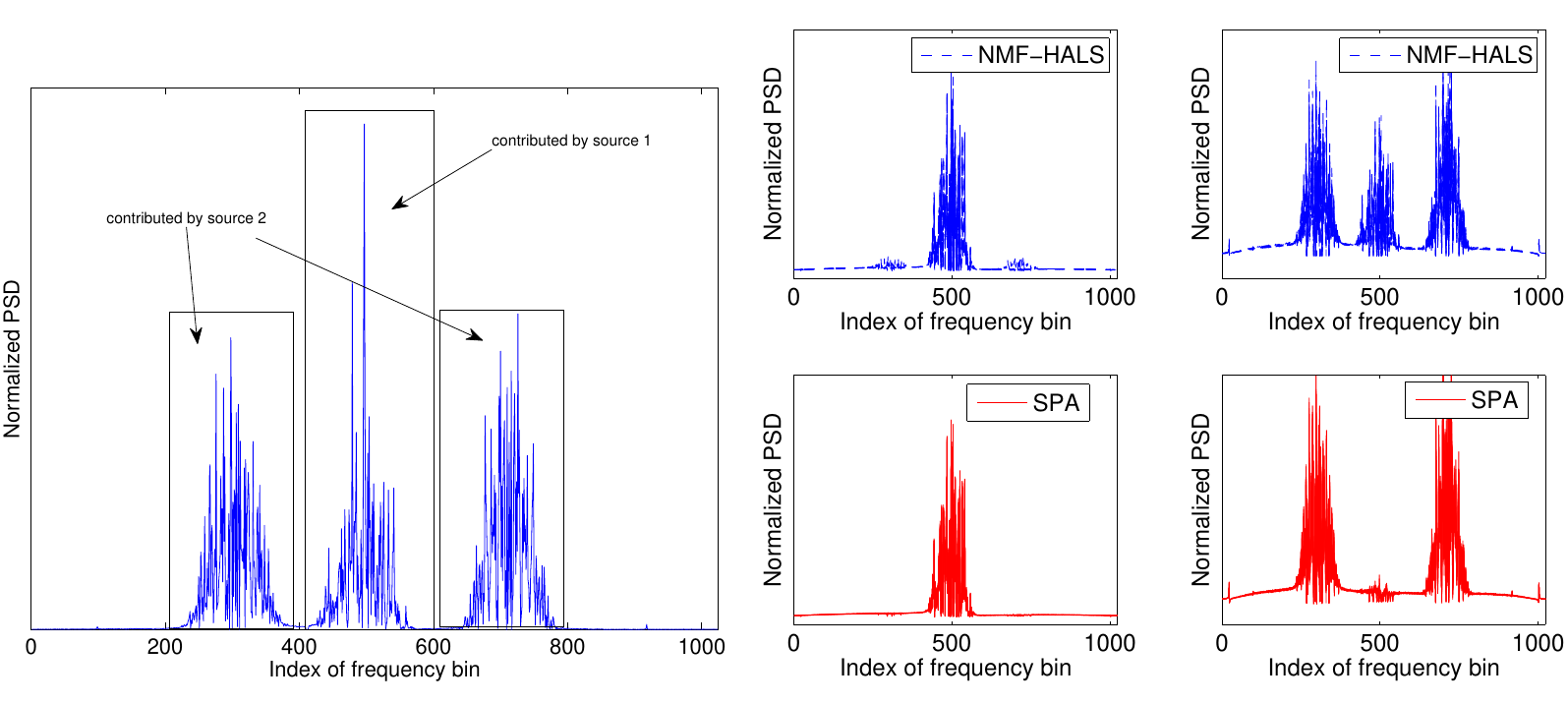}
		\vspace{-.55cm}
		\caption{Different outcomes for a spectra separation problem in cognitive radio. Two primary transmitters and two cognitive receivers are used in this experiment. The data was measured in the Communications Laboratory at University of Minnesota with USRP software defined radios; see \cite{fu2015power}. Left: mixture of the received power spectra observed at receiver 1. Right: the unmixed spectra by different NMF approaches. Figure reproduced (permission will be sought) from \cite{fu2015power}, $\copyright$ IEEE.}
		\label{fig:halsvsspa}
	\end{figure}

	
	Some experiment results are shown in Fig.~\ref{fig:halsvsspa}, where different NMF approaches are applied to separate power spectra of two sources. The received signals were measured in the communications laboratory at University of Minnesota. Fig.~\ref{fig:halsvsspa} shows the spectra separation results using the HALS algorithm (plain NMF fitting) \cite{cichocki2009fast} and SPA, respectively. One can see that the plain NMF does not work, since $\W$ here is dense and violates the necessary condition for plain NMF identifiability (to be precise,  $\W_\natural$ results from the Hadamard product of the wireless channel matrix and its conjugate (i.e.,$\W_\natural=\A^\ast \circledast\A$), which is nonnegative. However, it is in general dense almost surely since  $\A$ is well-approximated by a random matrix whose elements are drawn from some continuous distribution). On the other hand, SPA does not need to assume any zero patterns in the wireless channel, and the considered case obviously satisfies the separability condition (i.e., the power spectra of the two transmitters both have exclusive frequency bins that the other does not use).
	
	\section{Tri-factorization Model and Identifiability}
	The results that were developed for identifiability of VolMin and determinant minimization can be generalized to cover certain tri-factorization models, which find important applications in topic modeling, community detection, and HMM identification, as we have seen. Consider the following signal model:
	\[ \X = \C_\natural \E_\natural\C_\natural^\T .\]
	One way to handle the factorization problem is to first treat $\C_\natural\E_\natural=\W_\natural$ and $\H_\natural =\C_\natural$ and directly apply any of the previously introduced methods. However, this would be sup-optimal in noisy cases, since the structure of the data matrix is not yet fully exploited. In \cite{huafusid2016nips,fu2018anchor}, the authors proposed the following identification criterion:
	\begin{subequations}\label{eq:cec_v1}
		\begin{align}
		\minimize_{{\bm C},\E}&~|\det\left(\E\right)|\\
		{\rm subject~to}&~\X =\C\E\C^\T\\
		&~\bm 1^\T\C=\bm 1^\T,~\C\geq\bm 0.
		\end{align}
	\end{subequations}
	It was also shown in \cite{huafusid2016nips,fu2018anchor} that: 
	
	\vspace{.2cm}
	\noindent
	\boxed{
		\begin{minipage}{.98\linewidth}
			{\it Solving \eqref{eq:cec_v1} leads to $\E_\star =\bm \Pi\E_\natural\bm \Pi^\T$ and $\bm C_\star=\C_\natural\bm \Pi$ if $\C_\natural$ satisfies the sufficiently scattered condition and $\C_\natural$ and $\E_\natural$ have full rank, where $\bm \Pi$ is a permutation matrix.} 
		\end{minipage}
	}
	\vspace{.2cm}
	%

	\noindent
	The above can be proven utilizing the result that we have for the determinant-based bi-factorization criterion, by simply looking at the identifiability of $\C_\natural$ in the column or row subspace of $\X$. However, the practical implication is significant. For example, the identifiability of \eqref{eq:cec_v1} allows us to use the following implementation with identifiability support:
	\begin{subequations}\label{eq:fit_v1}
		\begin{align}
		\minimize_{{\bm C},\E}&~\|\X-\C\E\C^\T\|_F^2 +\lambda |\det(\E)| \\
		{\rm subject~to}&~\bm 1^\T\C=\bm 1^\T,~\C\geq \bm 0,
		\end{align}
	\end{subequations}
	Problem~\eqref{eq:fit_v1} accounts for noise and it directly works with the model $\X=\C\E\C^\T$ rather than treating $\C\E=\W$ and ignoring the symmetry.
	This way, one can also easily incorporate prior information on $\E_\natural$ and $\C_\natural$ (e.g., $\E_\natural$ is nonnegative and positive semidefinite in topic modeling) to enhance performance in noisy cases.
	In addition, $\E$ is usually a small matrix of size $R\times R$, and dealing with $\det(\E)$ can be relatively easier than dealing with $\det(\W^\T\W)$ where $\W=\C\E$ in algorithm design.		
	
	\section{Subspace Methods for Symmetric NMF}
	As we have seen, in some applications such as topic modeling and HMM identification, symmetric NMF naturally arises.
	The associated models, e.g., $\X=\W_\natural\W_\natural^\T$ and $\X=\C_\natural\E_\natural\C_\natural^\T$, pose very hard optimization problems, though. Unlike the asymmetric case, where the subproblems w.r.t. each of the block variables are convex or can be effectively approximated by convex problems, now the optimization criteria such as
	\begin{equation}\label{eq:symm}
	\minimize_{\bm W\geq{\bm 0}}~\|\X-\W\W^\T\|_F^2
	\end{equation}
	or Problem~\eqref{eq:fit_v1}, are fourth-order optimization problems.
	In recent years, much effort has been invested to handle Problem~\eqref{eq:symm}, e.g., using coordinate descent \cite{vandaele2016efficient} and judiciously designed BSUM \cite{shi2017inexact}. These are viable approaches but are considered `heavy' in implementation since they are essentially optimizing the problem w.r.t. each element (or row) of $\W$ cyclically. For Problem~\eqref{eq:fit_v1}, the problem is even harder since $\det(\E)$ is involved and more variables need to be optimized.
	
	An interesting way to handle symmetric factorization has a strong flavor of array processing, in particular, the subspace methods.
	Using the idea of subspace methods, the problem of identifying both the models $\X=\W_\natural\W_\natural^\T$ and $\X=\C_\natural\E_\natural\C_\natural^\T$ can be recast as problems that are easier to deal with---without losing identifiability. To see this, let us consider the tri-factorization model in \eqref{eq:cec_v1}, i.e.,
	$ \X=\C_\natural\E_\natural\C_\natural^\T.$ The idea is simple: In many applications $\X=\B\B^\T$ always holds since $\X\succeq \bm 0$  (e.g., the second-order statistics-based topic mining), where $\B\in\mathbb{R}^{M\times R}$ is a square root of $\X$. Then, it is easy to see that $\B^\T=\G_\natural\C_\natural^\T$ for a certain invertible $\bm G_\natural$ if ${\rm rank}(\X)={\rm rank}(\C_\natural)={\rm rank}(\E_\natural)=R$, since ${\rm range}(\B)={\rm range}(\X)={\rm range}(\C_\natural)$ under such circumstances. Therefore, the algorithms for determinant minimization can be directly applied for identifying $\G_\natural$ and $\C_\natural$---this is the idea in \cite{huafusid2016nips,fu2018anchor}.
	
	For the model $\X=\W_\natural\W_\natural^\T$, the same arguments still hold---i.e., $\B = \W_\natural \G_\natural$ holds for a certain invertible $\G_\natural$.
	Therefore, the determinant minimization-based can also be applied to identify $\W_\natural$, if $\W_\natural$ is sufficiently scattered. Nevertheless, since under the model $\X=\W_\natural\W_\natural^\T$, $\G_\natural\G_\natural^\T=\bm I$ holds, an even simpler reformulation can be employed:
	\[  \minimize_{\G\G^\T={\bm I},\W\geq{\bm 0}}~\|\B -\W\G\|_F^2,\]
	which can be handled by two-block coordinate descent, and each subproblem admits a very simple solution \cite{huang2014non}; to be more specific, the updates are as follows:
	\begin{align*}
	\bm G^{(t+1)} &\leftarrow \bm U\bm V^\T,\quad \U\bm \Sigma\bm V^\T \leftarrow {\rm svd}((\W^{(t)})^\T\B)\\
	\bm W^{(t+1)}&\leftarrow \max\{ \bm B(\bm G^{(t+1)})^\T,\bm 0 \},
	\end{align*}
	which are both easy to implement.
	
	The viability of the described algorithms for $\X=\C_\natural\E_\natural\C_\natural^\T$ and $\X=\W_\natural\W_\natural^\T$ are results of the fact that the identifiability of $\C_\natural$ and $\W_\natural$ in the two models has nothing to do with the structure of the respective $\G_\natural$'s, if $\C_\natural$ and $\W_\natural$ are sufficiently scattered. In other words, for complicated models, if one can identify ${\rm range}(\C_\natural)$ or ${\rm range}(\W_\natural)$ in a reliable way, the identifiability of the associated nonnegative factors are not lost---this insight can substantially simplify the identification procedure in some applications.

\end{document}